\documentclass[%
 reprint, %linenumbers,
 superscriptaddress,
 amsmath,amssymb,
 aps, %physrev,
 pra,
]{revtex4-2}

\usepackage{graphicx}% Include figure files
\usepackage{dcolumn}% Align table columns on decimal point
\usepackage{bm}% bold math
\usepackage{hyperref}% add hypertext capabilities
\hypersetup{
    colorlinks=true,     % Enable colored links
    linkcolor=blue,      % Color for internal links (e.g., table of contents)
    citecolor=blue,      % Color for citation links (e.g., references)
    urlcolor=blue,       % Color for URLs (e.g., web links)
    }
\usepackage[T1]{fontenc}
\usepackage{longtable}

\begin{document}

\preprint{APS/123-QED}

\title{\textbf{Electron scattering from polar hydrides using relativistic optical potential method} 
}% 
% Relativistic optical-potential calculations for electron scattering from polar hydrides

\author{Sudhanshu Arya}
% \email{22dr0229@iitism.ac.in}
\affiliation{%
Atomic and Molecular Physics Laboratory, Department of Physics, Indian Institute of Technology (ISM) Dhanbad, Jharkhand 826004, India
}%

\author{Bobby Antony}%
\email{Contact author: bobby@iitism.ac.in}
\affiliation{%
Atomic and Molecular Physics Laboratory, Department of Physics, Indian Institute of Technology (ISM) Dhanbad, Jharkhand 826004, India
}%

\date{\today}% It is always \today, today,
             %  but any date may be explicitly specified

\begin{abstract}
We extend the relativistic spherical complex optical-potential method with group additivity developed in our recent work [S. Arya and B. Antony, \textit{RSC Adv.} \textbf{16}, 13548--13558 (2026)] to molecules with permanent dipole moments. The short-range electronic collision is described by a central Dirac partial-wave calculation, while the missing anisotropic long-range dipole interaction is restored through a rotationally resolved first-Born contribution. The rotational thresholds, state-to-state transition-dipole strengths, and thermal populations are obtained from the HITRAN2024 spectroscopic database and thermally reweighted at the adopted rotational temperature. Both excitation and superelastic
de-excitation channels are included. Electron scattering from benchmark polar hydrides ($\rm H_2O$, $\rm H_2S$, $\rm NH_3$, and $\rm PH_3$) is investigated over the incident-energy range 0.1--10,000 eV. We report vibrationally elastic, differential, integral, and momentum-transfer cross sections, together with a total cross section that also contains electronically inelastic loss from the quasifree absorption potential. The calculations reproduce the broad experimental and recommended trends over a wide energy range. The largest deviations are confined mainly to the low-energy and resonance-sensitive regions. Overall, the agreement improves substantially from the tens-of-eV region upward. The method therefore retains the low computational cost and wide energy coverage of the optical-potential approach while adding the long-range rotational physics needed for polar molecules.
\end{abstract}

%\keywords{Suggested keywords}%Use showkeys class option if keyword
                              %display desired
\maketitle

%\tableofcontents

% ============================================================================
\section{\label{sec:intro}Introduction}
% ============================================================================
Electron-molecule collision cross sections are basic inputs in plasma kinetics, gas discharges, atmospheric and planetary models, radiation transport, and electron-driven chemistry \cite{JonathanJPB2024}. In such systems, the electron-energy distribution is linked to elastic momentum transfer and to inelastic production of excited states, ions, and radicals \cite{WinsteadAP1996,RosettaA2017}. Hence, uncertainties in these cross sections propagate directly into swarm parameters, reaction rates, and plasma-chemical predictions \cite{BrungerPR2002,BartschatPNAS2016}. This requirement is especially demanding when a model must span several orders of magnitude in incident energy. The difficulty is that no single detailed close-coupling calculation is equally convenient from the threshold to the keV range for all targets. Close-coupling methods are the best reference near thresholds and resonances, but optical-potential and additivity approaches are much cheaper when a broad energy range and several observables are needed \cite{Madison1996,JonathanPR2010,BartschatPRP2017,ZammitJPB2017,AryaJAP2025}.

Polar molecules constitute a qualitatively more difficult class. A neutral molecule with a permanent electric dipole moment $\bm \mu_D$ produces an asymptotic interaction proportional to $r^{-2}$, much longer ranged than the induced-dipole polarization interaction, which behaves as $-\alpha_d/(2r^4)$ \cite{TakayanagiAAMP1970,NorcrossAAMP1982,ItikawaTCA2000}. Spherical averaging can represent the short-range static, exchange, correlation-polarization, and absorptive interactions efficiently, but it removes the anisotropic interaction from the permanent molecular dipole. That missing long-range coupling matters most at low incident energies and small angles, and it can dominate dipole-allowed rotational excitation. A useful broad-range model intended for polar targets therefore must restore this long-range rotational physics without giving up the numerical efficiency of the central optical-potential calculation. The goal of the present work is to provide that common description.

In our recent work, we developed a relativistic spherical complex optical-potential formulation \cite{AryaRSCA2026}, referred to here as Dirac--SCOP, in which group-additive relativistic atomic densities are transformed into an effective central molecular potential and the resulting scattering problem is solved with a Dirac partial-wave expansion \cite{SalvatCPC2015}. Methane and silane were used as nonpolar benchmarks. Among the local exchange forms tested there, the modified Furness--McCarthy (mFM) potential \cite{GianturcoJPB1987} gave the most consistent overall results. Although relativistic corrections to cross sections of light targets were generally modest, the method offered a single computational route over a broad energy range and was explicitly intended as a basis for extension to more anisotropic and polar systems. The main advantage of the formulation is that the same calculation can give differential (DCS), integral (ICS), momentum-transfer (MTCS), absorption (ACS), and total (TCS) cross sections over a wide energy range. The present work extends that framework to molecules with permanent dipole moments by adding the missing permanent-dipole part through state-resolved rotational first-Born cross sections built from spectroscopic transition data.

The four familiar polar hydrides--water (1.857 D), hydrogen sulfide (0.977 D), ammonia (1.476 D), and phosphine (0.580 D)--are selected as targets in its gaseous form \cite{CCCBDB2022}. Water ($\rm H_2O$) and hydrogen sulfide ($\rm H_2S$) are bent asymmetric tops, whereas ammonia ($\rm NH_3$) and phosphine ($\rm PH_3$) are pyramidal symmetric-top-like rotors. Their permanent dipole moments, rotational spacings, electronic sizes, polarizabilities, and inelastic thresholds span sufficiently different values to separate the effects of long-range permanent-dipole scattering from short-range exchange and induced polarization (Table \ref{tab:target-parameters}). The literature survey projects interesting observations: $\rm H_2O$ and $\rm NH_3$ have recommended data sets, $\rm H_2S$ has a strong set of DCS and TCS measurements, and $\rm PH_3$ has good TCS data but very little absolute DCS information. 

Water is the best documented target in the set. Electron-water scattering cross sections are used in radiation and track-structure calculations, plasma modeling, radiation chemistry, and planetary or interstellar applications. Its strong permanent dipole also makes the forward angular region very important. As a result, transmission TCS measurements and elastic integral cross sections can depend strongly on angular acceptance and how much of the forward cone is included in it. These motivations were already emphasized in the early absolute elastic work of Johnstone-Newell \cite{JohnstoneJPB1991} and in the later backward-angle measurements of Cho \emph{et al.} \cite{ChoJPB2004}, while Mu\~noz \emph{et al.} \cite{MunozPRA2007} stressed the need for a continuous cross-section set extending from low energies to the keV regime for radiation-damage modeling. Absolute TCS measurements span roughly 0.5 eV to 5 keV, while elastic DCS measurements are available for shorter ranges of a few electronvolts to about 200 eV \cite{JohnstoneJPB1991,ChoJPB2004,MunozPRA2007,SueokaJPB1986,SzmytkowskiCPL1987,KhakooPRA2008,MatsuiEPJD2016,SongJPCRD2021}. State-resolved rotational $R$-matrix calculations provide low-energy data \cite{FaureMNRAS2004}. Recently, swarm-validated and broad-range theoretical datasets have also been reported \cite{BuddeJPD2022,BuddeJPD2023,ShorifuddozaEPJD2025}. The water ICS and transmission TCS are very sensitive to the forward cone, whereas the MTCS is much less sensitive because of the $(1-\cos\theta)$ weighting \cite{KhakooPRA2008,SongJPCRD2021}. This is one of the focuses of the present work.

Hydrogen sulfide is relevant in the study of atmospheric pollution, cometary and interstellar chemistry, and plasma or materials processing. For $\rm H_2S$, transmission TCS measurements cover the few-eV region to several keV, and absolute DCS measurements are available from the low-energy regime to about 500 eV \cite{SzmytkowskiCPL1986,ZeccaPRA1992,GulleyJPB1993,RawatPRA2003,SzmytkowskiRPC2003,ChoJKPS2005,BrescansinJPB2008}. Jones \emph{et al.} \cite{JonesPRA2008} measured integral and backward-scattering cross sections down to 25 meV, directly probing the regime dominated by elastic and rotationally inelastic scattering, which is an important region for the present work. Optical-potential and rotating-dipole calculations cover complementary ranges \cite{JainPRA1990,YuanZPDAMC1993,LimbachiyaPRA2011,MahatoA2020}, and a recent relativistic independent-atom study extends elastic DCS, ICS, and MTCS calculations up to 1 MeV \cite{MeenaJPB2024}. These data make $\rm H_2S$ useful for testing the transition from low-energy dipole and resonance effects to intermediate-energy absorption and high-energy scattering.

Ammonia has an equally broad literature, but the main issues are somewhat different. TCS measurements extend from about 1 eV to 5 keV, while absolute elastic DCS, ICS, and MTCS measurements are available from a few electronvolts to 1 keV \cite{SueokaJPB1987,AlleJPB1992,ZeccaPRA1992,GarciaJPB1996,AriyasingheNIMPRSB2004,HomemPRA2014,DingerPRA2025}. Sueoka \emph{et al.} \cite{SueokaJPB1987} measured the TCS from 1 to 400 eV. Alle \emph{et al.} \cite{AlleJPB1992} reported the first broad set of absolute vibrationally elastic DCS from 2 to 30 eV, together with ICS and MTCS, and observed a narrow Feshbach resonance at $5.59\pm0.05$ eV. Zecca \emph{et al.} \cite{ZeccaPRA1992} measured TCSs from 75 to 4000 eV. Garc\'ia and Manero \cite{GarciaJPB1996} extended precision measurements to 5 keV, and Ariyasinghe \emph{et al.} \cite{AriyasingheNIMPRSB2004} later measured cross section in the 400-4000 eV. The two latter datasets sit above the Zecca \emph{et al.} values beyond about 1.2 keV, so the high-energy experimental scale itself has a real uncertainty. Jones \emph{et al.} \cite{JonesPRA2008} provide high-resolution integral scattering data from 20 meV to 10 eV. Theoretical work includes spherical optical-potential calculations from 0.1 to 1 keV \cite{JainPRA1989}, low-energy rotationally summed $R$-matrix calculations \cite{MunjalPRA2006}, broad relativistic IAM calculations to 1 MeV \cite{AkterMP2022}, and the recent comprehensive $R$-matrix dataset of Chen \emph{et al.} \cite{ChenPSST2023}. Itikawa's \cite{ItikawaJPCRD2017} evaluation remains the main recommended reference and stresses that experimentally reported ``elastic'' cross sections are generally rotationally unresolved and that low-energy transmission TCS can miss very-forward dipole scattering.

Phosphine is also important in semiconductor and microelectronic processing, where electron-impact ionization and rate coefficients enter discharge models. The experimental database for $\rm PH_3$ is thin. Absolute TCS measurements cover 0.5-370 eV and 90-3500 eV \cite{AriyasinghePRA2003,SzmytkowskiJPB2004} ranges, but absolute electron-$\rm PH_3$ DCS measurements remain scarce. Theoretical data therefore play a larger role. These include rotating-dipole calculations from 0.1 to 50 eV \cite{YuanZPDAMC1993}, Schwinger-multichannel DCS and ICS calculations from 10 to 30 eV \cite{BettegaJCP1996}, low-energy static-exchange-plus-polarization and $R$-matrix calculations \cite{BettegaJPB2004,MunjalJPB2007}, broad total-cross-section calculations to 5 keV \cite{JainPRA1992}, the combined $R$-matrix and SCOP study of Limbachiya \emph{et al.} \cite{LimbachiyaPRA2011}, and later optical-potential DCS, ICS, MTCS, and TCS studies \cite{KaurPRA2015,AouchicheRJPCA2019,MahatoJPB2020}. This makes $\rm PH_3$ a useful test of whether the present method can separate a relatively weak rotational dipole contribution from resonance, polarization, and absorption effects.

Because the literature for these targets is extensive, the discussion here is limited to the studies that are most useful for interpreting the present calculations. For reference, Table~\ref{tab:literature-summary} in Appendix~\ref{app:literature-map} gives a compact summary of the principal experimental, theoretical, and evaluated datasets for all four molecules, together with the method, energy range, and reported observables. $\rm H_2O$ and $\rm NH_3$ have larger permanent dipoles and are strong tests of the rotational correction. $\rm H_2S$ and $\rm PH_3$ have more diffuse valence charge around the heavier $\rm S$ and $\rm P$ atoms, so exchange, induced polarization, and absorption become especially useful tests. These targets set strict benchmarks for scattering calculations and provide distinct tests for the proposed extension regarding polar molecules. A unified method can evaluate all four targets against forward-angle dipole physics, including elastic scattering, momentum transfer, electronic loss, and high-energy scattering. The present calculations cover 0.1--10,000 eV energy-range. We report vibrationally elastic DCS, ICS, and MTCS. The TCS also includes the %electronically 
inelastic loss from the quasifree absorption potential. The fixed-nuclei point-dipole limit and its forward-divergence problem are explained in Appendix~\ref{app:fixed-nuclei}. % for context.

% ============================================================================
\section{\label{sec:method}Theoretical Methodology}
% ============================================================================
% All equations are given in atomic units ($\hbar=m_e=e=1$) unless otherwise stated. The speed of light is denoted by $c$.
% ----------------------------------------------------------------------------
\subsection{\label{subsec:optical}Overview of the relativistic optical-potential framework}
% ----------------------------------------------------------------------------
The calculation is separated into a central electronic scattering problem and a long-range rotational dipole contribution. The short-range collision is described with the relativistic group-additivity SCOP formulation introduced in Ref. \cite{AryaRSCA2026}. In atomic units ($\hbar=m_e=e=1$) the complex, local, energy-dependent electronic optical potential is
\begin{equation}
U(r,E)=V_{\rm st}(r)+V_{\rm ex}(r,E)+V_{\rm cp}(r)+iW_{\rm ab}(r,E),
\label{eq:optical}
\end{equation}
where $V_{\rm st}$ is the electrostatic interaction of the projectile with the molecular nuclei and electron density, $V_{\rm ex}$ represents exchange with the bound electrons, $V_{\rm cp}$ describes short-range correlation and induced-dipole polarization, and $W_{\rm ab}$ removes flux from the elastic channel into electronically inelastic channels.

For each target, relativistic atomic charge densities \cite{DesclauxCPC1975} are placed at the molecular geometry and spherically averaged about the selected group center. For each of the four compact hydrides the molecule is represented as one effective scattering group. If $n_A^{(0)}$ is the spherical density of atom $A$ located at $\bm R_A$, the central molecular density can be written schematically as \cite{AryaRSCA2026}:
\begin{equation}
n(r)=\sum_A\frac{1}{4\pi}\int d\Omega_{\bm r} n_A^{(0)}\!\left(|\bm r-\bm R_A|\right).
\label{eq:density}
\end{equation}

We keep the same potential choices that gave the best overall performance in the earlier nonpolar benchmark study so that the main change here is restricted to the permanent-dipole term. In particular, exchange is represented by the modified Furness--McCarthy local form used in Ref. \cite{GianturcoJPB1987}, based on the semiclassical Furness--McCarthy \cite{FurnessJPB1973} construction. The correlation-polarization interaction \cite{PerdewPRB1981} joins a local short-range correlation energy to the asymptotic induced-dipole form $-\alpha/(2r^4)$, with the molecular dipole polarizability $\alpha$ % entering explicitly 
\cite{ZhangJPB1992}. Electronically inelastic loss is described by the quasifree-scattering absorption model of Staszewska \emph{et al.} \cite{StaszewskaAPS1983}, in which a local imaginary potential is built from the target density, projectile energy, and Pauli-blocked quasifree electron scattering. Table \ref{tab:target-parameters} lists the dipole moment $\mu_D$, numerical polarizabilities $\alpha$ and ionization thresholds $\Delta_e$ used in the calculations.

Spherical averaging keeps the monopole electrostatic field and the isotropic induced-polarization response, but it cannot keep the anisotropic $1/r^2$ field of a permanent dipole. It is added separately through explicit rotational transitions. This keeps permanent-dipole scattering distinct from induced polarization and lets us change the rotational temperature or state distribution without rebuilding the central optical potential.

\begin{table}%[t]
\caption{Target properties \cite{CCCBDB2022} used in the optical-potential calculation. $\mu_D$ is the permanent dipole moment, $\alpha$ is the isotropic static dipole polarizability, and $\Delta_e$ is the effective electronic-loss threshold used in the quasifree absorption potential.}
\label{tab:target-parameters}
\begin{ruledtabular}
\begin{tabular}{lcccc}
Target & Geometry (group) & $\mu_D$ (D) & $\alpha$ (\AA$^3$) & $\Delta_e$ (eV) \\
\hline
$\rm H_2O$ & bent,      ($C_{2v}$) & 1.857 & 1.501 & 12.600 \\
$\rm H_2S$ & bent,      ($C_{2v}$) & 0.977 & 3.631 & 10.500 \\
$\rm NH_3$ & pyramidal, ($C_{3v}$) & 1.476 & 2.103 & 10.820 \\
$\rm PH_3$ & pyramidal, ($C_{3v}$) & 0.580 & 4.237 & 10.590 \\
\end{tabular}
\end{ruledtabular}
\end{table}

% ----------------------------------------------------------------------------
\subsection{\label{subsec:rotational}Rotationally resolved first-Born dipole contribution}
% ----------------------------------------------------------------------------
Let the incident electron have kinetic energy $E$, and let $i$ and $f$ denote the initial and final rotational levels with internal energies $\varepsilon_i$ and $\varepsilon_f$. We define the rotational energy transfer
\begin{equation}
\Delta E_{if}=\varepsilon_f-\varepsilon_i,
\label{eq:deltaE}
\end{equation}
so that $\Delta E_{if}>0$ denotes excitation and $\Delta E_{if}<0$ denotes superelastic de-excitation. The channel is open when $E>\Delta E_{if}$, with
\begin{equation}
k_i=k(E),\qquad k_f=k(E-\Delta E_{if}),
\label{eq:ki-kf}
\end{equation}
The momentum transfer for scattering angle $\theta$ is therefore
\begin{equation}
q_{if}^2(\theta)=k_i^2+k_f^2-2k_i k_f\cos\theta.
\label{eq:qif}
\end{equation}
Let $g_i$ and $g_f$ be the total statistical weights of states $i$ and $f$, $m_i$ and $m_f$ their magnetic sublevels, and $\mu_\lambda$ ($\lambda=0,\pm1$) the spherical components of the molecular electric-dipole operator. We define
\begin{equation}
\mathcal R_{if}=\frac{1}{g_i}
\sum_{m_i,m_f,\lambda}
\left|\langle f m_f|\mu_\lambda|i m_i\rangle\right|^2.
\label{eq:Rif-def}
\end{equation}
$\mathcal R_{if}$, expressed in atomic units $(ea_0)^2$, is the squared state-to-state transition dipole strength, averaged over the degenerate initial sublevels and summed over the final sublevels and dipole components. With this convention, the orientation-averaged first-Born differential cross section for an allowed rotational transition is \cite{TakayanagiAAMP1970,NorcrossAAMP1982}
\begin{equation}
\left(\frac{d\sigma_{if}}{d\Omega}\right)_{\rm B}
=\frac{4}{3}\frac{k_f}{k_i}
\frac{\mathcal R_{if}}{q_{if}^{2}(\theta)}.
\label{eq:born-rot}
\end{equation}
The factor $k_f/k_i$ is the final-to-initial projectile flux ratio. Equations \eqref{eq:qif} and \eqref{eq:born-rot} are evaluated only for energetically open transitions.

At a rotational temperature $T_{\rm rot}$, state $i$ has normalized population $P_i(T_{\rm rot})$. The total rotational contribution for a thermal ensemble is the incoherent sum over orthogonal final rotational channels,
\begin{equation}
\frac{d\sigma_{\rm rot}}{d\Omega}(E,T_{\rm rot})
=\sum_i \mathcal P_i(T_{\rm rot})
\sum_{f\ne i}^{\rm open}
\left(\frac{d\sigma_{if}}{d\Omega}\right)_{\rm B}.
\label{eq:rot-sum}
\end{equation}

The second sum in Eq. \eqref{eq:rot-sum} contains all retained final rotational states accessible from $i$. Cross sections, rather than amplitudes, are summed because different rotational final states are orthogonal. This is the same physical separation used in rotating-dipole and Born-closure treatments of polar molecules \cite{YuanZPDAMC1993,ItikawaTCA2000,RawatPRA2003}, but here the sum is constructed from individual spectroscopic transitions rather than from one effective rotational line.

The rotational integral and momentum-transfer cross sections are obtained from Eq. \eqref{eq:rot-sum} as
\begin{align}
\sigma_{\rm rot}(E)&=2\pi\int_0^\pi
\frac{d\sigma_{\rm rot}}{d\Omega}\sin\theta\,d\theta,
\label{eq:rot-ics}\\
\sigma_{\rm mt}^{\rm rot}(E)&=2\pi\int_0^\pi
(1-\cos\theta)\frac{d\sigma_{\rm rot}}{d\Omega}
\sin\theta\,d\theta.
\label{eq:rot-mt}
\end{align}
For a single upward transition, direct integration of Eq. \eqref{eq:born-rot} gives
\begin{equation}
\sigma_{if}=\frac{4\pi \mathcal R_{if}}{3k_i^2}
\ln{\left[\frac{(k_i+k_f)^2}{(k_i-k_f)^2}\right]}.
\label{eq:rot-ics-analytic}
\end{equation}
A useful feature follows immediately from Eqs. \eqref{eq:deltaE}--\eqref{eq:qif}. For a transition with finite energy transfer, $k_i\ne k_f$ and
\begin{equation}
q_{if}(0)=|k_i-k_f|>0,
\label{eq:q-forward}
\end{equation}
so the state-resolved rotational DCS is finite at $\theta=0$. Very small rotational spacings can still produce a sharply forward-peaked DCS.

The pure rotational transition list used in Eqs. \eqref{eq:born-rot}--\eqref{eq:rot-mt} is prepared from HITRAN2024 \cite{Hitran2024}. The HITRAN absorption-line intensity is \emph{not} inserted into Eq. \eqref{eq:born-rot}; it already contains thermal population, partition-function, isotopic-abundance, and stimulated-emission factors. Instead we use the temperature-independent Einstein coefficient $\mathcal A_{fi}$ for spontaneous emission from the upper state $f$ to the lower state $i$. With HITRAN total statistical weights $g_f$ and $g_i$, the initial-state-averaged transition strength is reconstructed as
\begin{equation}
\mathcal R_{if}(\mathrm{D}^2)=
\frac{\mathcal A_{fi}g_f}
{3.13618932\times10^{-7}\,\tilde\nu_{if}^{3}g_i},
\label{eq:hitran-A}
\end{equation}
where $\mathcal A_{fi}$ is in s$^{-1}$ and $\tilde\nu_{if}$ is in cm$^{-1}$. 

Because the central Dirac--SCOP target represents the equilibrium geometry of the ground vibrational state, the rotational Born contribution was restricted consistently to the $v=0$ rotational (or rotation--inversion) manifold. The population of the unique initial state $i$ is reconstructed independently at the adopted rotational temperature,
\begin{equation}
\mathcal P_i(T_{\rm rot})=
\frac{g_i\exp[-c_2\varepsilon_i/T_{\rm rot}]}
{\mathcal Q(T_{\rm rot})},
\label{eq:population}
\end{equation}
where $c_2=hc/k_B$ in $\mathrm{cm\,K}$ units, $\varepsilon_i$ is expressed as a spectroscopic term value in cm$^{-1}$, $\mathcal Q(T_{\rm rot})$ is the rotational partition function consistent with the HITRAN statistical-weight convention. All calculations reported here use $T_{\rm rot}=296$ K.

For each spectroscopic lower--upper pair $(i,f)$, both collision directions are included. The excitation channel uses $(+\Delta E_{if},R_{if},P_i)$. The reverse superelastic channel uses $(-\Delta E_{if},R_{fi},P_f)$, where
\begin{equation}
R_{fi} = \frac{g_i}{g_f}R_{if}.
\label{eq:Rreverse}
\end{equation}
Thus the thermal rotational contribution contains both excitation and de-excitation whenever the corresponding initial level is populated. For the modeled equilibrium ensemble the populations satisfy $\sum_i \mathcal P_i=1$.

% ----------------------------------------------------------------------------
\subsection{\label{subsec:dirac}Dirac partial-wave formulation and cross sections}
% ----------------------------------------------------------------------------
The central optical-potential scattering problem defined by Eq. \eqref{eq:optical} is solved with the RADIAL implementation of Salvat and Fern\'andez-Varea \cite{SalvatCPC2015}. For a real or complex central potential the spinor separates into radial large and small components and spin-angular functions labeled by the relativistic angular quantum number $\kappa$. The RADIAL package integrates the coupled large- and small-component equations and supplies free-state phase shifts for each partial wave. The corresponding partial-wave $S$-matrix is
\begin{equation}
S_\kappa(E)=\exp[2i\delta_\kappa(E)].
\label{eq:Smatrix}
\end{equation}
Writing $S_\ell^-\equiv S_{\kappa=-(\ell+1)}$ and $S_\ell^+\equiv S_{\kappa=\ell}$, the direct and spin-flip amplitudes are \cite{SalvatCPC1991,SalvatCPC2005}
\begin{align}
f(\theta)&=\frac{1}{2ik}\sum_{\ell}
\left[(\ell+1)(S_\ell^--1)+\ell(S_\ell^+-1)\right]P_\ell(\cos\theta),
\label{eq:famp}\\
g(\theta)&=\frac{1}{2ik}\sum_{\ell}
\left(S_\ell^+-S_\ell^-\right)P_\ell^1(\cos\theta).
\label{eq:gamp}
\end{align}
The central differential cross section is
\begin{equation}
\left(\frac{d\sigma}{d\Omega}\right)_{\rm c}
=|f(\theta)|^2+|g(\theta)|^2.
\label{eq:dcs-central}
\end{equation}
The corresponding central elastic integral cross section can be written in partial-wave form as
\begin{equation}
\sigma_i^{\rm c}=\frac{\pi}{k^2}\sum_{\ell}
\left[(\ell+1)|S_\ell^--1|^2+\ell|S_\ell^+-1|^2\right],
\label{eq:ics-central}
\end{equation}
The central momentum-transfer cross section is evaluated directly from Eq. \eqref{eq:dcs-central},
\begin{equation}
\sigma_{\rm mt}^{\rm c}=2\pi\int(1-\cos\theta)
\left(\frac{d\sigma}{d\Omega}\right)_{\rm c}
\sin\theta\,d\theta.
\label{eq:mt-central}
\end{equation}
The quantities reported as vibrationally elastic (VE) but rotationally state-unresolved are formed by adding the state-resolved rotational contribution of Sec. \ref{subsec:rotational} to the central result. Thus
\begin{equation}
\left(\frac{d\sigma}{d\Omega}\right)_{\rm VE}
=\left(\frac{d\sigma}{d\Omega}\right)_{\rm c}
+\frac{d\sigma_{\rm rot}}{d\Omega},
\label{eq:dcs-total}
\end{equation}
In the figures and tables, $\sigma_i$ or ICS is referred to as the (vibrationally elastic) integral cross section, with the understanding that the rotational contribution has been included.

Finally, the total cross section used in this work is
\begin{equation}
\sigma_t=\sigma_i^{\rm c}+\sigma_{\rm rot}+\sigma_{\rm ab}
=\sigma_i+\sigma_{\rm ab}.
\label{eq:tcs-final}
\end{equation}
where $\sigma_{\rm ab}$ represents summed electronically inelastic loss generated by the quasifree optical potential.

% ============================================================================
\section{\label{sec:results}Results and Discussion}
% ============================================================================

The total/integral cross sections are reported in units of $10^{-16}\ {\rm cm^2}$, while the DCS are in $10^{-16}\ {\rm cm^2\,sr^{-1}}$. The results reported here are using eqs.~\eqref{eq:dcs-total}--\eqref{eq:tcs-final}. The DCS, ICS, and MTCS are therefore vibrationally elastic. This distinction is important when comparing with attenuation measurements that may include unresolved vibrational excitation. 

A common pattern is visible across the four targets. At low incident energies the rotational term mainly changes the forward DCS and, through the angular integral, the ICS. The MTCS is less sensitive because its $(1-\cos\theta)$ weight suppresses the extreme forward cone. At higher energy the rotational correction becomes progressively less important and the comparison is controlled mainly by the central static, exchange, correlation--polarization, and absorption terms. This difference between the ICS and MTCS is used below as a diagnostic: a large ICS discrepancy accompanied by a much smaller MTCS discrepancy points primarily to the forward region, whereas a discrepancy that remains in the measured DCS over tens of degrees must also involve the short-range electronic scattering model.

% ----------------------------------------------------------------------------
\subsection{\label{subsec:result-h2o}Water}
% ----------------------------------------------------------------------------
\begin{figure*}
\includegraphics[width=0.32\linewidth]{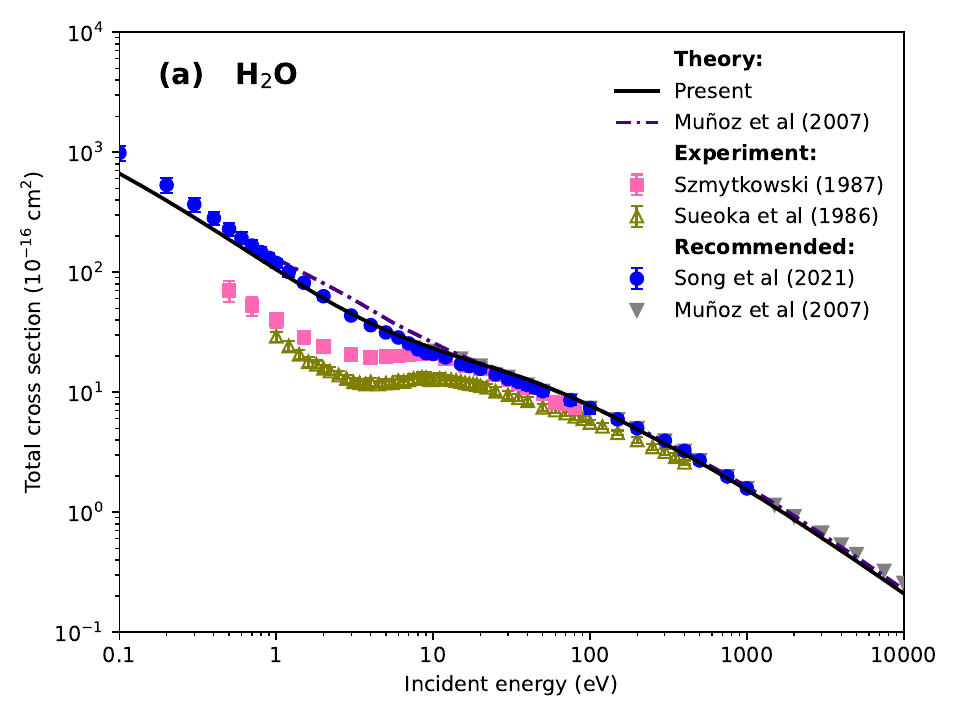}
\includegraphics[width=0.32\linewidth]{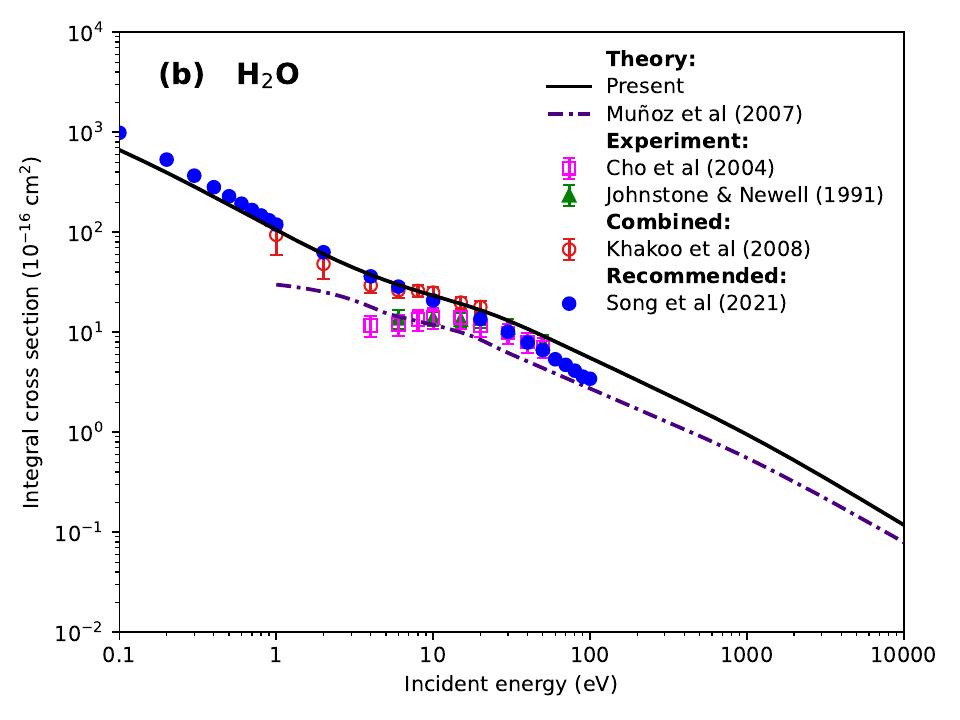}
\includegraphics[width=0.32\linewidth]{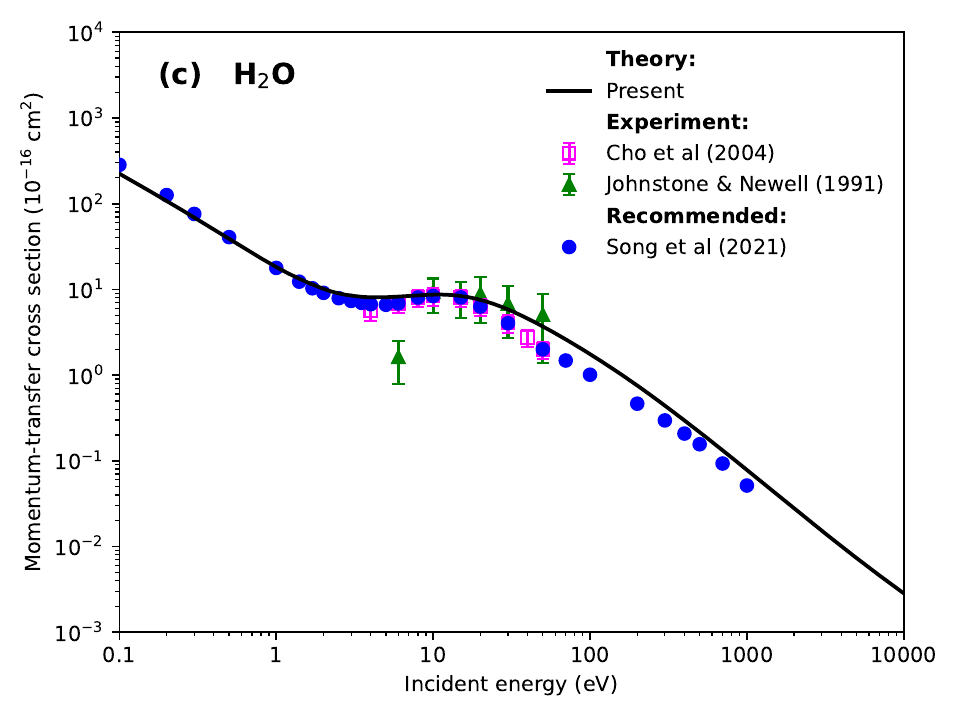}
\caption{\label{fig:h2o_cs}Total (a), integral (b), and momentum-transfer (c) cross sections for electron--$\mathrm{H_2O}$ scattering. The present calculation (solid line) is compared with the available data: filled triangle-up, Johnstone and Newell \cite{JohnstoneJPB1991}; open square, Cho  \emph{et al.} \cite{ChoJPB2004}; dash-dotted line and filled triangle-down, Mu\~noz \emph{et al.} \cite{MunozPRA2007}; open triangle-up, Sueoka  \emph{et al.} \cite{SueokaJPB1986}; filled square, Szmytkowski \cite{SzmytkowskiCPL1987}; open cicle, Khakoo  \emph{et al.} \cite{KhakooPRA2008}; filled circle, Song  \emph{et al.} \cite{SongJPCRD2021}.}
\end{figure*}
Figure \ref{fig:h2o_cs} compares the calculated (a) TCS, (b) ICS, and (c) MTCS of $\rm H_2O$ with the main wide-range datasets used in this work. The low-energy TCS rises rapidly as the incident energy is reduced. The present curve follows the scale and energy dependence of the Song \emph{et al.}'s \cite{SongJPCRD2021} recommended values much more closely than the transmission measurements of Sueoka \emph{et al.} \cite{SueokaJPB1986} and Szmytkowski \cite{SzmytkowskiCPL1987} in the low-energy region. This is physically reasonable for a strongly polar target. Forward-scattered electrons can remain inside the transmitted beam, so an attenuation experiment can miss a substantial part of the rotationally enhanced small-angle scattering. Song \emph{et al.} showed that the corresponding correction is modest near 100 eV but becomes very large below a few electronvolts. The present calculation includes forward rotational strength explicitly and therefore should not be expected to follow an uncorrected low-energy transmission curve. From the tens-of-eV region upward, the different experimental datasets move much closer together and the present TCS follows the measured and recommended trend through the intermediate-energy region and into the keV range. The comparison with the Mu\~noz \emph{et al.}'s \cite{MunozPRA2007} experiment and optical-potential calculations is especially useful above 50 eV, where the rotational correction is no longer the main source of uncertainty. In figure \ref{fig:h2o_cs}(b), Mu\~noz \emph{et al.}'s integral (elastic) cross section is without rotational contribution. Cho \emph{et al.} \cite{ChoJPB2004} removed the backward extrapolation by measuring to $180^\circ$, but the forward cone still had to be inferred. Johnstone \& Newell \cite{JohnstoneJPB1991} and Khakoo \emph{et al.} \cite{KhakooPRA2008} likewise obtained their ICS by integrating finite-angle DCSs with different forward extrapolations. The fact that the present ICS can be high relative to those direct integrations while the DCS comparison over measured angles is considerably better, as discussed below, indicates that a significant part of the difference is concentrated at very small angles rather than being a uniform normalization error. 

The momentum-transfer cross section supports this interpretation. It follows the Song \emph{et al.} recommendation and the Cho \emph{et al.} and Johnstone \emph{et al.} measurements much more closely than the ICS over their common range. Since $(1-\cos\theta)$ removes much of the weight of the forward peak, the MTCS is controlled more strongly by the angular distribution away from zero degrees. The simultaneous behavior of the ICS and MTCS therefore provides an important internal check on the polar extension: the model restores a large forward contribution without destroying the wider-angle momentum-transfer scale. At high energy, all three integrated quantities show the expected smooth and monotonous fall.

\begin{figure*}
\includegraphics[width=0.32\linewidth]{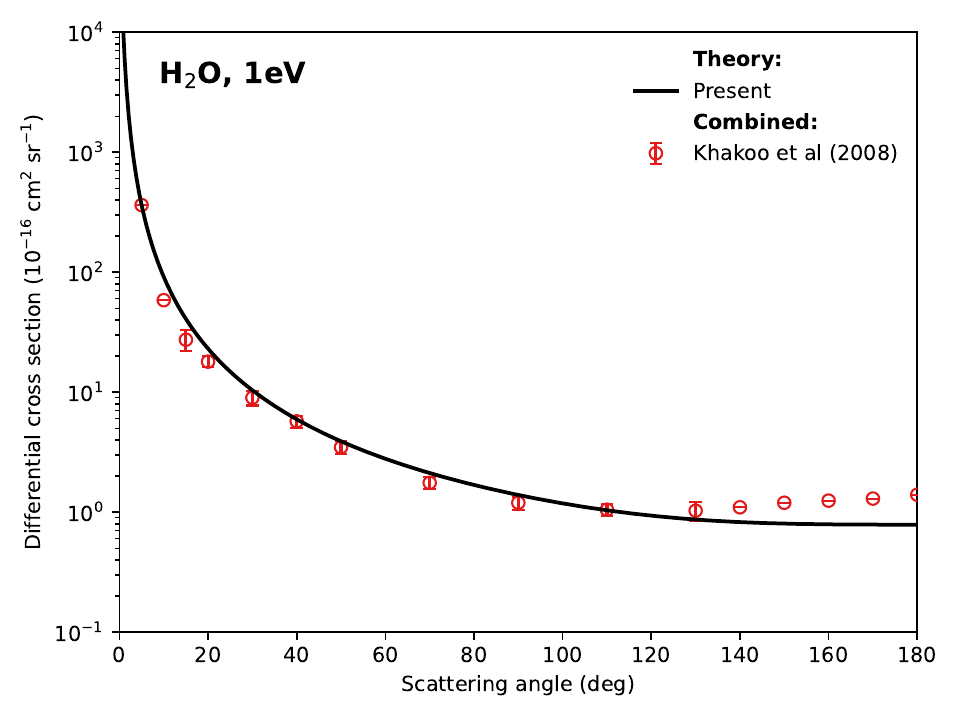}
\includegraphics[width=0.32\linewidth]{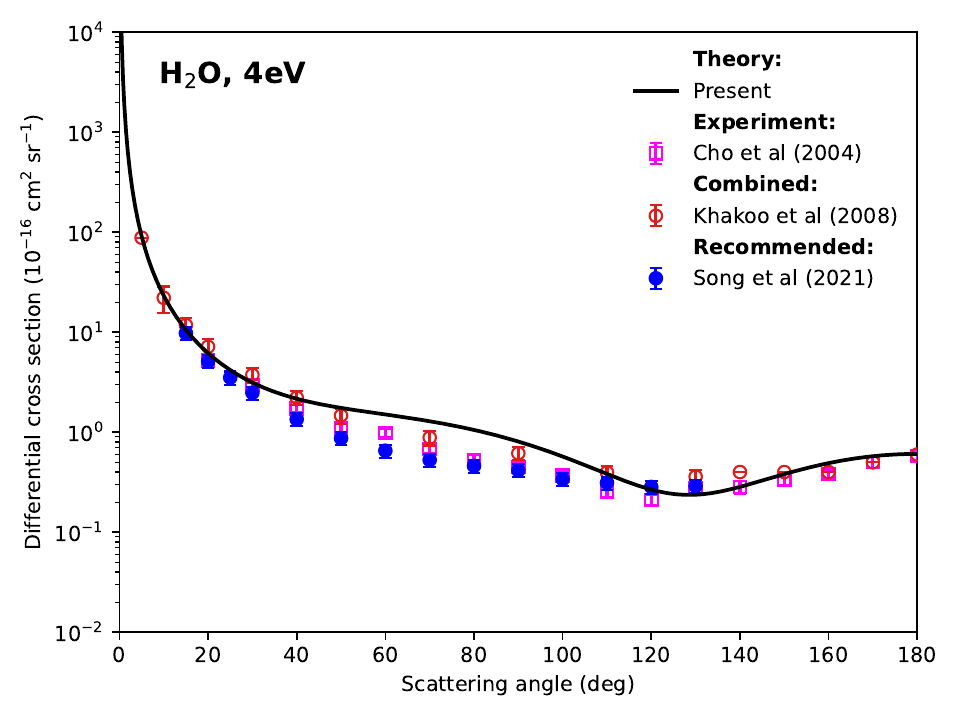}
\includegraphics[width=0.32\linewidth]{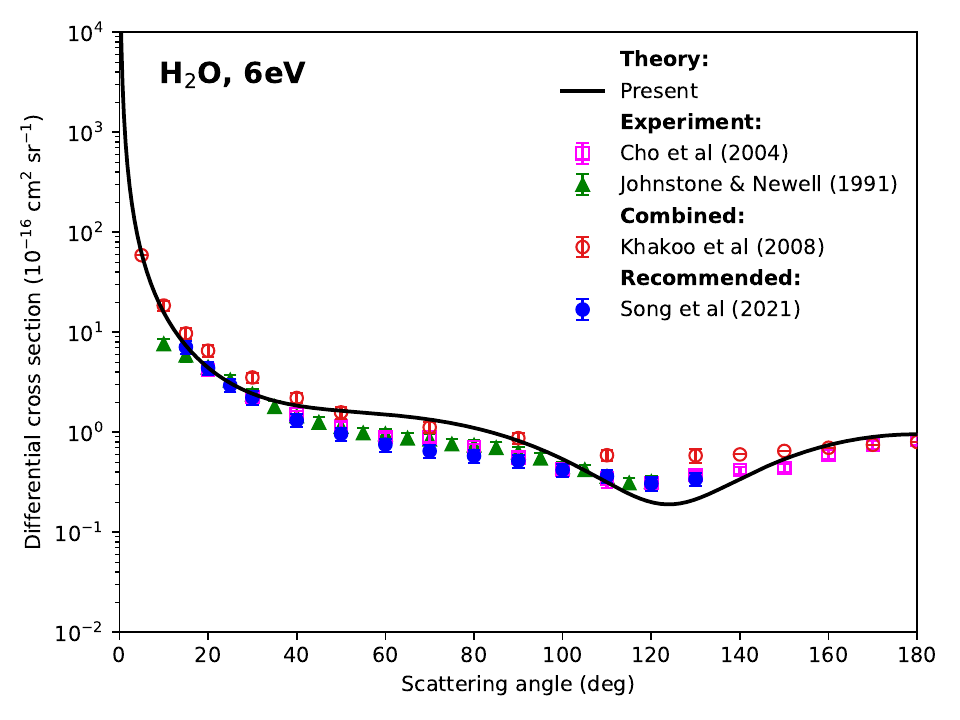}\\
\includegraphics[width=0.32\linewidth]{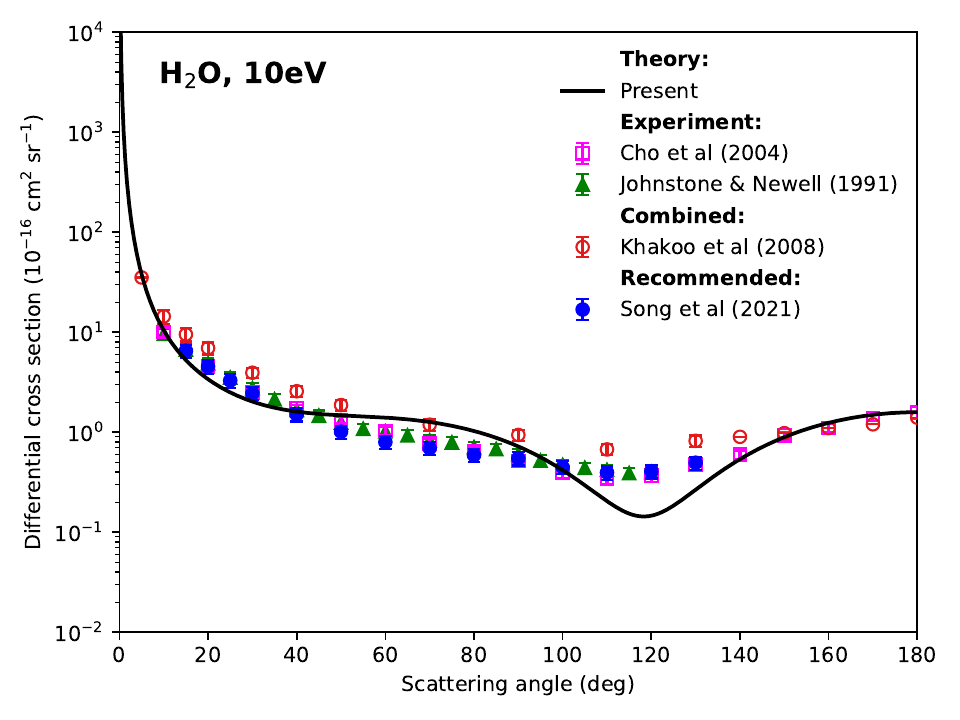}
\includegraphics[width=0.32\linewidth]{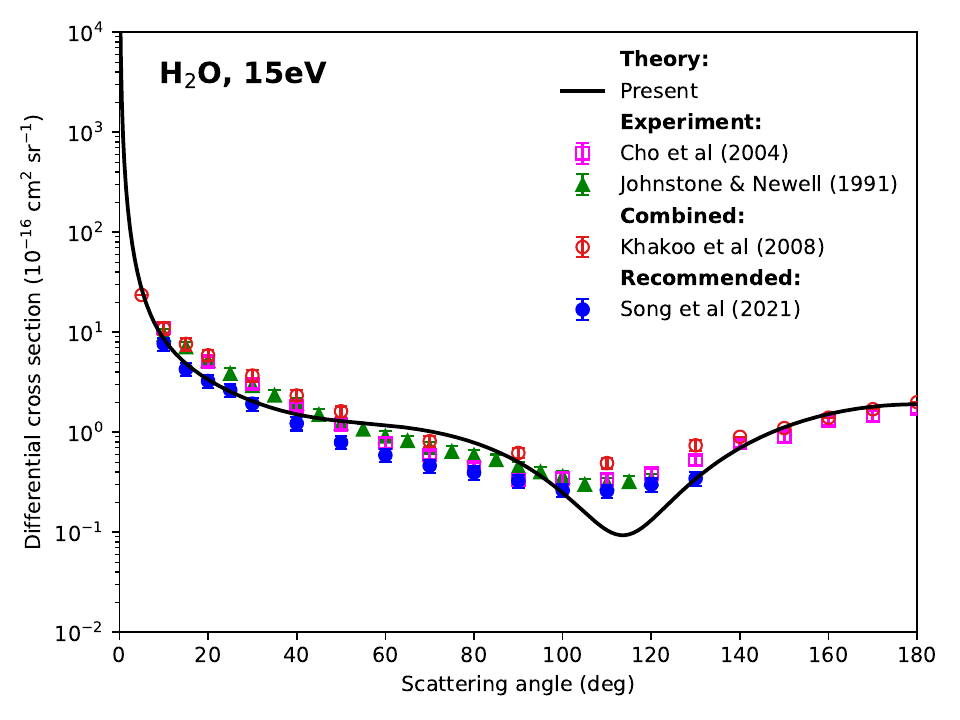}
\includegraphics[width=0.32\linewidth]{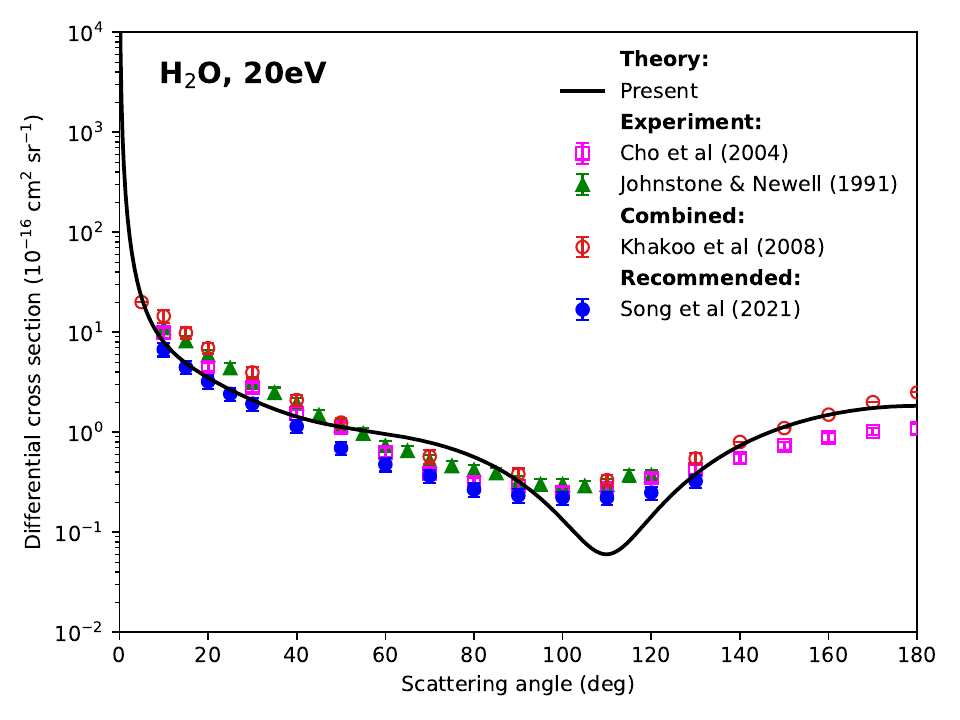}\\
\includegraphics[width=0.32\linewidth]{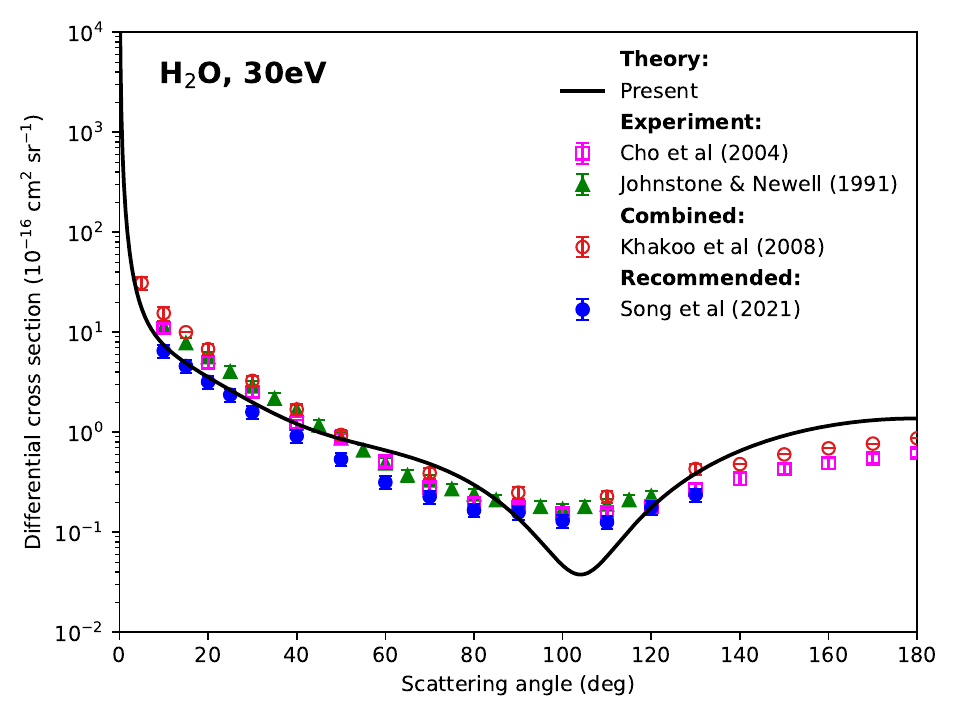}
\includegraphics[width=0.32\linewidth]{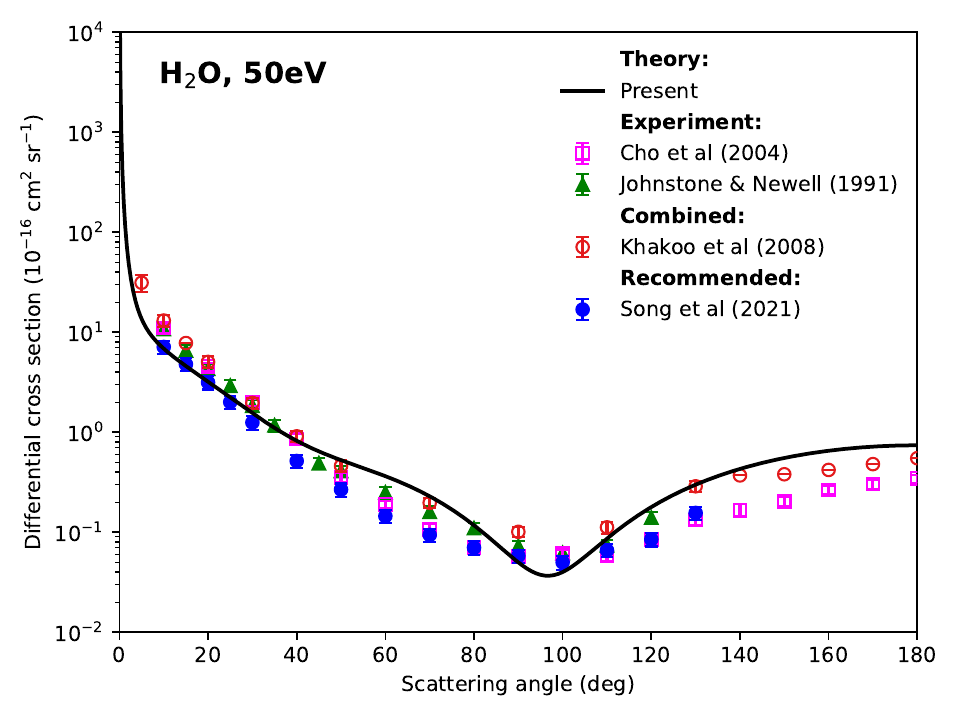}
\includegraphics[width=0.32\linewidth]{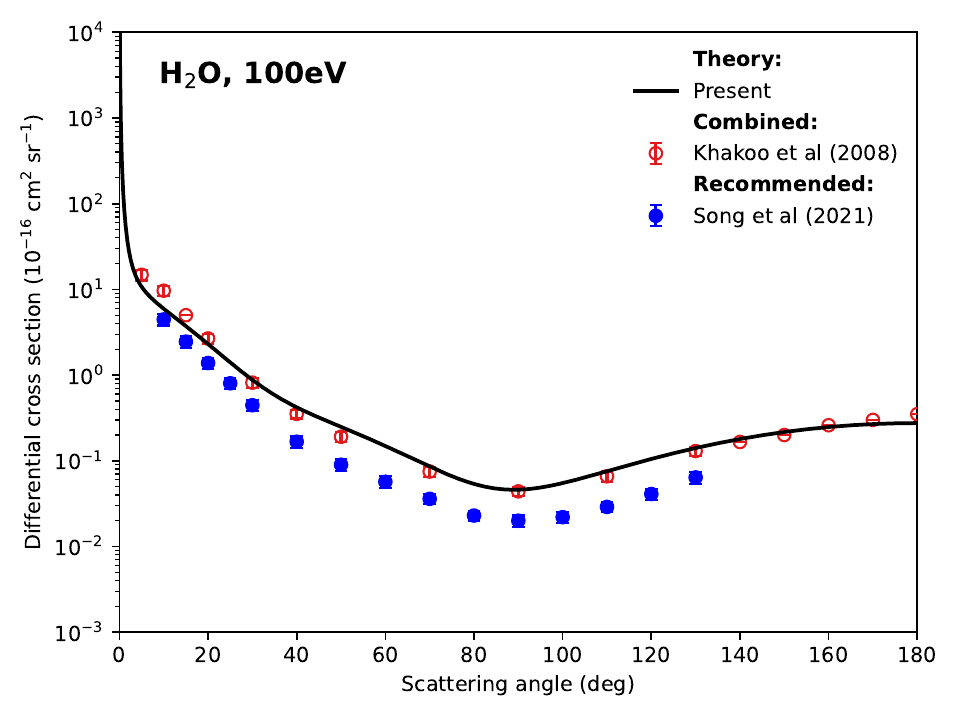}\\
\caption{\label{fig:h2o_dcs}Vibrationally elastic DCS for electron--$\rm H_2O$ scattering at selected incident energies, as labeled in each panel. Symbols are the same as Fig. \ref{fig:h2o_cs}.}
\end{figure*}
The angular distributions in Fig. \ref{fig:h2o_dcs} clarify where the remaining integrated differences arise. At 1 eV the present DCS is strongly forward peaked and follows the Khakoo \emph{et al.}'s combined theoretical and experimental dataset closely over most of the measured angular interval. The agreement at a finite angle is much better than one might infer from the spread of low-energy ICS values alone. This again shows why an ICS comparison for water cannot be separated from the treatment of the unmeasured forward cone. From 4 to 20 eV the calculation reproduces the main evolution of the angular distribution: a steep forward decrease, a broad mid/back-angle minimum, and a recovery toward $180^\circ$. The calculated curve is often somewhat higher than the experimental or recommended points through intermediate angles, and the minimum is generally deeper than the measured one. The position of the minimum nevertheless moves with energy in the same way as the data. At 10--20 eV the backward rise is also reproduced, although its magnitude may be slightly larger than the Cho and Song values at some energies. These differences are no longer controlled by the forward rotational cone and therefore provide a test of the central optical potential itself. The same trend becomes clearer at 30 and 50 eV. The present calculation reproduces the existence and approximate position of the angular minimum but tends to make it too deep and to recover somewhat strongly at backward angles. At 100 eV the calculated DCS shows a good overall comparison in terms of scale and shape but lies visibly above the Song-recommended values over a broad intermediate-angle interval. Overall, the calculation reproduces the main trend of the angular distribution.

% ----------------------------------------------------------------------------
\subsection{\label{subsec:result-h2s}Hydrogen sulfide}
% ----------------------------------------------------------------------------
\begin{figure*}
\includegraphics[width=0.32\linewidth]{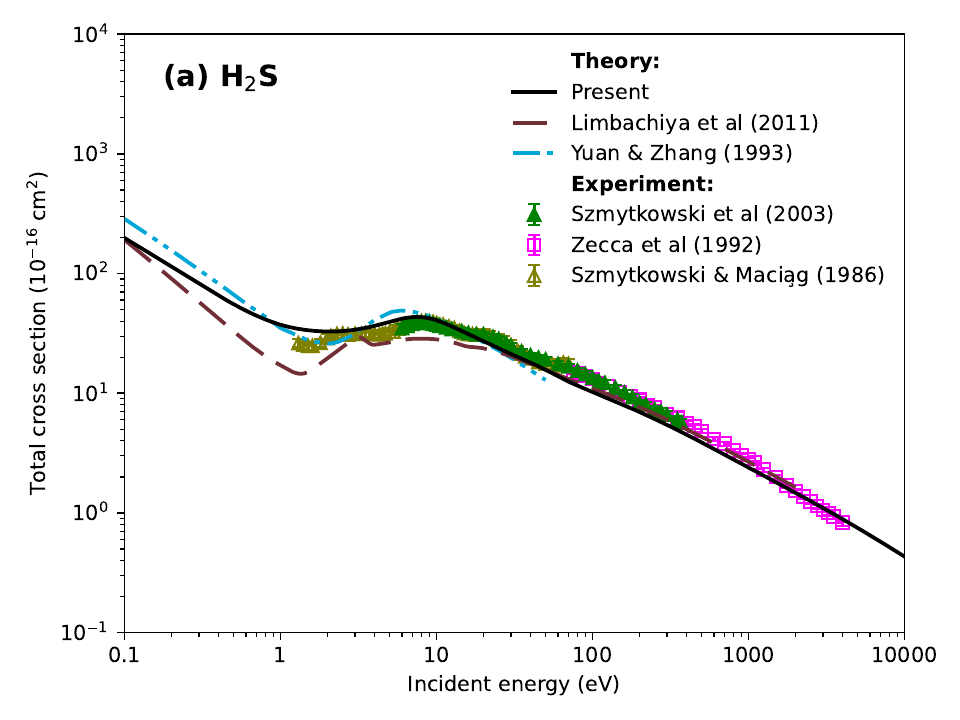}
\includegraphics[width=0.32\linewidth]{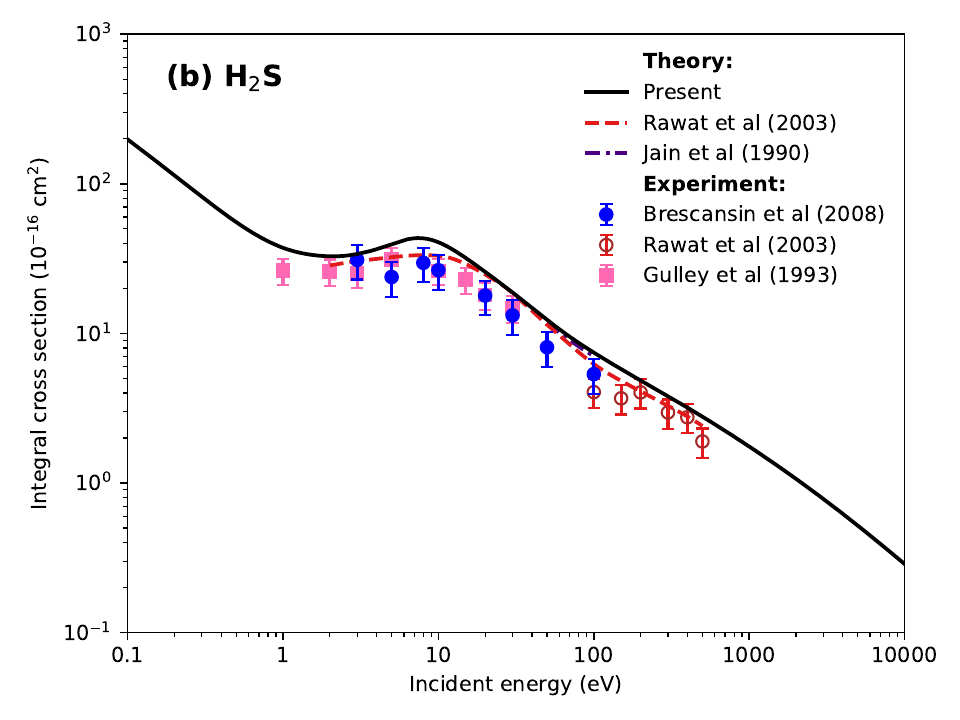}
\includegraphics[width=0.32\linewidth]{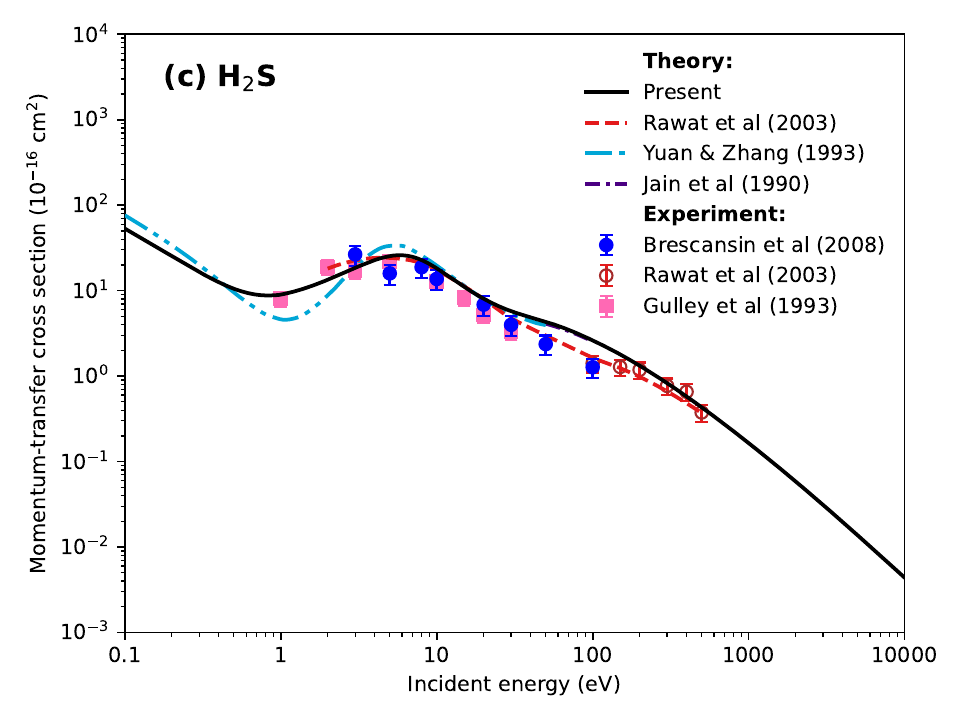}
\caption{\label{fig:h2s_cs}Total (a), integral (b), and momentum-transfer (c) cross sections for electron--$\mathrm{H_2S}$ scattering. The present calculation (solid line) is compared with the available data: long-dashed line, Limbachiya  \emph{et al.} \cite{LimbachiyaPRA2011}; short-dashed line and open circle, Rawat  \emph{et al.} \cite{RawatPRA2003}; dash-dot-dotted line, Yuan and Zhang \cite{YuanZPDAMC1993}; dash-dotted line, Jain \emph{et al.} \cite{JainPRA1990}; filled triangle-up, Szmytkowski \cite{SzmytkowskiRPC2003}; open triangle-up, Szmytkowski and Maci\k{a}g \cite{SzmytkowskiCPL1986}; filled circle, Brescansin  \emph{et al.} \cite{BrescansinJPB2008}; filled square, Gulley  \emph{et al.} \cite{GulleyJPB1993}; open square, Zecca  \emph{et al.} \cite{ZeccaPRA1992}.}
\end{figure*}
The integrated $\rm H_2S$ results are shown in Fig. \ref{fig:h2s_cs}. The TCS displays the characteristic low-energy decrease followed by a broad enhancement in the several-eV region and then a smooth high-energy fall. The present curve reproduces the broad experimental envelope as shown by the Szmytkowski--Maci\k{a}g \cite{SzmytkowskiCPL1986}, Szmytkowski \emph{et al.} \cite{SzmytkowskiRPC2003}, and Zecca \emph{et al.} \cite{ZeccaPRA1992} transmission measurements. It also follows the main energy dependence of the Limbachiya \emph{et al.} \cite{LimbachiyaPRA2011} and Yuan--Zhang \cite{YuanZPDAMC1993} calculations. The few-eV structure is smoother than some resonance-specific calculations and measurements. This is expected from a local spherical optical potential: the method is designed to reproduce the broad collision background, not to resolve every temporary-negative-ion features.

In the middle panel of Fig. \ref{fig:h2s_cs}, the present result is systematically above much of the Gulley, Brescansin, and Rawat experimental scale from the low-energy region into the tens-of-eV range \cite{GulleyJPB1993,RawatPRA2003,BrescansinJPB2008}. Part of this difference is again associated with the forward polar contribution, because the experiments do not cover the complete small-angle region. However, the separation persists farther into the intermediate-energy range than for a purely forward-angle effect. The central elastic interaction is therefore also somewhat too strong in this energy interval. At higher energy the present ICS approaches the older Jain and Rawat theoretical results and the relative spread becomes much smaller.

The MTCS gives a more favorable comparison. The calculated low-energy minimum and broad several-eV maximum are close to the experimental trends, and from roughly the several-eV region upward the present curve follows the Rawat, Brescansin, and Gulley data more closely than does the ICS. Yuan \& Zhang showed explicitly that the extreme forward cone makes a large contribution to the $\rm H_2S$ integral cross section below about 1 eV but a much smaller contribution to the MTCS \cite{YuanZPDAMC1993}. The difference in comparison of the ICS and MTCS as shown in Fig. \ref{fig:h2s_cs} is consistent with the inference as above.

\begin{figure*}
\includegraphics[width=0.32\linewidth]{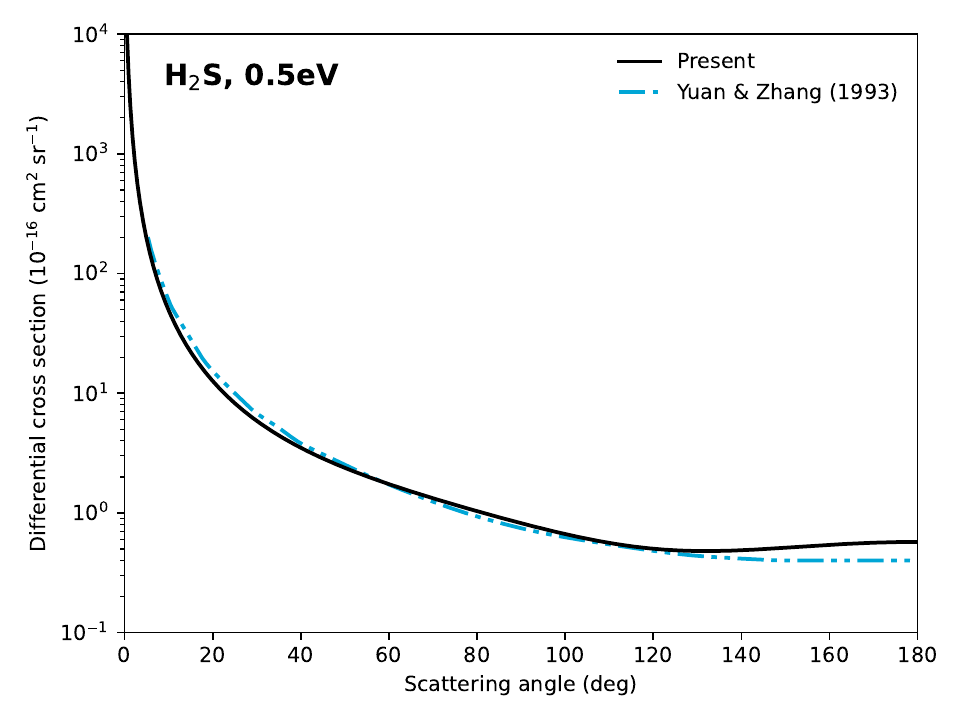}
\includegraphics[width=0.32\linewidth]{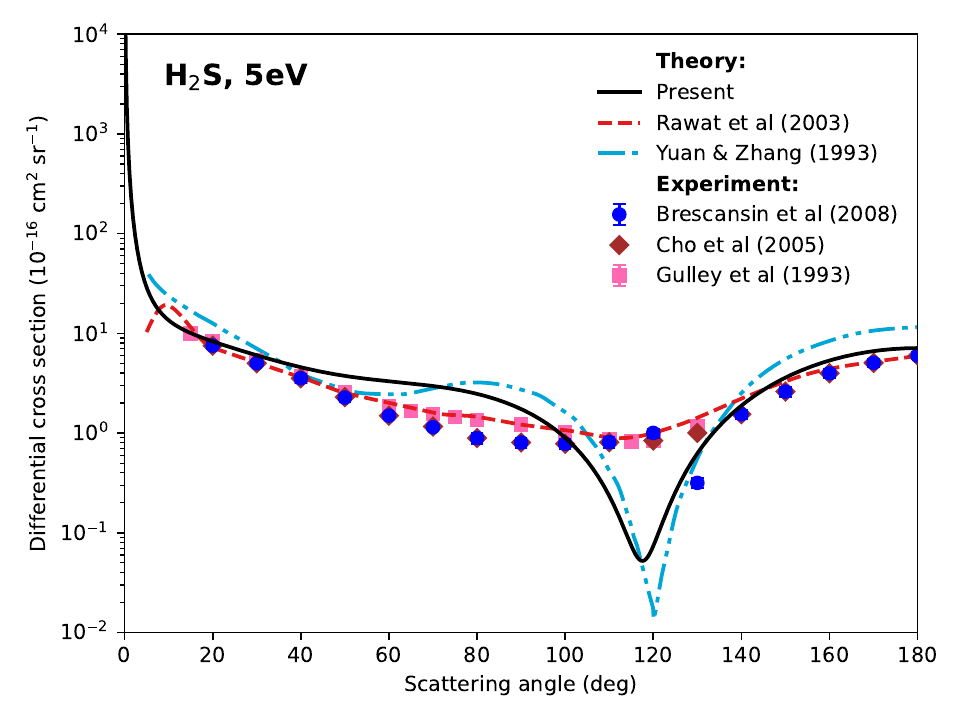}
\includegraphics[width=0.32\linewidth]{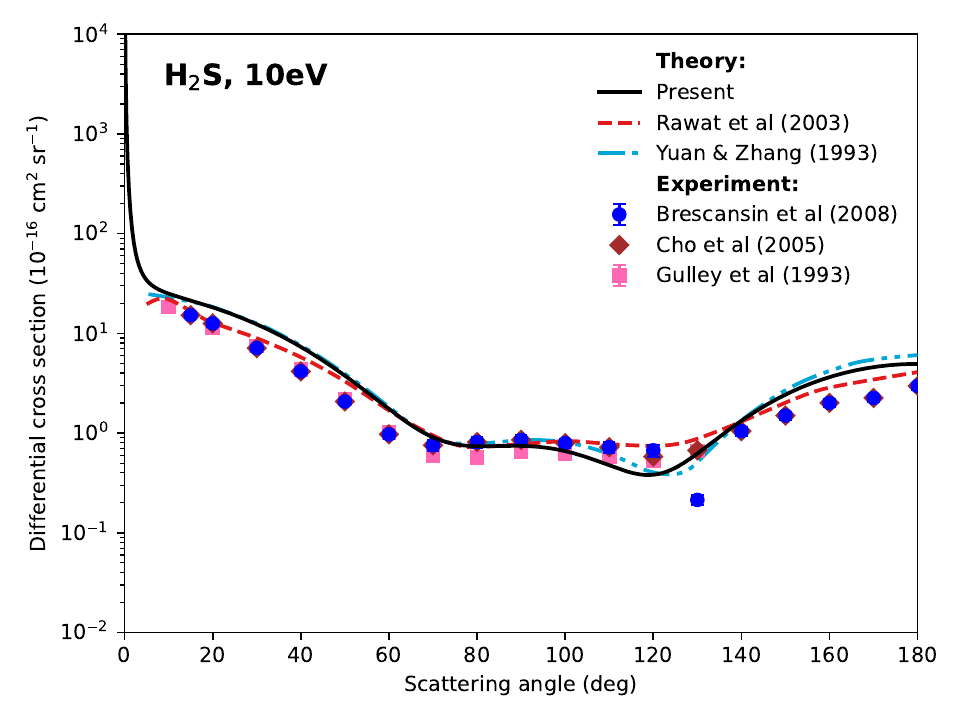}\\
\includegraphics[width=0.32\linewidth]{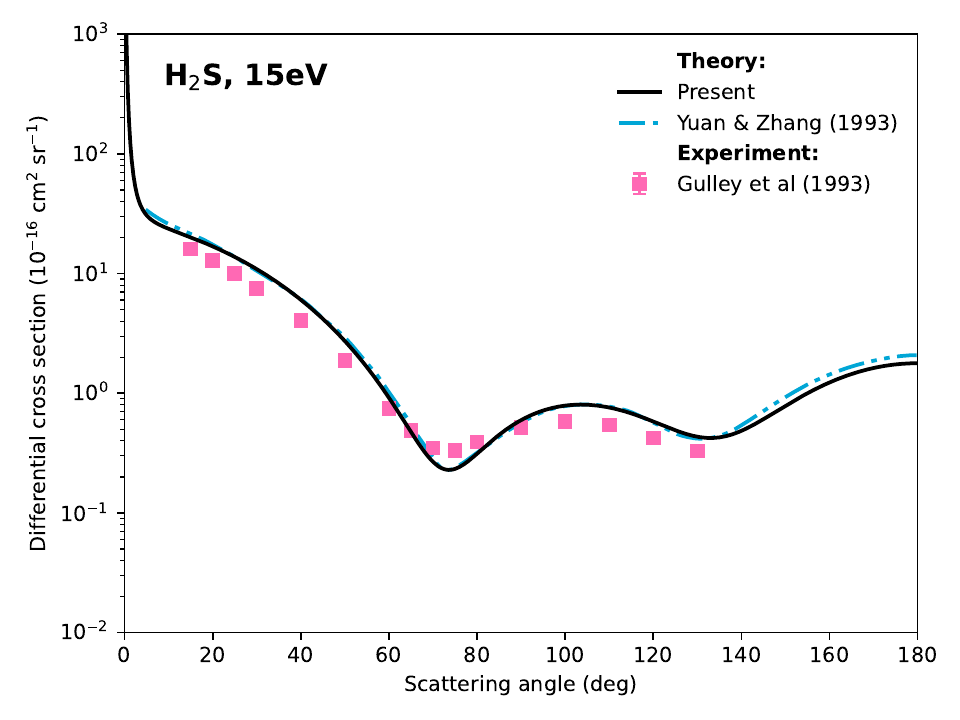}
\includegraphics[width=0.32\linewidth]{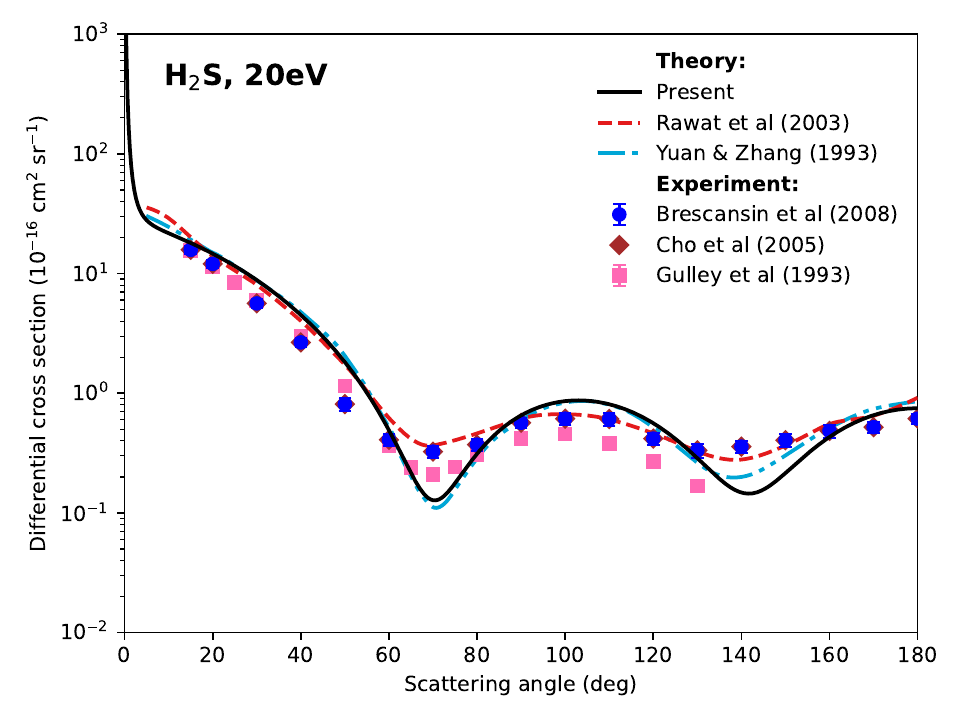}
\includegraphics[width=0.32\linewidth]{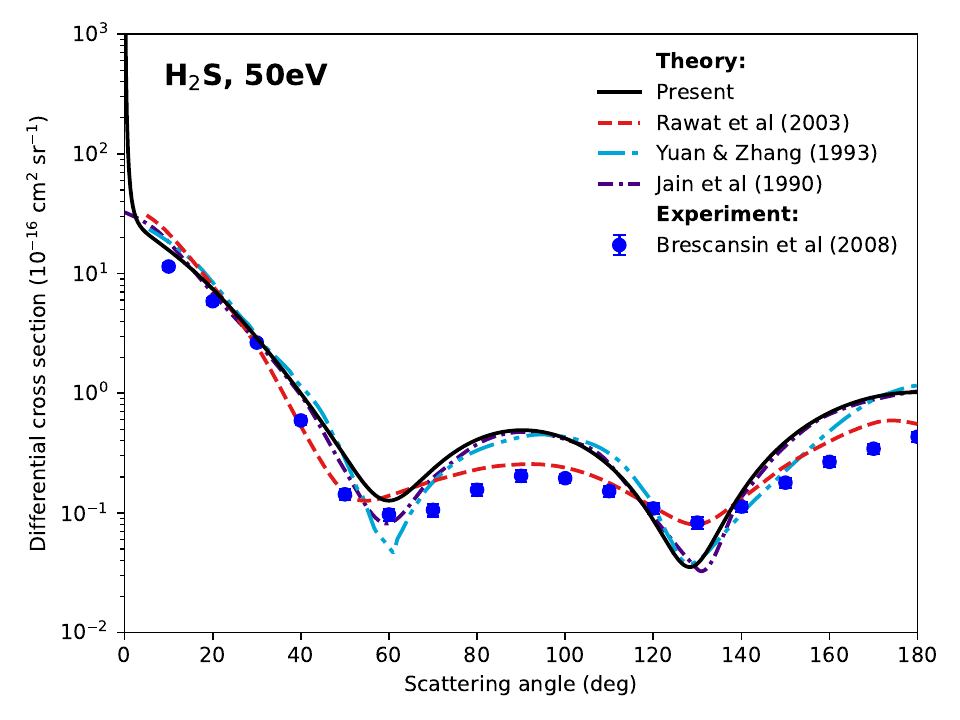}\\
\includegraphics[width=0.32\linewidth]{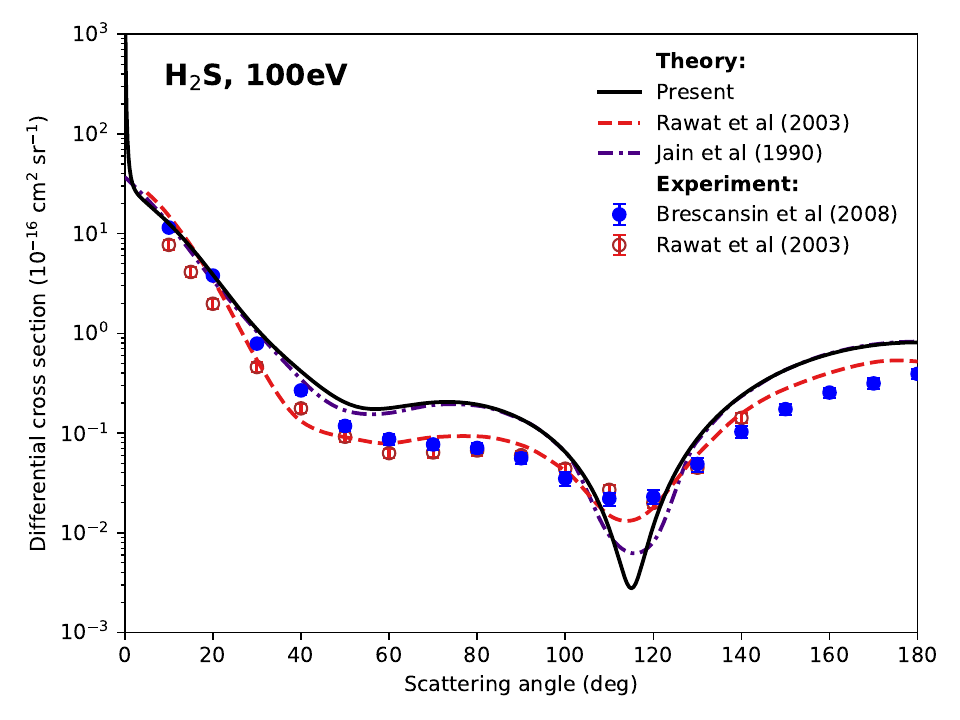}
\includegraphics[width=0.32\linewidth]{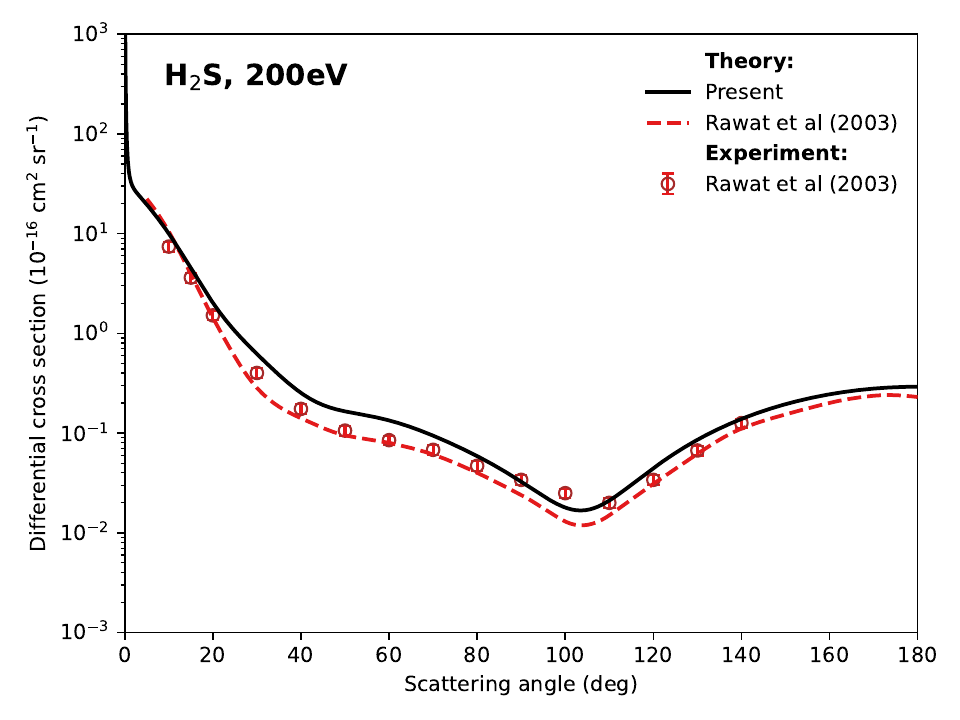}
\includegraphics[width=0.32\linewidth]{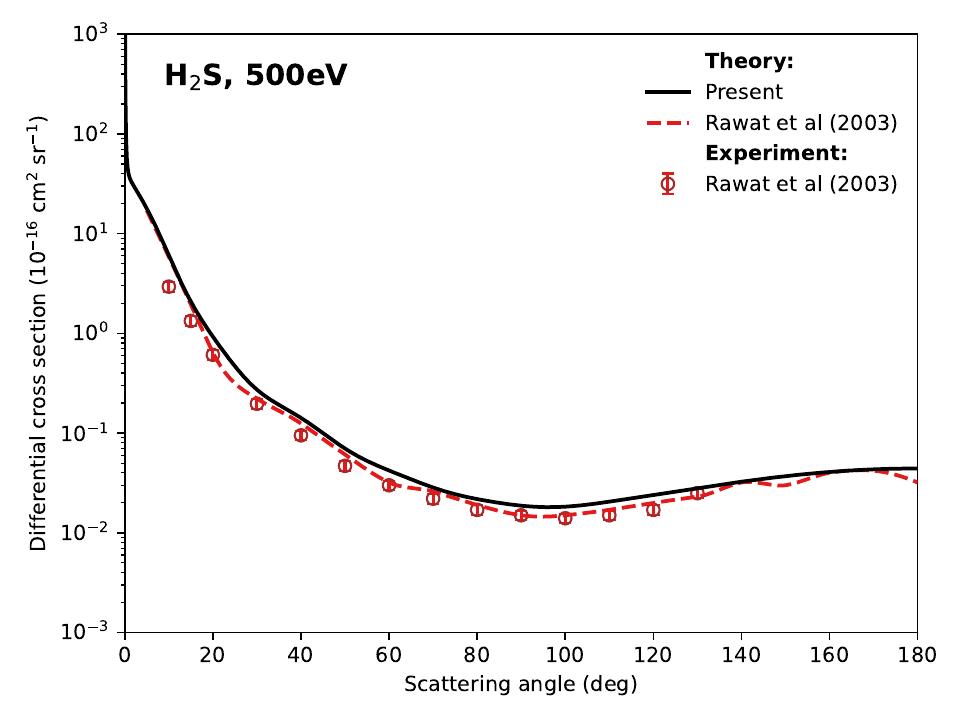}\\
\caption{\label{fig:h2s_dcs}Vibrationally elastic DCS for electron--$\rm H_2S$ scattering at selected incident energies, as labeled in each panel. Symbols are the same as Fig. \ref{fig:h2s_cs} except available data: filled diamond,  Cho  \emph{et al.} \cite{ChoJKPS2005}.}
\end{figure*}

The DCS comparison in Fig. \ref{fig:h2s_dcs} shows that the quality of the central angular distribution improves strongly with energy. At 0.5 eV the present calculation and the rotating-dipole result of Yuan and Zhang are very close over almost the complete angular range. Both show the expected strong forward enhancement and a slowly varying large-angle tail. This is a useful low-energy validation of the rotational contribution because the comparison is made before electronic absorption becomes important.

Between 5 and 20 eV the calculation reproduces the main angular structures as reported  by the measurements of Gulley and Cho and calculated by Rawat and Yuan--Zhang. At 5 eV, however, the present minima near the backward hemisphere is too deep and sharp. At 10 eV the broad minimum structure and backward recovery are reproduced, although the calculation is somewhat high at the smallest angles and differs around the second minimum. The 15 eV comparison is particularly good over the measured angular interval. At 20 eV, both observed minima are reproduced correctly in the calculation. These are precisely the details most sensitive to phase shifts from exchange and polarization.

At 50 and 100 eV the two-minimum structure is well established. The present curve reproduces the positions of the minima reasonably well, but it tends to lie above the Brescansin and Rawat data at intermediate and backward angles and makes the second minimum, near the $110^\circ$--$120^\circ$ region, deep at 100 eV. By 200 and 500 eV the agreement improves markedly. At 500 eV the present curve, the Rawat calculation, and the experiment are close over nearly the full angular range. This progression is consistent with the intended domain of the optical potential: detailed low- to intermediate-energy minima expose local-potential limitations, while the high-energy behavior is governed by smoother central scattering for which the model is more reliable.

% ----------------------------------------------------------------------------
\subsection{\label{subsec:result-nh3}Ammonia}
% ----------------------------------------------------------------------------
\begin{figure*}
\includegraphics[width=0.32\linewidth]{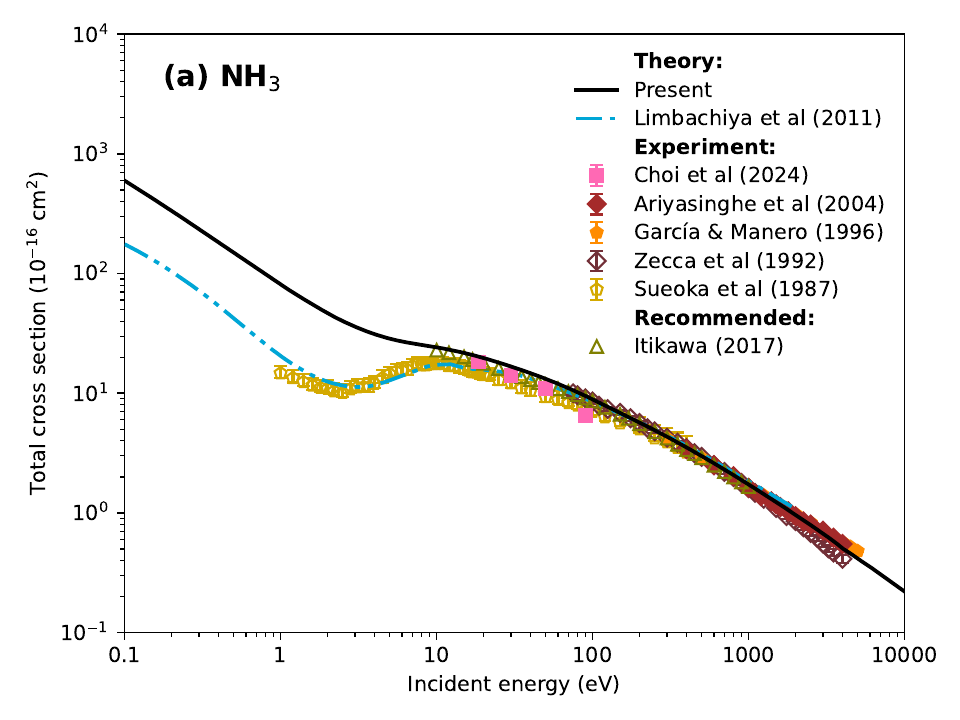}
\includegraphics[width=0.32\linewidth]{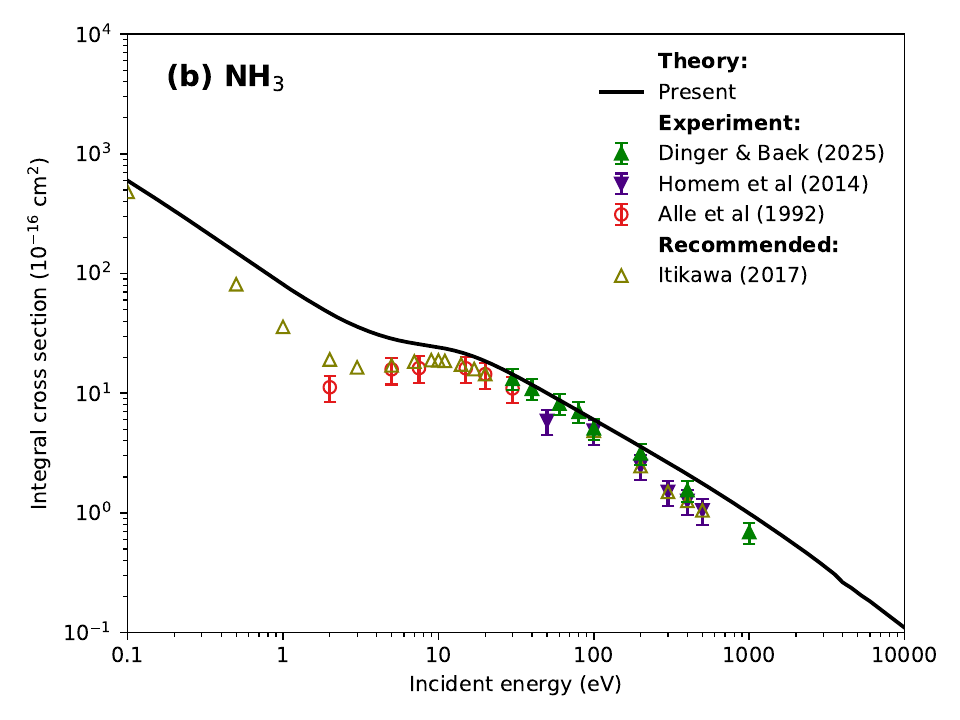}
\includegraphics[width=0.32\linewidth]{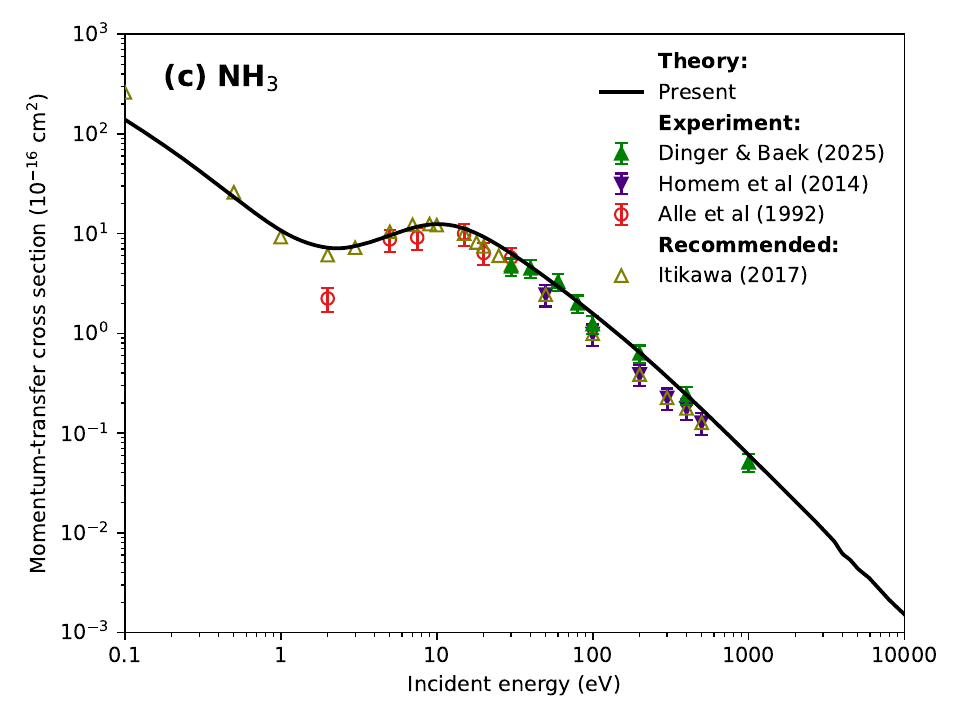}
\caption{\label{fig:nh3_cs}Total (a), integral (b), and momentum-transfer (c) cross sections for electron--$\mathrm{NH_3}$ scattering. The present calculation (solid line) is compared with the available data: dash-dot-dotted line, Limbachiya  \emph{et al.} \cite{LimbachiyaPRA2011}; filled square, Choi  \emph{et al.} \cite{ChoiEPJD2024}; filled diamond, Ariyasinghe  \emph{et al.} \cite{AriyasingheNIMPRSB2004}; filled pentagon, Garc\'{i}a and Manero \cite{GarciaJPB1996}; open diamond, Zecca  \emph{et al.} \cite{ZeccaPRA1992}; open pentagon, Sueoka  \emph{et al.} \cite{SueokaJPB1987}; open triangle-up, Itikawa \cite{ItikawaJPCRD2017}; filled triangle-up, Dinger and Baek \cite{DingerPRA2025}; filled triangle-down, Homem  \emph{et al.} \cite{HomemPRA2014}; open circle, Alle  \emph{et al.} \cite{AlleJPB1992}.}
\end{figure*}
Figure \ref{fig:nh3_cs} shows that the calculated TCS rises much more rapidly below about 10 eV than the direct transmission data of Sueoka \emph{et al.} and the later intermediate-energy measurements \cite{SueokaJPB1987,ChoiEPJD2024}. Part of this difference has a known experimental origin. Itikawa emphasized that the measured TCS of a strongly polar molecule can be underestimated when very-forward rotationally scattered electrons are accepted as transmitted particles, and therefore treated the low-energy $\rm NH_3$ transmission data with caution \cite{ItikawaJPCRD2017}. Choi \emph{et al.} found that a forward-angle correction remains significant even between 18.5 and 90 eV \cite{ChoiEPJD2024}. The present calculation includes this rotational forward contribution.

\begin{figure*}
\includegraphics[width=0.32\linewidth]{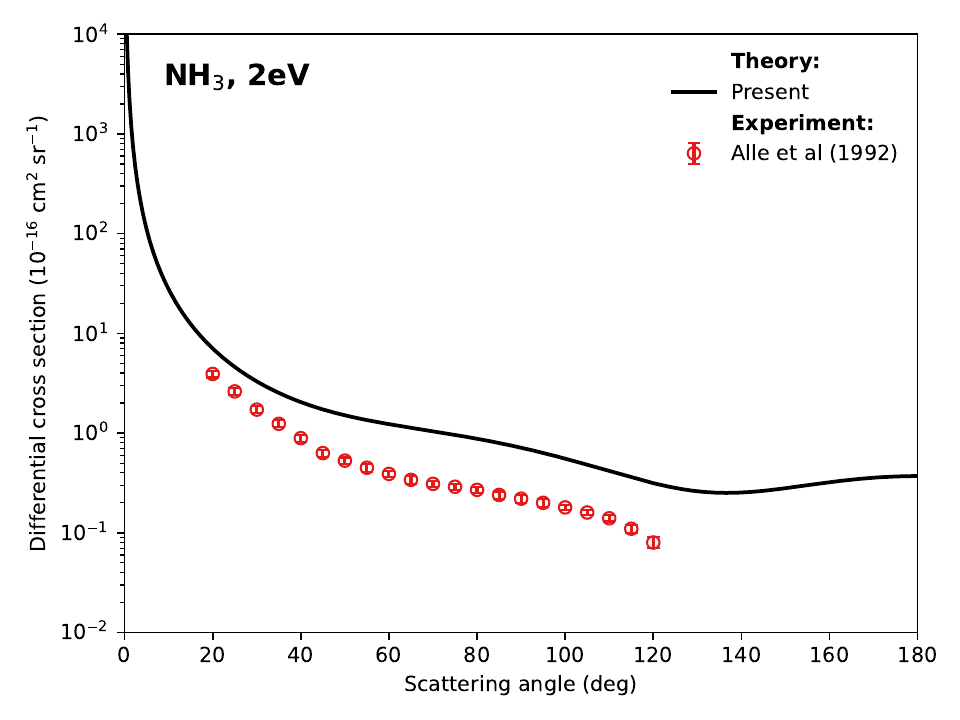}
\includegraphics[width=0.32\linewidth]{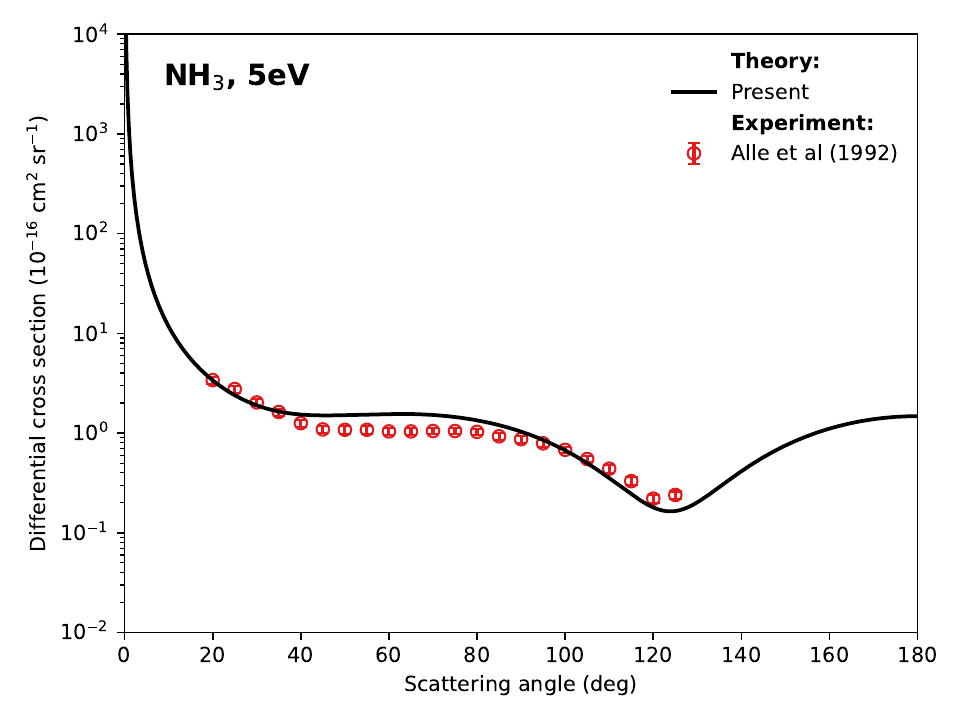}
\includegraphics[width=0.32\linewidth]{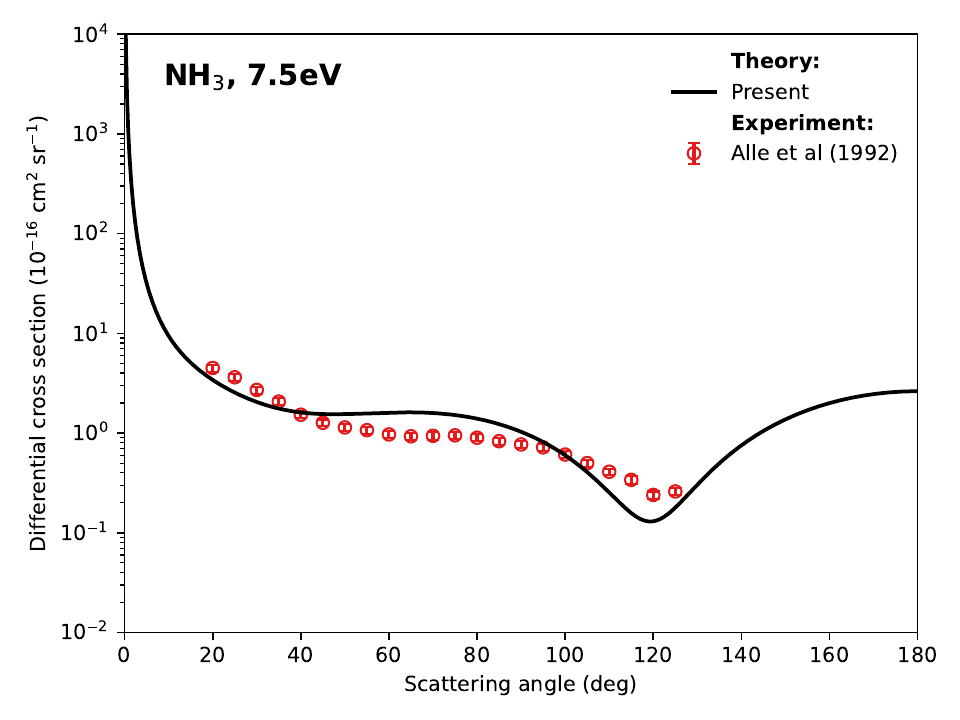}\\
\includegraphics[width=0.32\linewidth]{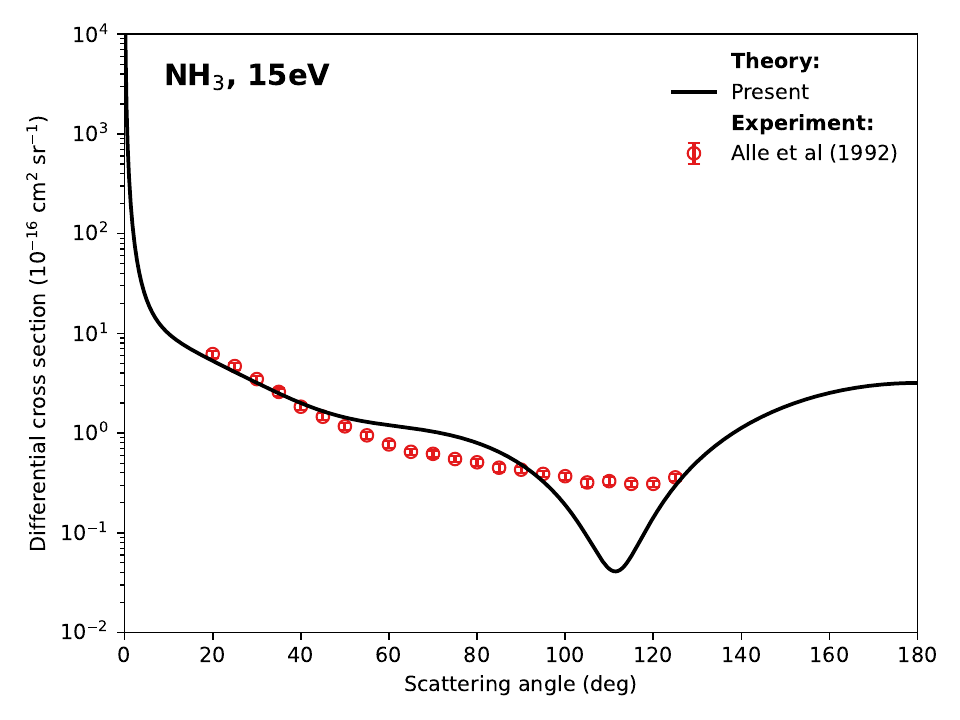}
\includegraphics[width=0.32\linewidth]{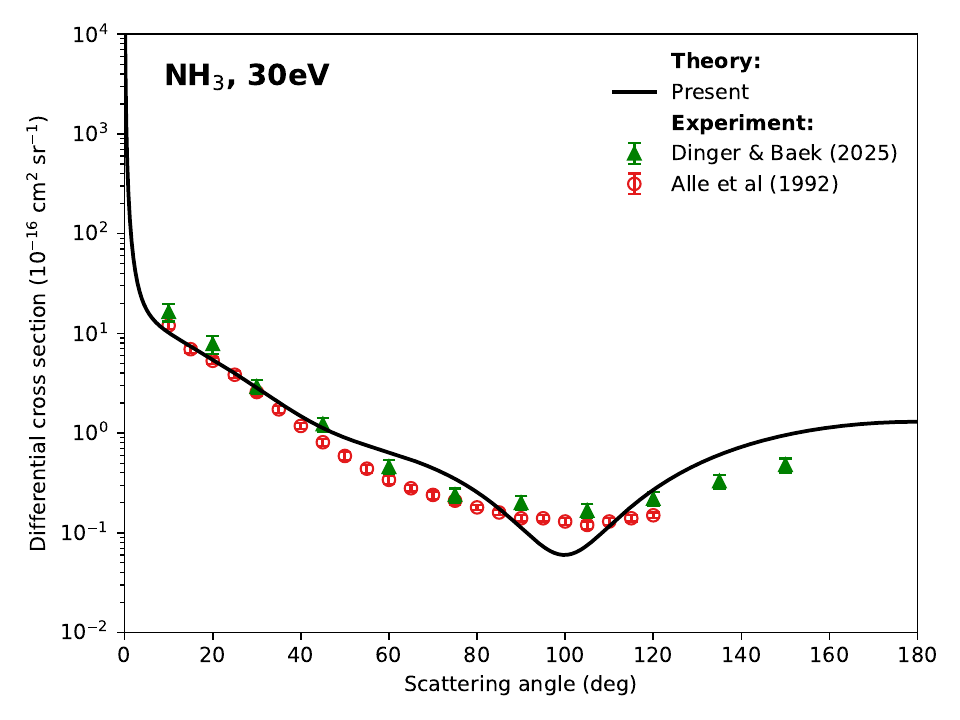}
\includegraphics[width=0.32\linewidth]{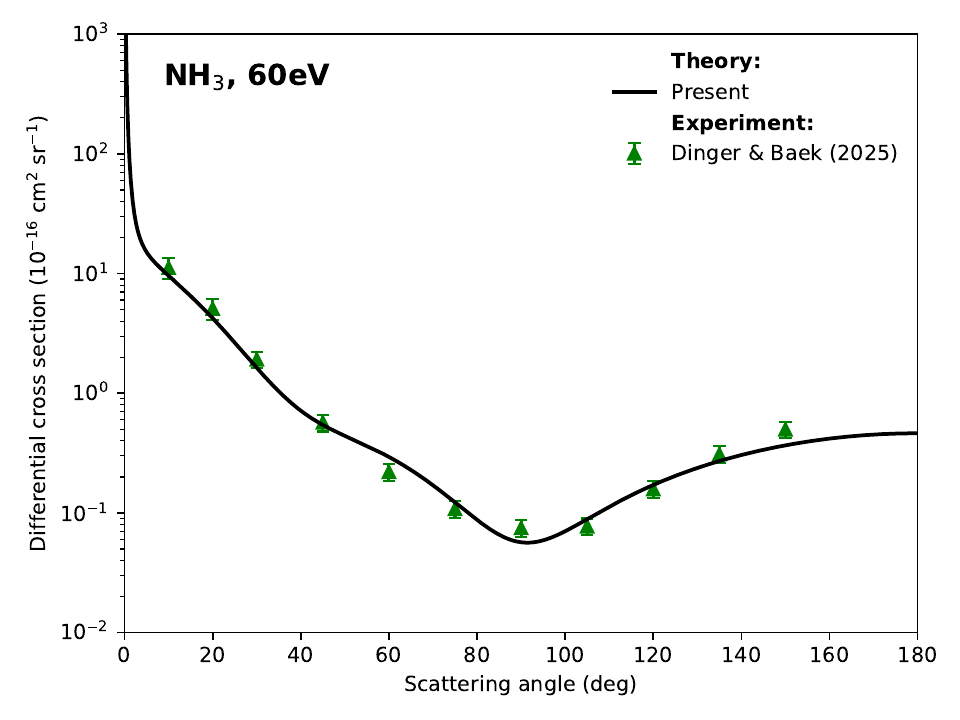}\\
\includegraphics[width=0.32\linewidth]{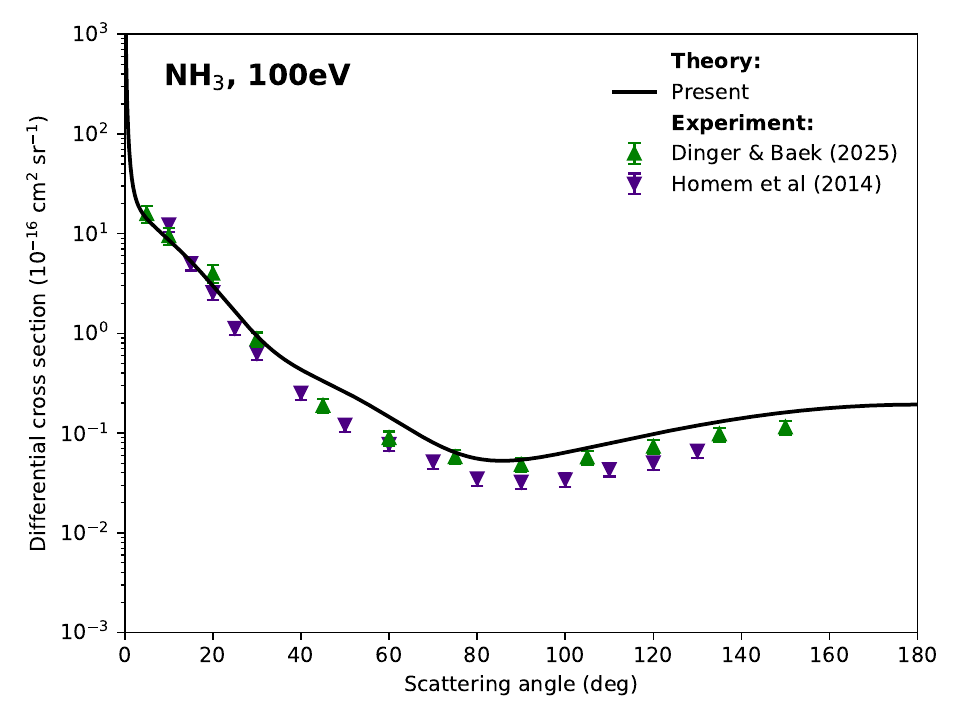}
\includegraphics[width=0.32\linewidth]{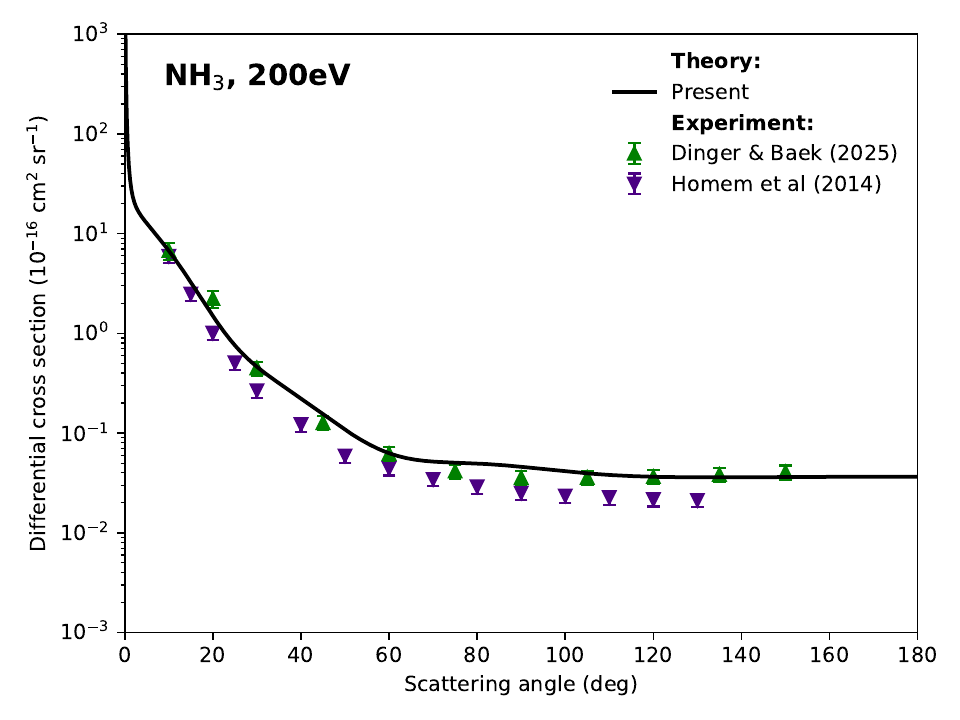}
\includegraphics[width=0.32\linewidth]{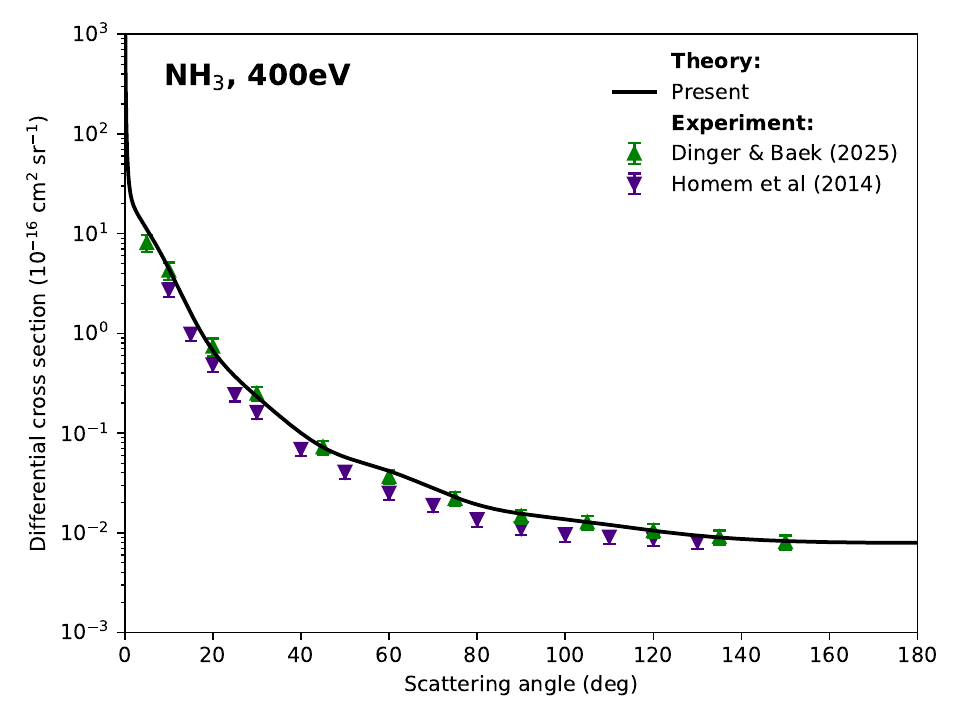}\\
\caption{\label{fig:nh3_dcs}Vibrationally elastic DCS for electron--$\rm NH_3$ scattering at selected incident energies, as labeled in each panel. Symbols are the same as Fig. \ref{fig:nh3_cs}.}
\end{figure*}

The low-energy overestimation cannot, however, be assigned entirely to experimental acceptance. The ICS in the middle panel is also well above the Alle, Homem, and Dinger--Baek scale at the lowest energies, and Fig. \ref{fig:nh3_dcs} shows that the present DCS itself is too large over the measured finite-angle range at 2 eV. This is a qualitatively different situation from the water case: the discrepancy is not confined to an unmeasured sub-degree cone. This discrepancy at 2 eV is also observed in different theoretical investigations including close-coupling approaches \cite{ChenPSST2023}. The narrow Feshbach resonance observed by Alle \emph{et al.} at $5.59\pm0.05$~eV is one example of dynamics that a smooth local optical potential is not designed to reproduce in detail \cite{AlleJPB1992}. It shows that the local central potential plus first-Born rotational term is incomplete in the near-threshold region, where channel coupling and resonance physics are important.

The comparison improves quickly with energy. In the TCS panel, the present result moves into the experimental/recommended band through the tens-of-eV region and follows the García--Manero, Ariyasinghe, Zecca, and Choi trends into the keV range \cite{GarciaJPB1996,AriyasingheNIMPRSB2004,ZeccaPRA1992,ChoiEPJD2024}. The high-energy datasets themselves are not perfectly consistent: García--Manero and Ariyasinghe lie above the Zecca scale at the upper end. The present curve passes through this experimental spread rather than following one dataset. In the ICS and MTCS panels, the difference from the experiment and the recommendation from Itikawa also decreases with energy. The MTCS is again the more stable quantity because forward scattering carries less weight.

The DCSs in Fig. \ref{fig:nh3_dcs} make the energy dependence of the discrepancy particularly clear. At 2 eV the present curve lies substantially above the Alle measurements across essentially the whole measured angular interval. This is the largest systematic finite-angle disagreement among the four targets. At 5 eV, in contrast, the measured and calculated DCSs are much closer over the forward and intermediate angles. The calculation follows the broad angular envelope around the energy of the Feshbach feature, but it should not be interpreted as resolving that narrow resonance. At 7.5 and 15 eV the overall shape is reproduced, including the decrease toward the backward-angle minimum, although the calculated minimum is deeper and the backward recovery is stronger than indicated by the data.

At 30 eV the forward and mid-angle magnitude is reasonable, but the present calculation rises more strongly in the backward region than the Alle and Dinger--Baek datasets. The 60 eV comparison is one of the best in the entire $\rm NH_3$ set: the calculated curve follows the Dinger--Baek data through the forward decrease, the broad minimum near $90^\circ$, and the backward recovery. At 100 and 200 eV the calculation remains close to the measured angular dependence, with a modest tendency to overestimate the Homem measurements around intermediate angles. By 400 eV the overall agreement in both shape and magnitude is very good. These results support a clear domain statement: the present broad-range model is very useful once the collision is outside the strongest near-threshold multichannel region, but it should not replace a close-coupling treatment for resonance-resolved $\rm NH_3$ scattering below a few electronvolts.

% ----------------------------------------------------------------------------
\subsection{\label{subsec:result-ph3}Phosphine}
% ----------------------------------------------------------------------------
\begin{figure*}
\includegraphics[width=0.32\linewidth]{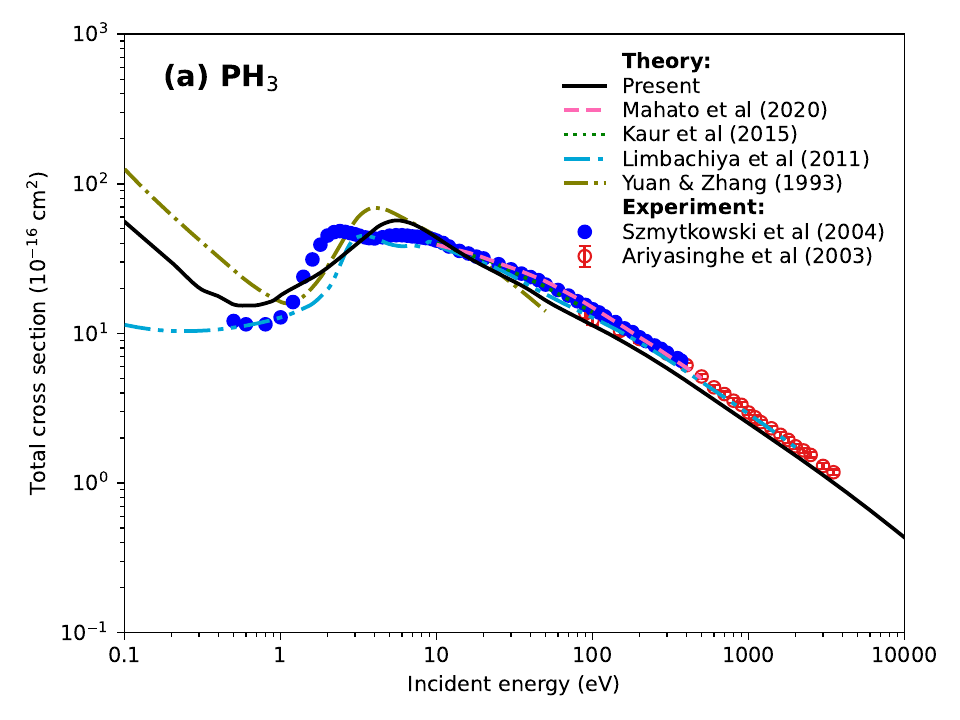}
\includegraphics[width=0.32\linewidth]{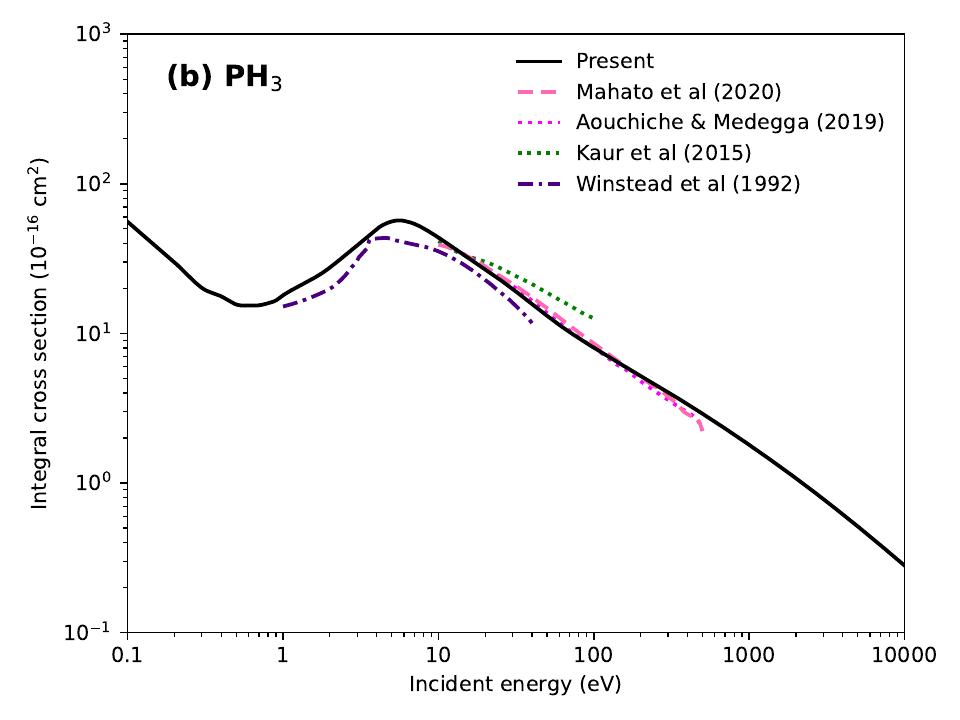}
\includegraphics[width=0.32\linewidth]{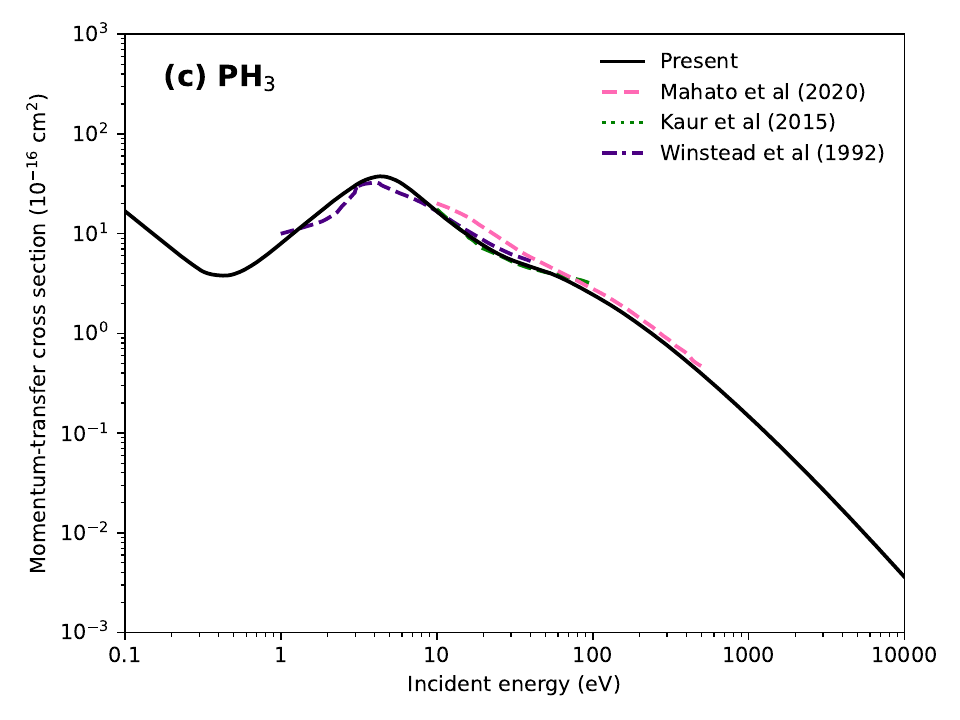}
\caption{\label{fig:ph3_cs}Total (a), integral (b), and momentum-transfer (c) cross sections for electron--$\mathrm{PH_3}$ scattering. The present calculation (solid line) is compared with the available data: short-dashed line, Mahato  \emph{et al.} \cite{MahatoJPB2020}; short-dotted green line, Kaur  \emph{et al.} \cite{KaurPRA2015}; dash-dot-dotted line, Limbachiya  \emph{et al.} \cite{LimbachiyaPRA2011}; dash-dotted line, Yuan and Zhang \cite{YuanZPDAMC1993}; short-dotted pink line, Aouchiche and Medegga \cite{AouchicheRJPCA2019}; short-dash-dotted line, Winstead  \emph{et al.} \cite{WinsteadZPDAMC1992}; filled circle, Szmytkowski  \emph{et al.} \cite{SzmytkowskiJPB2004}; open circle, Ariyasinghe  \emph{et al.} \cite{AriyasinghePRA2003}.}
\end{figure*}
Phosphine has the smallest permanent dipole moment of the four molecules, but its low-energy TCS contains pronounced resonance structure. Figure \ref{fig:ph3_cs} shows a shallow low-energy minimum followed by a broad maximum in the present calculation. The absolute measurements of Szmytkowski \textit{et al.}, however, exhibit a much sharper first maximum near 2.4 eV and a second broader structure in the 5--7 eV region \cite{SzmytkowskiJPB2004}. The present smooth optical-potential curve reproduces the broad envelope and the higher-energy hump but does not reproduce the sharp 2.4 eV structure. This difference is important: it is evidence for resonance dynamics in the central short-range collision. Kaur \emph{et al.} and Limbachiya \emph{et al.} likewise associate the few-eV structure with Ramsauer--Townsend and shape-resonance behavior \cite{KaurPRA2015,LimbachiyaPRA2011}.

Above about 10 eV the TCS comparison is much better. The present curve follows the Szmytkowski data through the intermediate-energy decline and joins smoothly onto the Ariyasinghe measurements up to the keV region \cite{AriyasinghePRA2003}. It also lies within the spread of the Kaur, Limbachiya, and Mahato optical-model calculations. This high-energy behavior is a useful check on the group-additive central potential because the rotational correction is a small fraction of the total scattering processes.

The ICS and MTCS cannot be tested against an equally complete absolute experimental set. Instead, the present results are compared with independent calculations. The ICS follows the Kaur, Mahato, and Aouchiche--Medegga trends closely over their respective overlap regions, and the MTCS likewise stays within the theoretical band defined by Kaur, Mahato, and the earlier Schwinger-multichannel work. The agreement becomes especially close above tens to hundreds of electronvolts. These comparisons support the numerical stability of the central calculation.

\begin{figure*}
\includegraphics[width=0.32\linewidth]{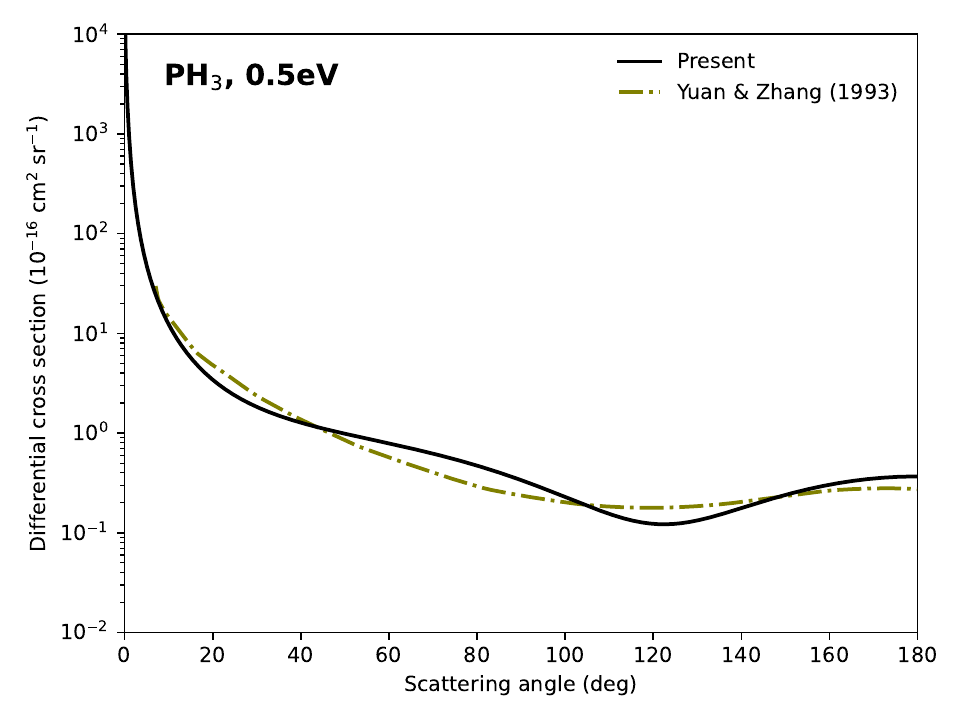}
\includegraphics[width=0.32\linewidth]{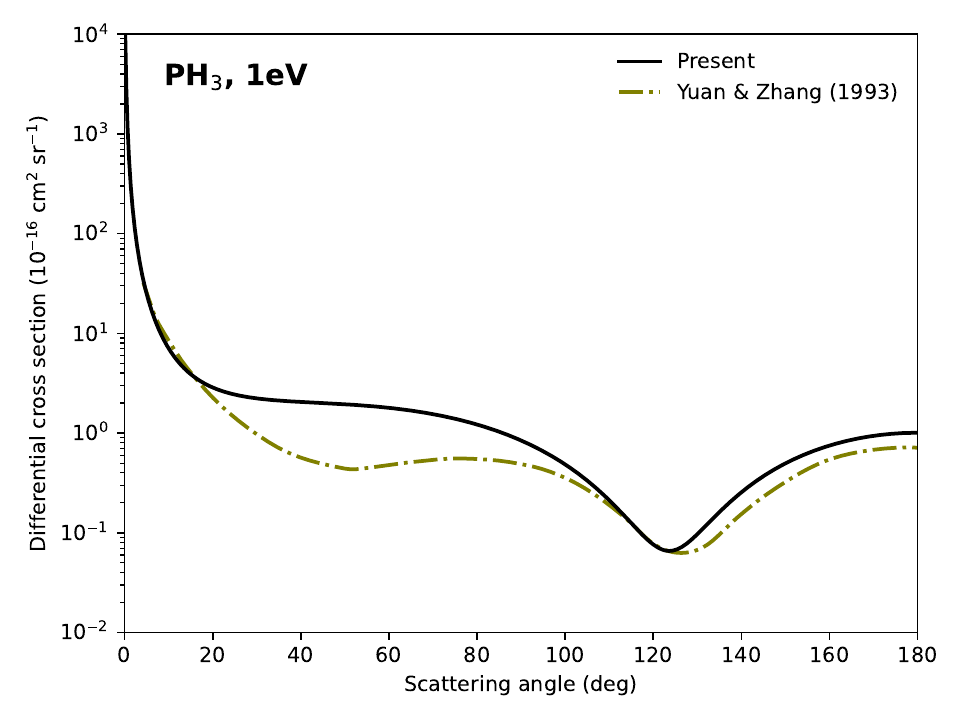}
\includegraphics[width=0.32\linewidth]{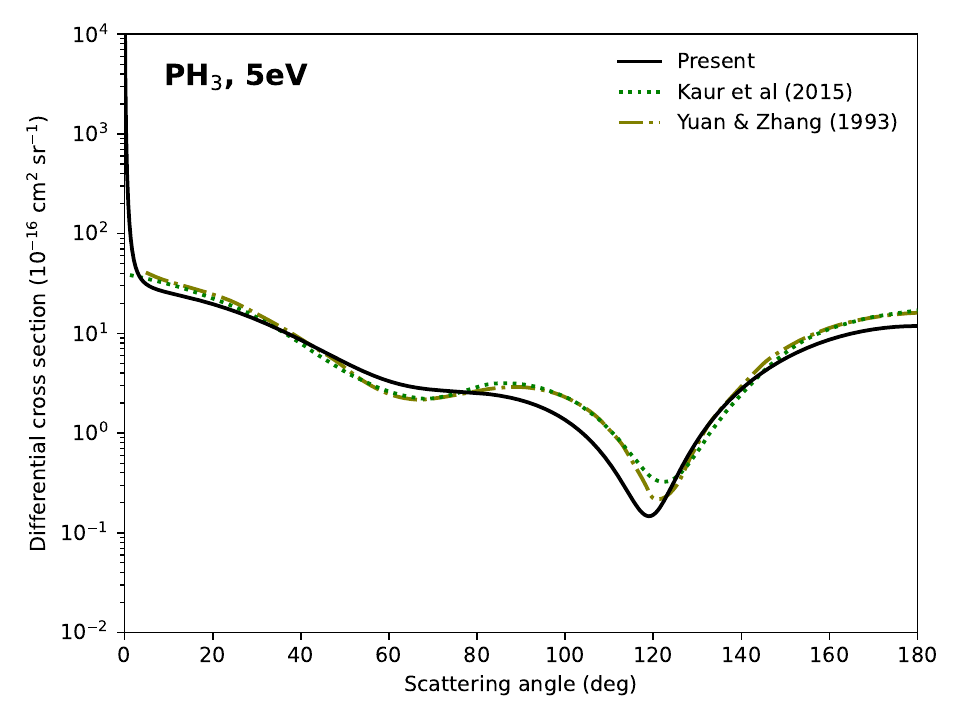}\\
\includegraphics[width=0.32\linewidth]{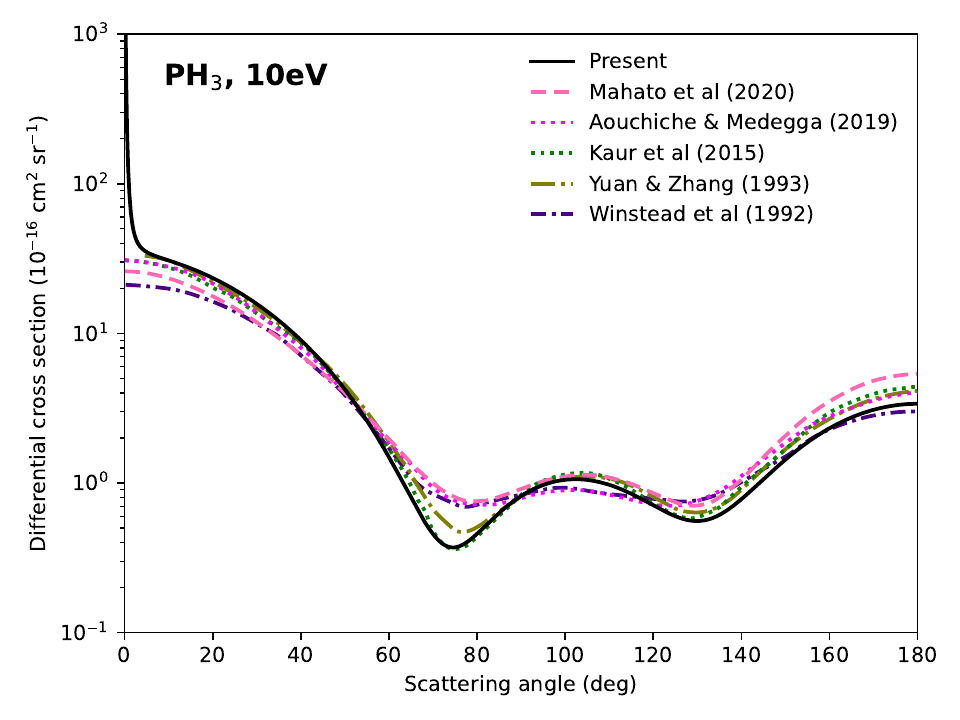}
\includegraphics[width=0.32\linewidth]{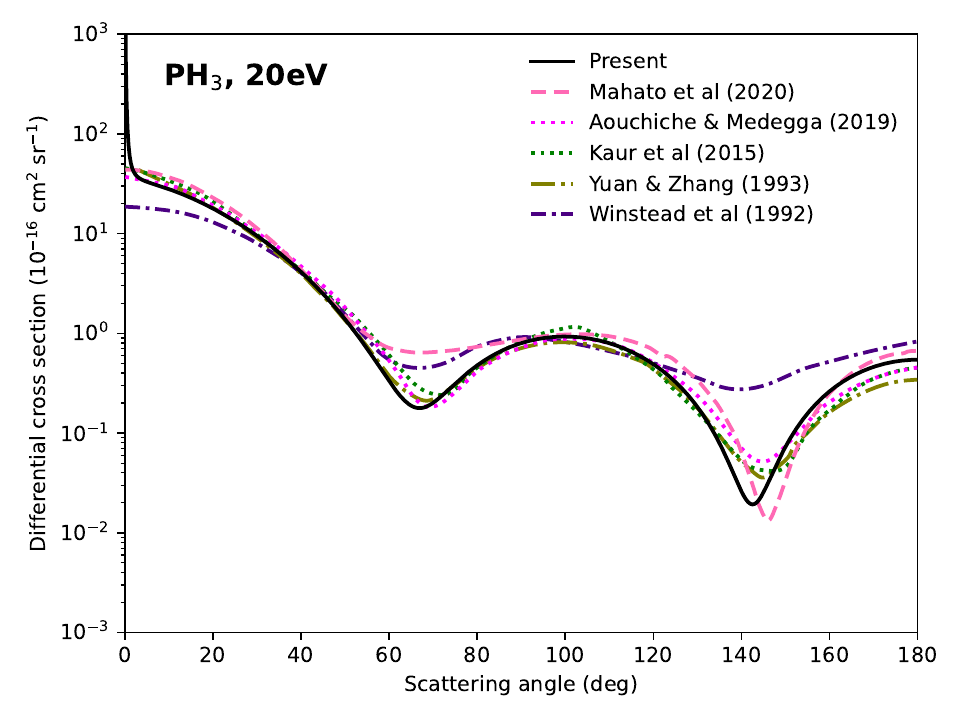}
\includegraphics[width=0.32\linewidth]{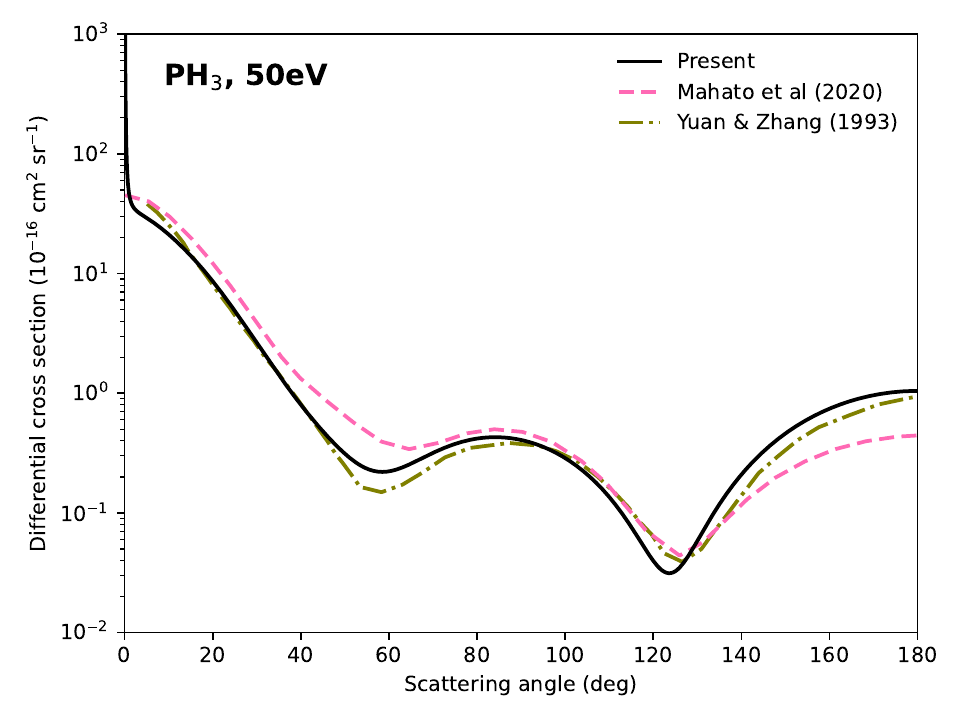}\\
\includegraphics[width=0.32\linewidth]{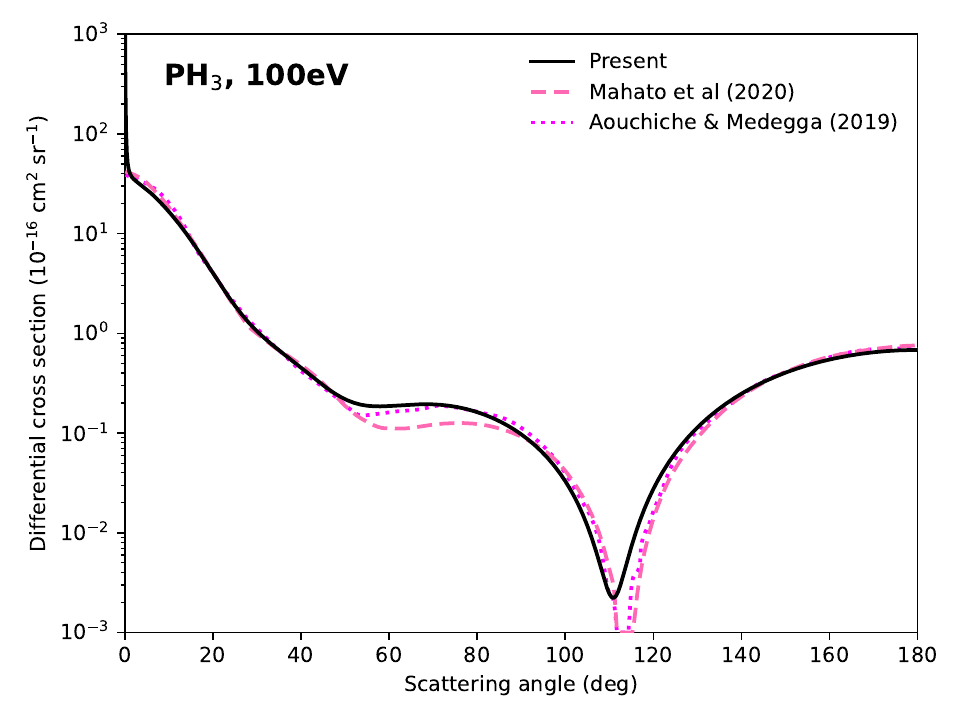}
\includegraphics[width=0.32\linewidth]{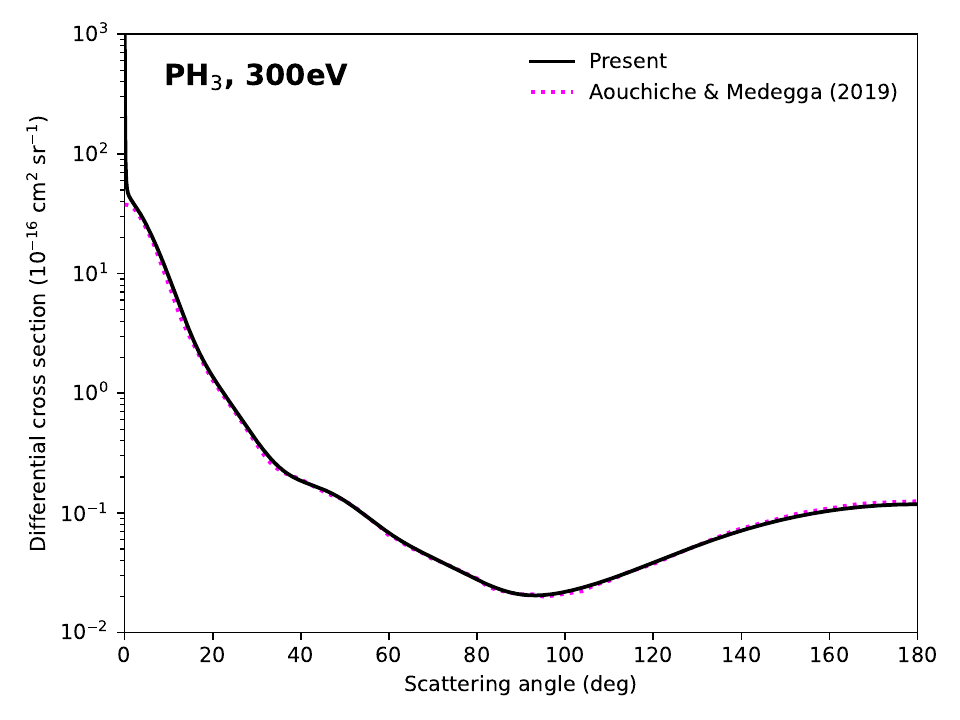}
\includegraphics[width=0.32\linewidth]{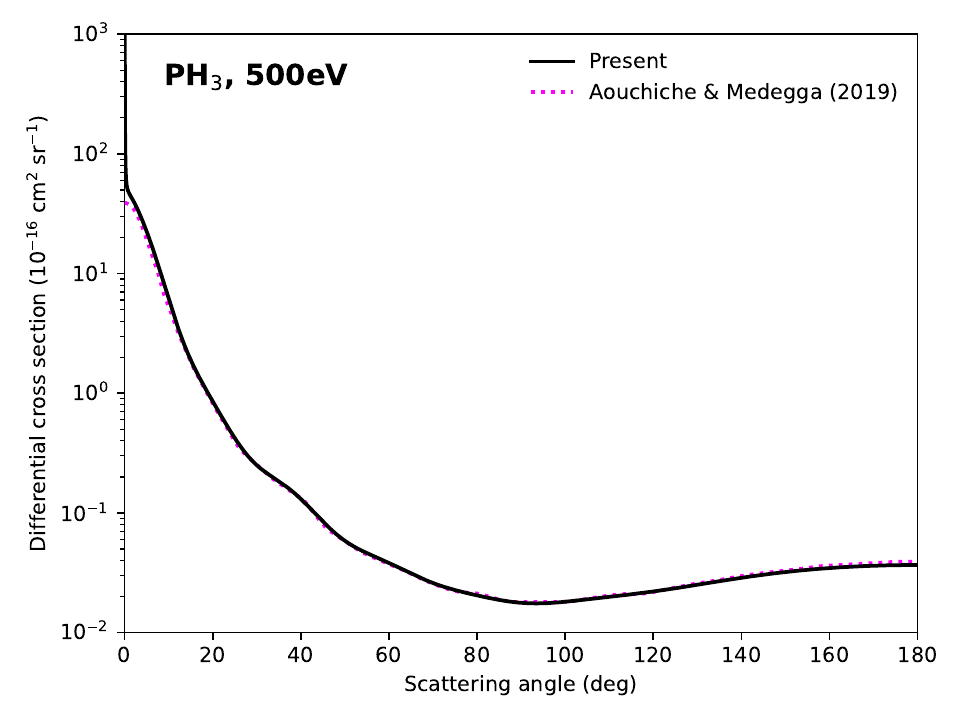}\\
\caption{\label{fig:ph3_dcs}Vibrationally elastic DCS for electron--$\rm PH_3$ scattering at selected incident energies, as labeled in each panel. Symbols are the same as Fig. \ref{fig:ph3_cs}.}
\end{figure*}

The $\rm PH_3$ DCSs depicted in Fig. \ref{fig:ph3_dcs} provide a useful theoretical cross-check over almost three orders of magnitude in energy. At 0.5 and 1 eV the present calculation and Yuan--Zhang share the same strong forward decrease and broad large-angle minimum, but the present result is higher over much of the intermediate and backward angular range. At 5 eV the agreement with Yuan--Zhang and Kaur is much closer, including the position of the principal backward-angle minimum. At 10 and 20 eV the available calculations form a relatively narrow band over most angles; the present curve lies within that band and reproduces the two-minimum angular structure.

At 50 eV, the models show a somewhat larger spread, especially between the two minima, but in the backward scattering region, the present result remains between the Yuan--Zhang and Mahato over most of the angular range. At 100 eV the present minimum near the backward region agrees well in position with Mahato and Aouchiche--Medegga, while its depth is slightly different. The agreement with Aouchiche--Medegga becomes nearly perfect at 300 and 500 eV. This high-energy agreement is expected because different local exchange and polarization choices become less important as the projectile energy rises. Since no absolute $\rm PH_3$ electron DCS measurement is available in the literature set used here, the 100--500 eV curves should be regarded as robust theoretical predictions awaiting direct experimental tests.

The four targets expose different limitations of the same approximation. For $\rm H_2O$, the most clearly visible integrated difference is between the ICS and MTCS: the ICS is very sensitive to the rotational forward cone, whereas the MTCS and finite-angle DCS show substantially better agreement. This is the behavior expected when long-range dipole scattering, rather than the entire central interaction, controls the remaining uncertainty. $\rm H_2S$ shows the same pattern more weakly. Its MTCS is generally closer to experiment than its ICS, while the detailed DCS minima reveal additional central-potential errors at intermediate energy. The agreement improves sharply by tens of electronvolts and is very good in several of the 60--400~eV angular comparisons. $\rm PH_3$ shows the complementary limitation: the broad optical potential gives a sensible smooth envelope but does not reproduce the sharp 2.4-eV experimental resonance structure. This is the clearest indication that resonance and close-coupling dynamics set the lower practical energy limit of the present local optical model for a specific target. Above about 10~eV, however, its integrated and differential cross sections converge well with both experiment and independent optical-potential calculations.

At high incident energy the four targets display a common convergence toward the available measurements and broad-range calculations. The rotational term becomes confined to very small angles and contributes little to MTCS or TCS, while the calculated DCSs are governed by the central potential. The reasonable comparisons for $\rm H_2S$ and $\rm NH_3$  at hundreds of electronvolts and the near-coincidence of the $\rm PH_3$ calculation with Aouchiche--Medegga at 300--500~eV are therefore meaningful tests of the underlying Dirac--SCOP treatment rather than of the rotational correction. Taken together, the results support the intended use of the method as a fast, continuous broad-range model. They also define where a more detailed calculation is needed: near individual rotational thresholds, narrow Feshbach resonances, strong shape resonances, or angular minima whose depth depends sensitively on nonlocal exchange and multichannel polarization.

% ============================================================================
\section{\label{sec:conclusion}Conclusion}
% ============================================================================
We have extended the relativistic group-additivity SCOP method from nonpolar targets to molecules with permanent dipole moments. The central Dirac--SCOP calculation is left unchanged. The missing anisotropic long-range dipole interaction is added as a rotationally resolved first-Born sum. The required rotational transition thresholds and state-to-state dipole strengths come from HITRAN database, and thermal populations are rebuilt at the chosen temperature. This keeps the polar correction tied to spectroscopic data. We applied the method to water, hydrogen sulfide, ammonia, and phosphine from 0.1 eV to 10 keV. The reported differential, integral, and momentum-transfer cross sections are vibrationally elastic and rotationally state-unresolved. The forward-angle contribution strongly affects the water and ammonia integral cross sections, and the sharp few-eV phosphine resonance structure is not resolved by the local optical potential. Overall, the calculations follow the main experimental and recommended trends across the four molecules. The largest remaining differences occur in low-energy region, where resonance structure, channel coupling, or experimental angular acceptance is likely to matter most. The main point of the work is practical. A state-resolved rotational Born sum can be added to the same relativistic optical-potential results over the full energy range without changing scattering formalisms. Water and ammonia show why the long-range rotational term is needed. Hydrogen sulfide and phosphine test whether the same calculation can also handle stronger induced polarization and heavier central atoms. The method remains fast enough for broad energy grids while giving a physically traceable treatment of the permanent dipole. There are limits. A local optical potential will not replace a detailed close-coupling calculation near a narrow Feshbach resonance, a strong shape resonance, or an individual rotational threshold. However, it could effectively resolve the forward angle cross-section problem.  Future work should test larger polar molecules, and study other effect more closely.

% \begin{acknowledgments}
% \end{acknowledgments}

\appendix

% ============================================================================
\section{\label{app:fixed-nuclei}Fixed-nuclei point-dipole limit and the forward-divergence problem}
% ============================================================================
All calculations in this article use the rotationally resolved treatment of Sec. \ref{subsec:rotational}. This appendix only explains why a fixed-nuclei point-dipole term becomes troublesome when it is integrated over angle.

At long range the interaction of an electron with a body-fixed permanent dipole $\bm \mu_D$ is
\begin{equation}
 V_d(\bm r)=\frac{\bm \mu_D\cdot\hat{\bm r}}{r^2}.
 \label{eq:fd-potential}
\end{equation}
In an orientation-averaged first-Born treatment with no rotational energy transfer, $k_f=k_i=k$ and the dipole DCS takes the familiar form \cite{TakayanagiAAMP1970,NorcrossAAMP1982}
\begin{equation}
 \left(\frac{d\sigma}{d\Omega}\right)_{\rm FN}
 =\frac{2\mu_D^2}{3k^2(1-\cos\theta)}.
 \label{eq:fn-dcs}
\end{equation}
Since $(1-\cos\theta)\simeq\theta^2/2$ at small $\theta$, Eq. \eqref{eq:fn-dcs} diverges as $1/\theta^2$. Its integral over a finite angular acceptance is
\begin{equation}
 \sigma_{\rm FN}(\theta_{\min},\theta_{\max})
 =\frac{8\pi \mu_D^2}{3k^2}
 \ln\!\left[
 \frac{\sin(\theta_{\max}/2)}{\sin(\theta_{\min}/2)}
 \right],
 \label{eq:fn-integral}
\end{equation}
which depends logarithmically on the minimum angle and has no finite $\theta_{\min}\rightarrow0$ limit.

The scale of the effect for $\rm H_2S$ was quantified explicitly by Yuan and Zhang \cite{YuanZPDAMC1993}. In their rotating-dipole calculation at 0.5 eV, scattering in the intervals $0^\circ$--$1^\circ$, $0^\circ$--$5^\circ$, and $0^\circ$--$10^\circ$ contributed approximately 30\%, 50\%, and 60\% of their total vibrationally elastic cross section. The same study found the momentum-transfer cross section to be much less sensitive to this missing forward cone because of its $(1-\cos\theta)$ weight. Thus a comparison of theoretical and transmission TCSs for a polar molecule must respect the experimental acceptance, while MTCS provides a substantially more robust test of the wider-angle dynamics.

The ammonia literature provides an equally direct experimental illustration. Itikawa \cite{ItikawaJPCRD2017} concluded that transmission-derived $\rm NH_3$ TCS below about 10 eV can carry a large uncertainty because the forward-scattered electrons are not cleanly separated from the incident beam, and therefore recommended the transmission TCS only above 10 eV. Thus the acceptance-angle issue is not confined to the sub-eV domain: for a strongly polar target such as $\rm NH_3$ it can remain experimentally visible well into the tens-of-eV range.

This is not just a numerical issue. A real gas-phase experiment has finite angular and energy resolution, and the long-range dipole interaction couples rotational channels that are not degenerate. Once a nonzero state-to-state energy transfer is retained, Eq. \eqref{eq:qif} gives $q_{if}(0)=|k_i-k_f|$ for an inelastic rotational transition; the physical rotational splitting therefore regularizes the forward denominator. This is why this work builds the polar contribution from spectroscopic rotational transitions rather than assigning a universal small-angle cutoff to Eq. \eqref{eq:fn-dcs}. Close to individual thresholds, however, first-Born dynamics can still be incomplete, as shown explicitly for water by the $R$-matrix calculations of Faure \emph{et al.} \cite{FaureMNRAS2004}. The appendix result should therefore be read as a statement about the ill-defined full-angle fixed-nuclei integral, not as a claim that a first-Born rotational treatment is exact arbitrarily close to threshold.

% ============================================================================
\section{\label{app:literature-map}Review of previous electron-scattering studies}
% ============================================================================
The literature for the four targets is too extensive to reproduce usefully in the Introduction. Table \ref{tab:literature-summary} therefore provides an expanded chronological map of studies that report at least one observable directly comparable with the present work: elastic or vibrationally elastic DCS, ICS, MTCS, rotational excitation, optical-model absorption, or total scattering. In this sense the table is intended to be comprehensive for the observables considered here rather than an exhaustive bibliography of every electron-collision channel.

\begingroup
\scriptsize
\setlength{\tabcolsep}{1pt}
\renewcommand{\arraystretch}{1.07}

% \newlength{\textwidth}
% \setlength{\textwidth}{\columnwidth}
\newcommand{\LSource}[1]{\parbox[t]{0.275\textwidth}{\raggedright #1}}
\newcommand{\LType}[1]{\parbox[t]{0.060\textwidth}{\centering #1}}
\newcommand{\LMethod}[1]{\parbox[t]{0.240\textwidth}{\raggedright #1}}
\newcommand{\LData}[1]{\parbox[t]{0.275\textwidth}{\raggedright #1}}
\newcommand{\LEnergy}[1]{\parbox[t]{0.125\textwidth}{\raggedright #1}}

\begin{longtable*}{lllll}
\caption{\label{tab:literature-summary}
Selected electron-scattering studies for the benchmark polar hydrides.
M, measurement; T, theory; E, evaluation.}\\
\hline\hline
\LSource{\textbf{Year and authors}} &
\LType{\textbf{Type}} &
\LMethod{\textbf{Method}} &
\LData{\textbf{Reported data}} &
\LEnergy{\textbf{Energy range}} \\
\hline
\endfirsthead

\multicolumn{5}{c}{\textbf{TABLE \thetable\ (continued)}}\\
\hline\hline
\LSource{\textbf{Year and authors}} &
\LType{\textbf{Type}} &
\LMethod{\textbf{Method}} &
\LData{\textbf{Reported data}} &
\LEnergy{\textbf{Energy range}} \\
\hline
\endhead

\hline
\multicolumn{5}{r}{\scriptsize Continued on next page}\\
\endfoot

\hline\hline
\endlastfoot

% ------------------------------------------------------------------
\multicolumn{5}{l}{\textbf{Water, $\mathrm{H_2O}$}}\\
\hline
\LSource{1982, Jung \emph{et al.} \cite{JungJPB1982}}
& \LType{M}
& \LMethod{Crossed beam \& EELS}
& \LData{Rot. \& rot.-vib. DCS}
& \LEnergy{0.5--6 eV} \\

\LSource{1985, Danjo--Nishimura \cite{DanjoJPSJ1985}}
& \LType{M}
& \LMethod{Crossed beam}
& \LData{El. DCS, ICS, MTCS}
& \LEnergy{4--200 eV} \\

\LSource{1986, Katase \emph{et al.} \cite{KataseJPB1986}}
& \LType{M/T}
& \LMethod{Crossed beam \& PW}
& \LData{Absolute el. DCS}
& \LEnergy{100--1000 eV} \\

\LSource{1986, Sueoka \emph{et al.} \cite{SueokaJPB1986}}
& \LType{M}
& \LMethod{TOF transmission}
& \LData{$e^\pm$ TCS}
& \LEnergy{1--400 eV} \\

\LSource{1987, Zecca \emph{et al.} \cite{ZeccaJPB1987}}
& \LType{M}
& \LMethod{Ramsauer-type technique}
& \LData{TCS}
& \LEnergy{81--3000 eV} \\

\LSource{1987, Szmytkowski \cite{SzmytkowskiCPL1987}}
& \LType{M}
& \LMethod{Linear transmission}
& \LData{TCS}
& \LEnergy{0.5--80 eV} \\

\LSource{1987, Gianturco--Scialla \cite{GianturcoJCP1987}}
& \LType{T}
& \LMethod{SCE+FNA+Born closure}
& \LData{El. DCS, MTCS}
& \LEnergy{2--20 eV} \\

\LSource{1987, Shyn--Cho \cite{ShynPRA1987}}
& \LType{M}
& \LMethod{Modulated crossed beam}
& \LData{Vib.-el. DCS, ICS, MTCS}
& \LEnergy{2.2--20 eV} \\

\LSource{1988, Nishimura--Yano \cite{NishimuraJPSJ1988}}
& \LType{M}
& \LMethod{Linear transmission}
& \LData{TCS}
& \LEnergy{7--500 eV} \\

\LSource{1990, Saglam--Aktekin \cite{SaglamJPB1990}}
& \LType{M}
& \LMethod{Linear transmission}
& \LData{Absolute TCS}
& \LEnergy{25--300 eV} \\

\LSource{1991, Johnstone--Newell \cite{JohnstoneJPB1991}}
& \LType{M}
& \LMethod{Crossed beam}
& \LData{Vib.-el. DCS, ICS, MTCS}
& \LEnergy{6--50 eV} \\

\LSource{1991, Saglam--Aktekin \cite{SaglamJPB1991}}
& \LType{M}
& \LMethod{Linear transmission}
& \LData{Absolute TCS}
& \LEnergy{4--20 eV} \\

\LSource{1992, Shyn--Grafe \cite{ShynPRA1992}}
& \LType{M}
& \LMethod{Modulated crossed beam}
& \LData{El. DCS, ICS, MTCS}
& \LEnergy{30--200 eV} \\

\LSource{1992, Rescigno--Lengsfield \cite{RescignoZPD1992}}
& \LType{T}
& \LMethod{Complex Kohn (FN)}
& \LData{El. DCS, MTCS}
& \LEnergy{2--20 eV} \\

\LSource{1993, Okamoto \emph{et al.} \cite{OkamotoJPB1993}}
& \LType{T}
& \LMethod{FNA-PW}
& \LData{Vib.-el. DCS, ICS, MTCS}
& \LEnergy{6--50 eV} \\

\LSource{1998, Gianturco \emph{et al.} \cite{GianturcoJCP1998}}
& \LType{T}
& \LMethod{SCE-FNA}
& \LData{El. \& rot.-inel. DCS, MTCS}
& \LEnergy{2--50 eV} \\

\LSource{1999, Varella \emph{et al.} \cite{Varella111JCP1999}}
& \LType{T}
& \LMethod{SMCPP \& Born closure}
& \LData{El. DCS, ICS, MTCS}
& \LEnergy{2--30 eV} \\

\LSource{2004, Faure \emph{et al.} \cite{FaureJPB2004}}
& \LType{T}
& \LMethod{$R$-matrix \& Born closure}
& \LData{Vib.-el. \& rot.-inel. DCS, ICS, MTCS}
& \LEnergy{$<7$ eV} \\

\LSource{2005, Machado \emph{et al.} \cite{MachadoEPJD2005}}
& \LType{T}
& \LMethod{OP \& SVM+DWA}
& \LData{El. DCS, ICS, MTCS; RE}
& \LEnergy{2--500 eV} \\

\LSource{2005, Itikawa--Mason \cite{ItikawaJPCRD2005}}
& \LType{E}
& \LMethod{Critical evaluation}
& \LData{Recommended CS set}
& \LEnergy{Quantity dependent} \\

\LSource{2006, \v{C}ur\'{i}k \emph{et al.} \cite{CurikPRL2006}}
& \LType{M/T}
& \LMethod{Cold-$e^-$ exp. \& QDT}
& \LData{Rotational CS}
& \LEnergy{0.017--0.25 eV} \\

\LSource{2007, Mu\~noz \emph{et al.} \cite{MunozPRA2007}}
& \LType{M/T}
& \LMethod{Transmission \& OP-FBA}
& \LData{tot., el., inel., ion., rot. CS}
& \LEnergy{M: 50--5000 eV; T: 1--10000 eV} \\

\LSource{2008, Silva \emph{et al.} \cite{SilvaPRL2008}}
& \LType{M}
& \LMethod{Relative flow}
& \LData{Rot.-avg. el. DCS, ICS}
& \LEnergy{1--100 eV} \\

\LSource{2008, Khakoo \emph{et al.} \cite{KhakooPRA2008}}
& \LType{M/T}
& \LMethod{Relative flow \& SMC}
& \LData{El. DCS, ICS, MTCS}
& \LEnergy{1--100 eV} \\

\LSource{2011, Vinodkumar \emph{et al.} \cite{VinodkumarEPJD2011}}
& \LType{T}
& \LMethod{$R$-matrix \& SCOP}
& \LData{Total ECS}
& \LEnergy{0.01 eV--2 keV} \\

\LSource{2016, Matsui \emph{et al.} \cite{MatsuiEPJD2016}}
& \LType{M}
& \LMethod{Angle-resolved EELS}
& \LData{El. \& exc. DCS, ICS}
& \LEnergy{2--100 eV} \\

\LSource{2019, Kadokura \emph{et al.} \cite{KadokuraPRL2019}}
& \LType{M}
& \LMethod{HR transmission}
& \LData{TCS, DCS}
& \LEnergy{3--100 eV} \\

\LSource{2021, Song \emph{et al.} \cite{SongJPCRD2021}}
& \LType{E}
& \LMethod{Critical evaluation}
& \LData{Recommended CS}
& \LEnergy{Quantity dependent} \\

\LSource{2022, Budde \emph{et al.} \cite{BuddeJPD2022}}
& \LType{E/T}
& \LMethod{Swarm fit \& 2-term BE}
& \LData{Complete LXCat CS set}
& \LEnergy{Quantity dependent} \\

\LSource{2023, Budde \emph{et al.} \cite{BuddeJPD2023}}
& \LType{T}
& \LMethod{LoKI-B \& MC simulation}
& \LData{Anisotropic CSs}
& \LEnergy{Quantity dependent} \\

\LSource{2023, Triggiani \emph{et al.} \cite{TriggianiFM2023}}
& \LType{T}
& \LMethod{DHF+HGF \& LS}
& \LData{Total elastic CS}
& \LEnergy{$10^{-3}$--$10^{3}$ eV} \\

\LSource{2025, Shorifuddoza \emph{et al.} \cite{ShorifuddozaEPJD2025}}
& \LType{T}
& \LMethod{IAM \& IAMS (ELSEPA)}
& \LData{DCS, ICS, MTCS, VCS, INCS, TCS, TICS}
& \LEnergy{1--10000 eV} \\

% ------------------------------------------------------------------
\hline
\multicolumn{5}{l}{\textbf{Hydrogen sulfide, $\mathrm{H_2S}$}}\\
\hline
\LSource{1984, Jain--Thompson \cite{JainJPB1984}}
& \LType{T}
& \LMethod{SCE(SEP)-Adiabatic FN}
& \LData{Rot. ICS; vib. DCS, ICS}
& \LEnergy{0.5--10 eV} \\

\LSource{1986, Szmytkowski--Maci\k{a}g \cite{SzmytkowskiCPL1986}}
& \LType{M}
& \LMethod{Linear transmission}
& \LData{TCS}
& \LEnergy{1.3--70 eV} \\

\LSource{1990, Jain \emph{et al.} \cite{JainPRA1990}}
& \LType{T}
& \LMethod{Spherical OP}
& \LData{DCS, ICS, MTCS}
& \LEnergy{50--1000 eV} \\

\LSource{1991, Gianturco \cite{GianturcoJPB1991}}
& \LType{T}
& \LMethod{Model pot. + Adiabatic NA}
& \LData{El., rot. DCS}
& \LEnergy{1--30 eV} \\

\LSource{1992, Zecca \emph{et al.} \cite{ZeccaPRA1992}}
& \LType{M}
& \LMethod{Transmission}
& \LData{TCS}
& \LEnergy{75--4000 eV} \\

\LSource{1992, Jain--Baluja \cite{JainPRA1992}}
& \LType{T}
& \LMethod{Complex OP}
& \LData{Total el. + inel. CS}
& \LEnergy{10--5000 eV} \\

\LSource{1993, Yuan--Zhang \cite{YuanZPDAMC1993}}
& \LType{T}
& \LMethod{PW + rotating-dipole FBA}
& \LData{Vib.-el. DCS, ICS, MTCS}
& \LEnergy{0.1--50 eV} \\

\LSource{1993, Gulley \emph{et al.} \cite{GulleyJPB1993}}
& \LType{M}
& \LMethod{Crossed beam}
& \LData{Vib.-el. DCS, ICS, MTCS; vib. exc.}
& \LEnergy{1--30 eV} \\

\LSource{1994, Greer--Thompson \cite{GreenerJPB1994}}
& \LType{T}
& \LMethod{SCE exact ex. + pol.}
& \LData{Rot.-summed DCS}
& \LEnergy{2--10 eV} \\

\LSource{1995, Machado \emph{et al.} \cite{MachadoJMS1995}}
& \LType{T}
& \LMethod{SVM (FN-SE) + Born}
& \LData{El. DCS, ICS, MTCS}
& \LEnergy{2--50 eV} \\

\LSource{1995, Jiang \emph{et al.} \cite{JiangPRA1995}}
& \LType{T}
& \LMethod{Complex OP + AR}
& \LData{TCS}
& \LEnergy{10--1000 eV} \\

\LSource{1996, Nishimura--Itikawa \cite{NishimuraJPB1996}}
& \LType{T}
& \LMethod{Rot.-sudden/vib. CC}
& \LData{Vib.-el. DCS/ICS; vib. exc.}
& \LEnergy{3--30 eV} \\

\LSource{1997, Joshipura--Vinodkumar \cite{JoshipuraZPD1997}}
& \LType{T}
& \LMethod{Atomic OP + AR}
& \LData{TCS; summed inel.}
& \LEnergy{50--5000 eV} \\

\LSource{1999, Varella \emph{et al.} \cite{Varella111JCP1999}}
& \LType{T}
& \LMethod{SMCPP + Born closure}
& \LData{El. DCS, ICS, MTCS}
& \LEnergy{5--30 eV} \\

\LSource{2001, Karwasz \emph{et al.} \cite{KarwaszRNC2001}}
& \LType{E}
& \LMethod{Critical review}
& \LData{TCS \& integral}
& \LEnergy{Broad} \\

\LSource{2003, Rawat \emph{et al.} \cite{RawatPRA2003}}
& \LType{M/T}
& \LMethod{Relative flow + OP/SVM-DWA}
& \LData{DCS, ICS, MTCS}
& \LEnergy{M: 100--500 eV; T: 0.5--500 eV} \\

\LSource{2003, Szmytkowski \emph{et al.} \cite{SzmytkowskiRPC2003}}
& \LType{M}
& \LMethod{Modified transmission}
& \LData{TCS}
& \LEnergy{6--370 eV} \\

\LSource{2005, Cho \emph{et al.} \cite{ChoJKPS2005}}
& \LType{M}
& \LMethod{Crossed beam}
& \LData{DCS}
& \LEnergy{3--30 eV} \\

\LSource{2006, Shi \emph{et al.} \cite{ShiCPL2006}}
& \LType{T}
& \LMethod{GS additivity rule (HF)}
& \LData{TCS}
& \LEnergy{30--5000 eV} \\

\LSource{2007, Gupta--Baluja \cite{GuptaEPJD2007}}
& \LType{T}
& \LMethod{$R$-matrix \& Born}
& \LData{DCS, ICS, MTCS; exc.}
& \LEnergy{0.025--15 eV} \\

\LSource{2008, Jones \emph{et al.} \cite{JonesPRA2008}}
& \LType{M}
& \LMethod{HR transmission}
& \LData{Integral CS}
& \LEnergy{0.025--10 eV} \\

\LSource{2008, Brescansin \emph{et al.} \cite{BrescansinJPB2008}}
& \LType{M/T}
& \LMethod{Crossed beam \& OP \& SVM-DWA}
& \LData{DCS; ICS, MTCS, ACS, TCS}
& \LEnergy{M: 30--100 eV; T: 1--500 eV} \\

\LSource{2011, Limbachiya \emph{et al.} \cite{LimbachiyaPRA2011}}
& \LType{T}
& \LMethod{$R$-matrix \& SCOP}
& \LData{TCS (el. + exc.}
& \LEnergy{0.01 eV--2 keV} \\

\LSource{2014, Aouchiche \emph{et al.} \cite{AouchicheNIMPRSB2014}}
& \LType{T}
& \LMethod{PW complex OP}
& \LData{El. DDCS, ICS}
& \LEnergy{10 eV--10 keV} \\

\LSource{2020, Mahato \emph{et al.} \cite{MahatoA2020}}
& \LType{T}
& \LMethod{Analytic-static COP}
& \LData{DCS, ICS, MTCS, ACS, TCS}
& \LEnergy{10--500 eV} \\

\LSource{2024, Meena--Purohit \cite{MeenaJPB2024}}
& \LType{T}
& \LMethod{SSIAM + Dirac/OP}
& \LData{el. DCS, ICS, MTCS, VCS}
& \LEnergy{1 eV--1 MeV} \\

% ------------------------------------------------------------------
\hline
\multicolumn{5}{l}{\textbf{Ammonia, $\mathrm{NH_3}$}}\\
\hline
\LSource{1974, Itikawa \cite{ItikawaADNDT1974}}
& \LType{E}
& \LMethod{Early evaluation}
& \LData{MTCS}
& \LEnergy{0.01--10 eV} \\

\LSource{1983, Jain--Thompson \cite{JainJPB1983}}
& \LType{T}
& \LMethod{FN + adiabatic rotation}
& \LData{Rot. el./inel.; summed MTCS}
& \LEnergy{0.01--10 eV} \\

\LSource{1987, Sueoka \emph{et al.} \cite{SueokaJPB1987}}
& \LType{M}
& \LMethod{TOF transmission}
& \LData{TCS}
& \LEnergy{1--400 eV} \\

\LSource{1988, Jain \cite{JainJPB1988}}
& \LType{T}
& \LMethod{Spherical COP}
& \LData{TCS}
& \LEnergy{10--3000 eV} \\

\LSource{1989, Jain \emph{et al.} \cite{JainPRA1989}}
& \LType{T}
& \LMethod{Spherical OP}
& \LData{DCS, ICS, MTCS}
& \LEnergy{100--1000 eV} \\

\LSource{1989, Szmytkowski \emph{et al.} \cite{SzmytkowskiJPB1989}}
& \LType{M}
& \LMethod{Linear transmission}
& \LData{TCS}
& \LEnergy{1--80 eV} \\

\LSource{1989, Pritchard \emph{et al.} \cite{PritchardPRA1989}}
& \LType{T}
& \LMethod{FN-SE Schwinger}
& \LData{DCS, MTCS}
& \LEnergy{2.5--20 eV} \\

\LSource{1991, Gianturco \cite{GianturcoJPB1991}}
& \LType{T}
& \LMethod{FNA + model ex./pol.}
& \LData{El./rot. ICS}
& \LEnergy{0.1--20 eV} \\

\LSource{1992, Rescigno \emph{et al.} \cite{RescignoPRA1992}}
& \LType{T}
& \LMethod{Complex Kohn (SE/pol.-SCF)}
& \LData{El. DCS}
& \LEnergy{1--20 eV} \\

\LSource{1992, Alle \emph{et al.} \cite{AlleJPB1992}}
& \LType{M}
& \LMethod{Crossed beam}
& \LData{Vib.-el. DCS, ICS, MTCS}
& \LEnergy{2--30 eV} \\

\LSource{1992, Yuan--Zhang \cite{YuanPRA1992}}
& \LType{T}
& \LMethod{FBA/central model}
& \LData{Vib.-el. DCS, MTCS, TCS}
& \LEnergy{0.5--20 eV} \\

\LSource{1992, Zecca \emph{et al.} \cite{ZeccaPRA1992}}
& \LType{M}
& \LMethod{Linear transmission}
& \LData{TCS}
& \LEnergy{75--4000 eV} \\

\LSource{1996, Garc\'{i}a--Manero \cite{GarciaJPB1996}}
& \LType{M}
& \LMethod{Transmission beam}
& \LData{TCS}
& \LEnergy{300--5000 eV} \\

\LSource{1997, Liu \emph{et al.} \cite{LiuZPD1997}}
& \LType{T}
& \LMethod{Semiempirical OP + AR}
& \LData{TCS}
& \LEnergy{10--1000 eV} \\

\LSource{1999, Varella \emph{et al.} \cite{Varella110JCP1999}}
& \LType{T}
& \LMethod{SMCPP + ANR/Born}
& \LData{El./rot.-inel. DCS, ICS}
& \LEnergy{7.5--30 eV} \\

\LSource{2001, Karwasz \emph{et al.} \cite{KarwaszRNC2001}}
& \LType{E}
& \LMethod{Critical review}
& \LData{Integral/TCS}
& \LEnergy{Broad} \\

\LSource{2001, Ribeiro \emph{et al.} \cite{RibeiroCPC2001}}
& \LType{T}
& \LMethod{Continued fractions}
& \LData{DCS}
& \LEnergy{6--30 eV} \\

\LSource{2004, Ariyasinghe \emph{et al.} \cite{AriyasingheNIMPRSB2004}}
& \LType{M}
& \LMethod{Linear transmission}
& \LData{TCS}
& \LEnergy{400--4000 eV} \\

\LSource{2005, Lino \cite{LinoRMF2005}}
& \LType{T}
& \LMethod{Schwinger variational}
& \LData{DCS}
& \LEnergy{8.5--30 eV} \\

\LSource{2006, Munjal--Baluja \cite{MunjalPRA2006}}
& \LType{T}
& \LMethod{$R$-matrix + Born}
& \LData{El. DCS/ICS/MTCS; exc.}
& \LEnergy{0.025--20 eV} \\

\LSource{2008, Brescansin \emph{et al.} \cite{BrescansinIJQC2008}}
& \LType{T}
& \LMethod{OP + SVM/CF}
& \LData{DCS, ICS, MTCS}
& \LEnergy{1--30 eV} \\

\LSource{2008, Jones \emph{et al.} \cite{JonesPRA2008}}
& \LType{M}
& \LMethod{HR transmission}
& \LData{Integral CS}
& \LEnergy{0.02--10 eV} \\

\LSource{2011, Limbachiya \emph{et al.} \cite{LimbachiyaPRA2011}}
& \LType{T}
& \LMethod{$R$-matrix + SCOP}
& \LData{TCS (el. + exc.)}
& \LEnergy{0.01 eV--2 keV} \\

\LSource{2014, Homem \emph{et al.} \cite{HomemPRA2014}}
& \LType{M/T}
& \LMethod{Relative flow + SCE/Pad\'e}
& \LData{DCS, ICS, MTCS; calc. ACS/TCS}
& \LEnergy{<50--500 eV} \\

\LSource{2015, Kaur \emph{et al.} \cite{KaurPRA2015}}
& \LType{T}
& \LMethod{Spherical COP}
& \LData{DCS, ICS, MTCS, TCS}
& \LEnergy{0.1--100 eV} \\

\LSource{2017, Itikawa \cite{ItikawaJPCRD2017}}
& \LType{E}
& \LMethod{Critical evaluation}
& \LData{Recommended CSs}
& \LEnergy{Quantity dependent} \\

\LSource{2020, Mahato \emph{et al.} \cite{MahatoJPB2020}}
& \LType{T}
& \LMethod{Analytic-static COP}
& \LData{DCS, ICS, MTCS, ACS, TCS}
& \LEnergy{10--500 eV} \\

\LSource{2022, Akter \emph{et al.} \cite{AkterMP2022}}
& \LType{T}
& \LMethod{IAM(SC) + COP/Dirac}
& \LData{DCS, ICS, MTCS, TCS, inel, ion}
& \LEnergy{1 eV--1 MeV} \\

\LSource{2023, Chen \emph{et al.} \cite{ChenPSST2023}}
& \LType{T}
& \LMethod{Molecular $R$-matrix}
& \LData{El. DCS/ICS/MTCS; ion., exc., diss., vib.}
& \LEnergy{Quantity dependent} \\

\LSource{2023, Snoeckx \emph{et al.} \cite{SnoeckxPSST2023}}
& \LType{T}
& \LMethod{R-matrix method}
& \LData{El./inel. CS set}
& \LEnergy{Quantity dependent} \\

\LSource{2024, Choi \emph{et al.} \cite{ChoiEPJD2024}}
& \LType{M}
& \LMethod{Transmission + Forw corr.}
& \LData{TCS}
& \LEnergy{18.5--90 eV} \\

\LSource{2025, Dinger--Baek \cite{DingerPRA2025}}
& \LType{M/T}
& \LMethod{Crossed beam-Relative flow; IAM-SCAR+I, POLYDCS, DWBA}
& \LData{El. DCS; ionization DDCS}
& \LEnergy{30 eV--1 keV} \\

% ------------------------------------------------------------------
\hline
\multicolumn{5}{l}{\textbf{Phosphine, $\mathrm{PH_3}$}}\\
\hline
\LSource{1992, Winstead \emph{et al.} \cite{WinsteadZPDAMC1992}}
& \LType{T}
& \LMethod{SMC}
& \LData{El. CS; shape resonance}
& \LEnergy{1--40 eV} \\

\LSource{1992, Jain--Baluja \cite{JainPRA1992}}
& \LType{T}
& \LMethod{COP Model}
& \LData{Total (el. + inel.) CS}
& \LEnergy{10--5000 eV} \\

\LSource{1993, Yuan--Zhang \cite{YuanZPDAMC1993}}
& \LType{T}
& \LMethod{PW + rot-dipole FBA}
& \LData{Vib.-el. DCS, ICS, MTCS}
& \LEnergy{0.1--50 eV} \\

\LSource{1996, Bettega \emph{et al.} \cite{BettegaJCP1996}}
& \LType{T}
& \LMethod{SMCPP}
& \LData{El. DCS, ICS}
& \LEnergy{10--30 eV} \\

\LSource{1999, Varella \emph{et al.} \cite{Varella110JCP1999}}
& \LType{T}
& \LMethod{SMCPP + ANR/Born}
& \LData{El./rot.-inel. DCS, ICS}
& \LEnergy{7.5--30 eV} \\

\LSource{2001, Karwasz \emph{et al.} \cite{KarwaszRNC2001}}
& \LType{E}
& \LMethod{Critical review}
& \LData{Integral-data review}
& \LEnergy{Broad} \\

\LSource{2003, Ariyasinghe \emph{et al.} \cite{AriyasinghePRA2003}}
& \LType{M}
& \LMethod{Gas-cell transmission}
& \LData{TCS}
& \LEnergy{90--3500 eV} \\

\LSource{2004, Szmytkowski \emph{et al.} \cite{SzmytkowskiJPB2004}}
& \LType{M}
& \LMethod{Linear transmission}
& \LData{TCS}
& \LEnergy{0.5--370 eV} \\

\LSource{2004, Bettega--Lima \cite{BettegaJPB2004}}
& \LType{T}
& \LMethod{SMCPP (SE + pol.)}
& \LData{El. ICS}
& \LEnergy{0.5--8 eV} \\

\LSource{2007, Munjal--Baluja \cite{MunjalJPB2007}}
& \LType{T}
& \LMethod{$R$-matrix + Born}
& \LData{El. CS}
& \LEnergy{0.025--15 eV} \\

\LSource{2011, Limbachiya \emph{et al.} \cite{LimbachiyaPRA2011}}
& \LType{T}
& \LMethod{$R$-matrix \& SCOP}
& \LData{TCS (el + exc)}
& \LEnergy{0.01 eV--2 keV} \\

\LSource{2015, Kaur \emph{et al.} \cite{KaurPRA2015}}
& \LType{T}
& \LMethod{Spherical COP}
& \LData{DCS, ICS, MTCS, TCS}
& \LEnergy{0.1--100 eV} \\

\LSource{2019, Aouchiche--Medegga \cite{AouchicheRJPCA2019}}
& \LType{T}
& \LMethod{PW spherical OP}
& \LData{El. DCS, ICS}
& \LEnergy{10 eV--20 keV} \\

\LSource{2020, Mahato \emph{et al.} \cite{MahatoJPB2020}}
& \LType{T}
& \LMethod{Analytic-static COP}
& \LData{$e^\pm$ DCS, ICS, MTCS, ACS, TCS}
& \LEnergy{10--500 eV} \\

\end{longtable*}
\endgroup

% The \nocite command causes all entries in a bibliography to be printed out whether or not they are actually referenced in the text. This is appropriate for the sample file to show the different styles of references, but authors most likely will not want to use it.
\nocite{*}

\bibliography{ref}% Produces the bibliography via BibTeX.

@PREAMBLE{
 "\providecommand{\noopsort}[1]{}" 
 # "\providecommand{\singleletter}[1]{#1}%" 
}

@ARTICLE{AryaRSCA2026,
   author       = "Arya, Sudhanshu and Antony, Bobby",
   title        = "{Improved electron-molecule scattering calculations with the relativistic optical-potential method}",
   journal      = "RSC Advances",
   volume       = "16",
   number       = "15",
   pages        = "13548-13558",
   month        = "03",
   year         = "2026",
   issn         = "2046-2069",
   doi          = "10.1039/d6ra00742b",
   url          = "https://doi.org/10.1039/d6ra00742b",
}

@ARTICLE{AryaJAP2025,
   author       = "Arya, Sudhanshu and Antony, Bobby",
   title        = "{Electron scattering from oxetane and thietane and their isomers}",
   journal      = "Journal of Applied Physics",
   volume       = "137",
   number       = "22",
   pages        = "224903",
   month        = "06",
   year         = "2025",
   issn         = "0021-8979",
   doi          = "10.1063/5.0265551",
   url          = "https://doi.org/10.1063/5.0265551",
}

@MISC{CCCBDB2022,
   author       = "Russell D. Johnson",
   title        = "{NIST Computational Chemistry Comparison and Benchmark Database}",
   year         = "2022",
   doi          = "10.18434/T47C7Z",
   note         = "{Available at \url{http://cccbdb.nist.gov/}, (accessed June 25, 2026})",
   howpublished = "{NIST Standard Reference Database Number 101}",
   month        = "22~May",
}

@ARTICLE{DesclauxCPC1975,
   author       = "J.P. Desclaux",
   title        = "{A multiconfiguration relativistic DIRAC-FOCK program}",
   journal      = "Computer Physics Communications",
   volume       = "9",
   number       = "1",
   pages        = "31-45",
   year         = "1975",
   issn         = "0010-4655",
   doi          = "https://doi.org/10.1016/0010-4655(75)90054-5",
   url          = "https://www.sciencedirect.com/science/article/pii/0010465575900545",
}

@ARTICLE{SalvatCPC1991,
   author       = "Francesc Salvat and Ricardo Mayol",
   title        = "{Accurate numerical solution of the Schrödinger and Dirac wave equations for central fields}",
   journal      = "Computer Physics Communications",
   volume       = "62",
   number       = "1",
   pages        = "65-79",
   year         = "1991",
   issn         = "0010-4655",
   doi          = "https://doi.org/10.1016/0010-4655(91)90122-2",
   url          = "https://www.sciencedirect.com/science/article/pii/0010465591901222",
}

@ARTICLE{SalvatCPC2005,
   author       = "Francesc Salvat and Aleksander Jablonski and Cedric J. Powell",
   title        = "{ELSEPA--Dirac partial-wave calculation of elastic scattering of electrons and positrons by atoms, positive ions and molecules}",
   journal      = "Computer Physics Communications",
   volume       = "165",
   number       = "2",
   pages        = "157-190",
   year         = "2005",
   issn         = "0010-4655",
   doi          = "https://doi.org/10.1016/j.cpc.2004.09.006",
   url          = "https://www.sciencedirect.com/science/article/pii/S0010465504004795",
}

@ARTICLE{FurnessJPB1973,
   author       = "J B Furness and I E McCarthy",
   title        = "{Semiphenomenological optical model for electron scattering on atoms}",
   journal      = "Journal of Physics B: Atomic and Molecular Physics",
   volume       = "6",
   number       = "11",
   pages        = "2280",
   month        = "11",
   year         = "1973",
   doi          = "10.1088/0022-3700/6/11/021",
   url          = "https://doi.org/10.1088/0022-3700/6/11/021",
}

@ARTICLE{GianturcoJPB1987,
   author       = "F A Gianturco and S Scialla",
   title        = "{Local approximations of exchange interaction in electron-molecule collisions: the methane molecule}",
   journal      = "Journal of Physics B: Atomic and Molecular Physics",
   volume       = "20",
   number       = "13",
   pages        = "3171",
   month        = "07",
   year         = "1987",
   doi          = "10.1088/0022-3700/20/13/024",
   url          = "https://doi.org/10.1088/0022-3700/20/13/024",
}

@ARTICLE{ZhangJPB1992,
   author       = "Xianzhou Zhang and Jinfeng Sun and Yufang Liu",
   title        = "{A new approach to the correlation polarization potential-low-energy electron elastic scattering by He atoms}",
   journal      = "Journal of Physics B: Atomic, Molecular and Optical Physics",
   volume       = "25",
   number       = "8",
   pages        = "1893",
   month        = "04",
   year         = "1992",
   doi          = "10.1088/0953-4075/25/8/021",
   url          = "https://dx.doi.org/10.1088/0953-4075/25/8/021",
}

@ARTICLE{PerdewPRB1981,
   author       = "Perdew, J. P. and Zunger, Alex",
   title        = "{Self-interaction correction to density-functional approximations for many-electron systems}",
   journal      = "Phys. Rev. B",
   volume       = "23",
   issue        = "10",
   pages        = "5048-5079",
   month        = "05",
   year         = "1981",
   publisher    = "American Physical Society",
   doi          = "10.1103/PhysRevB.23.5048",
   url          = "https://link.aps.org/doi/10.1103/PhysRevB.23.5048",
}

@ARTICLE{StaszewskaAPS1983,
   author       = "Staszewska, Grażyna and Schwenke, David W. and Thirumalai, Devarajan and Truhlar, Donald G.",
   title        = "{Quasifree-scattering model for the imaginary part of the optical potential for electron scattering}",
   journal      = "Phys. Rev. A",
   volume       = "28",
   issue        = "5",
   pages        = "2740-2751",
   month        = "11",
   year         = "1983",
   publisher    = "American Physical Society",
   doi          = "10.1103/PhysRevA.28.2740",
   url          = "https://link.aps.org/doi/10.1103/PhysRevA.28.2740",
}

@ARTICLE{SalvatCPC2015,
   author       = "Francesc Salvat and José M. Fernández-Varea",
   title        = "{RADIAL: A Fortran subroutine package for the solution of the radial Schrödinger and Dirac wave equations}",
   journal      = "Computer Physics Communications",
   volume       = "240",
   pages        = "165-177",
   year         = "2019",
   issn         = "0010-4655",
   doi          = "https://doi.org/10.1016/j.cpc.2019.02.011",
   url          = "https://www.sciencedirect.com/science/article/pii/S0010465519300633",
}

@ARTICLE{JonathanJPB2024,
   author       = "Jonathan Tennyson",
   title        = "{Electron-molecule collision calculations: a primer}",
   journal      = "Journal of Physics B: Atomic, Molecular and Optical Physics",
   volume       = "57",
   number       = "23",
   pages        = "233001",
   month        = "11",
   year         = "2024",
   publisher    = "IOP Publishing",
   doi          = "10.1088/1361-6455/ad4243",
   url          = "https://dx.doi.org/10.1088/1361-6455/ad4243",
}

@ARTICLE{RosettaA2017,
   author       = "Marinković, Bratislav P. and Bredehöft, Jan Hendrik and Vujčić, Veljko and Jevremović, Darko and Mason, Nigel J.",
   title        = "{Rosetta Mission: Electron Scattering Cross Sections—Data Needs and Coverage in BEAMDB Database}",
   journal      = "Atoms",
   volume       = "5",
   number       = "4",
   year         = "2017",
   issn         = "2218-2004",
   doi          = "10.3390/atoms5040046",
   url          = "https://www.mdpi.com/2218-2004/5/4/46",
}

@ARTICLE{BartschatPNAS2016,
   author       = "Klaus Bartschat and Mark J. Kushner",
   title        = "{Electron collisions with atoms, ions, molecules, and surfaces: Fundamental science empowering advances in technology}",
   journal      = "Proceedings of the National Academy of Sciences",
   volume       = "113",
   number       = "26",
   pages        = "7026-7034",
   year         = "2016",
   doi          = "10.1073/pnas.1606132113",
   url          = "https://www.pnas.org/doi/abs/10.1073/pnas.1606132113",
}

@ARTICLE{JonathanPR2010,
   author       = "Jonathan Tennyson",
   title        = "{Electron–molecule collision calculations using the R-matrix method}",
   journal      = "Physics Reports",
   volume       = "491",
   number       = "2",
   pages        = "29-76",
   year         = "2010",
   issn         = "0370-1573",
   doi          = "https://doi.org/10.1016/j.physrep.2010.02.001",
   url          = "https://www.sciencedirect.com/science/article/pii/S0370157310000451",
}

@ARTICLE{BrungerPR2002,
   author       = "Michael J. Brunger and Stephen J. Buckman",
   title        = "{Electron-molecule scattering cross-sections. I. Experimental techniques and data for diatomic molecules}",
   journal      = "Physics Reports",
   volume       = "357",
   number       = "3",
   pages        = "215-458",
   year         = "2002",
   issn         = "0370-1573",
   doi          = "https://doi.org/10.1016/S0370-1573(01)00032-1",
   url          = "https://www.sciencedirect.com/science/article/pii/S0370157301000321",
}

@ARTICLE{BartschatPRP2017,
   author       = "Bartschat, Klaus and Tennyson, Jonathan and Zatsarinny, Oleg",
   title        = "{Quantum-Mechanical Calculations of Cross Sections for Electron Collisions With Atoms and Molecules}",
   journal      = "Plasma Processes and Polymers",
   volume       = "14",
   number       = "1-2",
   pages        = "1600093",
   year         = "2017",
   doi          = "https://doi.org/10.1002/ppap.201600093",
   url          = "https://onlinelibrary.wiley.com/doi/abs/10.1002/ppap.201600093",
}

@INCOLLECTION{WinsteadAP1996,
   author       = "Carl Winstead and Vincent Mckoy",
   title        = "{Highly Parallel Computational Techniques for Electron-Molecule Collisions}",
   booktitle    = "{Advances in Atomic, Molecular, and Optical Physics}",
   editor       = "Benjamin Bederson and Herbert Walther",
   pages        = "183-219",
   year         = "1996",
   publisher    = "{Academic Press}",
   doi          = "https://doi.org/10.1016/S1049-250X(08)60210-4",
   url          = "https://www.sciencedirect.com/science/article/pii/S1049250X08602104",
   issn         = "1049-250X",
   series       = "Advances In Atomic, Molecular, and Optical Physics",
   volume       = "36",
}

@ARTICLE{ZammitJPB2017,
   author       = "Zammit, Mark C and Fursa, Dmitry V and Savage, Jeremy S and Bray, Igor",
   title        = "{Electron- and positron- molecule scattering: development of the molecular convergent close-coupling method}",
   journal      = "Journal of Physics B: Atomic, Molecular and Optical Physics",
   volume       = "50",
   number       = "12",
   pages        = "123001",
   month        = "05",
   year         = "2017",
   publisher    = "IOP Publishing",
   doi          = "10.1088/1361-6455/aa6e74",
   url          = "https://doi.org/10.1088/1361-6455/aa6e74",
}

@INBOOK{Madison1996,
   author       = "Madison, Don H. and Bartschat, Klaus",
   editor       = "Bartschat, Klaus",
   title        = "{The Distorted-Wave Method for Elastic Scattering and Atomic Excitation}",
   pages        = "65--86",
   year         = "1996",
   publisher    = "{Springer Berlin Heidelberg}",
   address      = "Berlin, Heidelberg",
   isbn         = "978-3-642-61010-3",
   doi          = "10.1007/978-3-642-61010-3_4",
   url          = "https://doi.org/10.1007/978-3-642-61010-3_4",
   booktitle    = "Computational Atomic Physics: Electron and Positron Collisions with Atoms and Ions",
}

@ARTICLE{SongJPCRD2021,
   author       = "Song, Mi-Young and Cho, Hyuck and Karwasz, Grzegorz P. and Kokoouline, Viatcheslav and Nakamura, Yoshiharu and Tennyson, Jonathan and Faure, Alexandre and Mason, Nigel J. and Itikawa, Yukikazu",
   title        = "{Cross Sections for Electron Collisions with H2O}",
   journal      = "Journal of Physical and Chemical Reference Data",
   volume       = "50",
   number       = "2",
   pages        = "023103",
   month        = "05",
   year         = "2021",
   issn         = "0047-2689",
   doi          = "10.1063/5.0035315",
   url          = "https://doi.org/10.1063/5.0035315",
}

@ARTICLE{KhakooPRA2008,
   author       = "Khakoo, M. A. and Silva, H. and Muse, J. and Lopes, M. C. A. and Winstead, C. and McKoy, V.",
   title        = "{Electron scattering from ${\mathrm{H}}_{2}\mathrm{O}$: Elastic scattering}",
   journal      = "Phys. Rev. A",
   volume       = "78",
   number        = "5",
   pages        = "052710",
   month        = "Nov",
   year         = "2008",
   publisher    = "American Physical Society",
   doi          = "10.1103/PhysRevA.78.052710",
   url          = "https://doi.org/10.1103/PhysRevA.78.052710",
}

@ARTICLE{MunozPRA2007,
   author       = "Mu\~noz, A. and Oller, J. C. and Blanco, F. and Gorfinkiel, J. D. and Lim\~ao-Vieira, P. and Garc\'{\i}a, G.",
   title        = "{Electron-scattering cross sections and stopping powers in ${\mathrm{H}}_{2}\mathrm{O}$}",
   journal      = "Phys. Rev. A",
   volume       = "76",
   number        = "5",
   pages        = "052707",
   month        = "Nov",
   year         = "2007",
   publisher    = "American Physical Society",
   doi          = "10.1103/PhysRevA.76.052707",
   url          = "https://doi.org/10.1103/PhysRevA.76.052707",
}

@ARTICLE{ChoJPB2004,
   author       = "H Cho and Y S Park and H Tanaka and S J Buckman",
   title        = "{Measurements of elastic electron scattering by water vapour extended to backward angles}",
   journal      = "Journal of Physics B: Atomic, Molecular and Optical Physics",
   volume       = "37",
   number       = "3",
   pages        = "625",
   month        = "jan",
   year         = "2004",
   doi          = "10.1088/0953-4075/37/3/008",
   url          = "https://doi.org/10.1088/0953-4075/37/3/008",
}

@ARTICLE{JohnstoneJPB1991,
   author       = "W M Johnstone and W R Newell",
   title        = "{Absolute vibrationally elastic cross sections for electrons scattered from water molecules between 6 eV and 50 eV}",
   journal      = "Journal of Physics B: Atomic, Molecular and Optical Physics",
   volume       = "24",
   number       = "16",
   pages        = "3633--3643",
   month        = "aug",
   year         = "1991",
   doi          = "10.1088/0953-4075/24/16/015",
   url          = "https://doi.org/10.1088/0953-4075/24/16/015",
}

@ARTICLE{SzmytkowskiCPL1987,
   author       = "Czesław Szmytkowski",
   title        = "{Absolute total cross sections for electron-water vapour scattering}",
   journal      = "Chemical Physics Letters",
   volume       = "136",
   number       = "3",
   pages        = "363-367",
   year         = "1987",
   issn         = "0009-2614",
   doi          = "10.1016/0009-2614(87)80267-1",
   url          = "https://doi.org/10.1016/0009-2614(87)80267-1",
}

@ARTICLE{SueokaJPB1986,
   author       = "O Sueoka and S Mori and Y Katayama",
   title        = "{Total cross sections for electrons and positrons colliding with H2O molecules}",
   journal      = "Journal of Physics B: Atomic and Molecular Physics",
   volume       = "19",
   number       = "10",
   pages        = "L373--L378",
   month        = "may",
   year         = "1986",
   doi          = "10.1088/0022-3700/19/10/008",
   url          = "https://doi.org/10.1088/0022-3700/19/10/008",
}

@ARTICLE{LimbachiyaPRA2011,
   author       = "Limbachiya, Chetan and Vinodkumar, Minaxi and Mason, Nigel",
   title        = "{Calculation of electron-impact rotationally elastic total cross sections for NH${}_{3}$, H${}_{2}$S, and PH${}_{3}$ over the energy range from 0.01 eV to 2 keV}",
   journal      = "Phys. Rev. A",
   volume       = "83",
   number        = "4",
   pages        = "042708",
   month        = "Apr",
   year         = "2011",
   publisher    = "American Physical Society",
   doi          = "10.1103/PhysRevA.83.042708",
   url          = "https://doi.org/10.1103/PhysRevA.83.042708",
}

@ARTICLE{BrescansinJPB2008,
   author       = "Brescansin, L M and Machado, L E and Lee, M-T and Cho, H and Park, Y S",
   title        = "{Absorption effects in intermediate-energy electron scattering by hydrogen sulphide}",
   journal      = "Journal of Physics B: Atomic, Molecular and Optical Physics",
   volume       = "41",
   number       = "18",
   pages        = "185201",
   month        = "sep",
   year         = "2008",
   doi          = "10.1088/0953-4075/41/18/185201",
   url          = "https://doi.org/10.1088/0953-4075/41/18/185201",
}

@ARTICLE{ChoJKPS2005,
   author       = "Hyuck Cho and Sung-Jong Park and Yeunsoo Park",
   title        = "{Measurements of elastic electron scattering by hydrogen sulfide extended to backward angles}",
   journal      = "Journal of the Korean Physical Society",
   volume       = "46",
   number       = "2",
   pages        = "431-434",
   year         = "2005",
   issn         = "0374-4884",
   url          = "https://www.jkps.or.kr/journal/view.html?uid=6848&vmd=Full#n",
}

@ARTICLE{RawatPRA2003,
   author       = "Rawat, P. and Iga, I. and Lee, M.-T. and Brescansin, L. M. and Homem, M. G. P. and Machado, L. E.",
   title        = "{Cross sections for elastic electron--hydrogen sulfide collisions in the low- and intermediate-energy range}",
   journal      = "Phys. Rev. A",
   volume       = "68",
   number        = "5",
   pages        = "052711",
   month        = "Nov",
   year         = "2003",
   publisher    = "American Physical Society",
   doi          = "10.1103/PhysRevA.68.052711",
   url          = "https://doi.org/10.1103/PhysRevA.68.052711",
   numpages     = "9",
}

@ARTICLE{GulleyJPB1993,
   author       = "R J Gulley and M J Brunger and S J Buckman",
   title        = "{The scattering of low energy electrons from hydrogen sulphide}",
   journal      = "Journal of Physics B: Atomic, Molecular and Optical Physics",
   volume       = "26",
   number       = "17",
   pages        = "2913--2925",
   month        = "sep",
   year         = "1993",
   doi          = "10.1088/0953-4075/26/17/023",
   url          = "https://doi.org/10.1088/0953-4075/26/17/023",
}

@ARTICLE{YuanZPDAMC1993,
   author       = "Yuan, Jianmin and Zhang, Zhijie",
   title        = "{Electron scattering with {H2S} and {PH3} molecules}",
   journal      = "Zeitschrift f{\"u}r Physik D Atoms, Molecules and Clusters",
   volume       = "28",
   number       = "3",
   pages        = "207--214",
   month        = "sep",
   year         = "1993",
   publisher    = "Springer",
   doi          = "10.1007/BF01437887",
   url          = "https://doi.org/10.1007/BF01437887",
}

@ARTICLE{ZeccaPRA1992,
   author       = "Zecca, Antonio and Karwasz, Grzegorz P. and Brusa, Roberto S.",
   title        = "{Total-cross-section measurements for electron scattering by ${\mathrm{NH}}_{3}$, ${\mathrm{SiH}}_{4}$, and ${\mathrm{H}}_{2}$S in the intermediate-energy range}",
   journal      = "Phys. Rev. A",
   volume       = "45",
   number        = "5",
   pages        = "2777--2783",
   month        = "Mar",
   year         = "1992",
   publisher    = "American Physical Society",
   doi          = "10.1103/PhysRevA.45.2777",
   url          = "https://doi.org/10.1103/PhysRevA.45.2777",
}

@ARTICLE{JainPRA1990,
   author       = "Jain, Arvind Kumar and Tripathi, A. N. and Jain, Ashok",
   title        = "{Elastic scattering of electrons by ${\mathrm{H}}_{2}$S at 50--1000 eV}",
   journal      = "Phys. Rev. A",
   volume       = "42",
   number        = "11",
   pages        = "6912--6915",
   month        = "Dec",
   year         = "1990",
   publisher    = "American Physical Society",
   doi          = "10.1103/PhysRevA.42.6912",
   url          = "https://doi.org/10.1103/PhysRevA.42.6912",
}

@ARTICLE{SzmytkowskiCPL1986,
   author       = "Czesław Szmytkowski and Krzysztof Maci\k{a}g",
   title        = "{Absolute total electron-scattering cross section of H2S}",
   journal      = "Chemical Physics Letters",
   volume       = "129",
   number       = "3",
   pages        = "321-324",
   year         = "1986",
   issn         = "0009-2614",
   doi          = "10.1016/0009-2614(86)80220-2",
   url          = "https://doi.org/10.1016/0009-2614(86)80220-2",
}

@ARTICLE{DingerPRA2025,
   author       = "Dinger, M. and Baek, W. Y.",
   title        = "{Electron-impact total, multiple-differential elastic, and ionization cross sections of ammonia: A joint theoretical and experimental study}",
   journal      = "Phys. Rev. A",
   volume       = "112",
   number        = "4",
   pages        = "042810",
   month        = "Oct",
   year         = "2025",
   publisher    = "American Physical Society",
   doi          = "10.1103/4wbr-p9wk",
   url          = "https://doi.org/10.1103/4wbr-p9wk",
}

@ARTICLE{ChoiEPJD2024,
   author       = "Choi, Young Rock and Sinha, Nidhi and Song, Mi-Young and Kim, Dae Chul and Kim, Yonghyun and Park, Yeunsoo",
   title        = "{Total cross section measurements of electron scattering from NH3 in the intermediate-energy region}",
   journal      = "The European Physical Journal D",
   volume       = "78",
   number       = "5",
   pages        = "65",
   month        = "05",
   year         = "2024",
   doi          = "10.1140/epjd/s10053-024-00862-2",
   url          = "https://doi.org/10.1140/epjd/s10053-024-00862-2",
}

@ARTICLE{ItikawaJPCRD2017,
   author       = "Itikawa, Yukikazu",
   title        = "{Cross Sections for Electron Collisions with Ammonia}",
   journal      = "Journal of Physical and Chemical Reference Data",
   volume       = "46",
   number       = "4",
   pages        = "043103",
   month        = "12",
   year         = "2017",
   issn         = "0047-2689",
   doi          = "10.1063/1.5001918",
   url          = "https://doi.org/10.1063/1.5001918",
}

@ARTICLE{HomemPRA2014,
   author       = "Homem, M. G. P. and Iga, I. and de Souza, G. L. C. and Zanelato, A. I. and Machado, L. E. and Ferraz, J. R. and dos Santos, A. S. and Brescansin, L. M. and Lucchese, R. R. and Lee, M.-T.",
   title        = "{Electron collisions with ammonia and formamide in the low- and intermediate-energy ranges}",
   journal      = "Phys. Rev. A",
   volume       = "90",
   number        = "6",
   pages        = "062704",
   month        = "Dec",
   year         = "2014",
   publisher    = "American Physical Society",
   doi          = "10.1103/PhysRevA.90.062704",
   url          = "https://doi.org/10.1103/PhysRevA.90.062704",
   numpages     = "9",
}

@ARTICLE{AriyasingheNIMPRSB2004,
   author       = "W.M Ariyasinghe and T Wijeratne and P Palihawadana",
   title        = "{Total electron scattering cross sections of CH4 and NH3 molecules in the energy range 400–4000 eV}",
   journal      = "Nuclear Instruments and Methods in Physics Research Section B: Beam Interactions with Materials and Atoms",
   volume       = "217",
   number       = "3",
   pages        = "389-395",
   year         = "2004",
   issn         = "0168-583X",
   doi          = "10.1016/j.nimb.2003.11.079",
   url          = "https://doi.org/10.1016/j.nimb.2003.11.079",
}

@ARTICLE{GarciaJPB1996,
   author       = "G García and F Manero",
   title        = "{Total cross sections for electron scattering by molecules in the energy range 300 - 5000 eV}",
   journal      = "Journal of Physics B: Atomic, Molecular and Optical Physics",
   volume       = "29",
   number       = "17",
   pages        = "4017",
   month        = "sep",
   year         = "1996",
   doi          = "10.1088/0953-4075/29/17/022",
   url          = "https://doi.org/10.1088/0953-4075/29/17/022",
}

@ARTICLE{AlleJPB1992,
   author       = "D T Alle and R J Gulley and S J Buckman and M J Brunger",
   title        = "{Elastic scattering of low-energy electrons from ammonia}",
   journal      = "Journal of Physics B: Atomic, Molecular and Optical Physics",
   volume       = "25",
   number       = "7",
   pages        = "1533--1542",
   month        = "apr",
   year         = "1992",
   doi          = "10.1088/0953-4075/25/7/023",
   url          = "https://doi.org/10.1088/0953-4075/25/7/023",
}

@ARTICLE{SueokaJPB1987,
   author       = "O Sueoka and S Mori and Y Katayama",
   title        = "{Total cross sections for positron and electron collisions with NH3 and H2O molecules}",
   journal      = "Journal of Physics B: Atomic and Molecular Physics",
   volume       = "20",
   number       = "13",
   pages        = "3237--3246",
   month        = "jul",
   year         = "1987",
   doi          = "10.1088/0022-3700/20/13/028",
   url          = "https://doi.org/10.1088/0022-3700/20/13/028",
}

@ARTICLE{MahatoJPB2020,
   author       = "Mahato, Dibyendu and Sharma, Lalita and Srivastava, Rajesh",
   title        = "{An approach to study electron and positron scattering from NH3 and PH3 using the analytic static potential}",
   journal      = "Journal of Physics B: Atomic, Molecular and Optical Physics",
   volume       = "53",
   number       = "22",
   pages        = "225204",
   month        = "oct",
   year         = "2020",
   publisher    = "IOP Publishing",
   doi          = "10.1088/1361-6455/abb9f4",
   url          = "https://doi.org/10.1088/1361-6455/abb9f4",
}

@ARTICLE{AouchicheRJPCA2019,
   author       = "Aouchiche, H. and Medegga, F.",
   title        = "{Differential and Integral Cross Sections of Electron Elastic Scattering by PH3 Molecule in the Energy Ranging from 10 eV up to 20 keV}",
   journal      = "Russian Journal of Physical Chemistry A",
   volume       = "93",
   number       = "1",
   pages        = "116--124",
   month        = "01",
   year         = "2019",
   doi          = "10.1134/S0036024419010035",
   url          = "https://doi.org/10.1134/S0036024419010035",
}

@ARTICLE{KaurPRA2015,
   author       = "Kaur, Gurpreet and Jain, Arvind Kumar and Mohan, Harsh and Singh, Parjit S. and Sharma, Sunita and Tripathi, A. N.",
   title        = "{Studies of cross sections for collisions of electrons from hydride molecules: $\mathrm{N}{\mathrm{H}}_{3}$ and $\mathrm{P}{\mathrm{H}}_{3}$}",
   journal      = "Phys. Rev. A",
   volume       = "91",
   number        = "2",
   pages        = "022702",
   month        = "Feb",
   year         = "2015",
   publisher    = "American Physical Society",
   doi          = "10.1103/PhysRevA.91.022702",
   url          = "https://doi.org/10.1103/PhysRevA.91.022702",
   numpages     = "10",
}

@ARTICLE{SzmytkowskiJPB2004,
   author       = "Czesław Szmytkowski and Łukasz Kłosowski and Alicja Domaracka and Michał Piotrowicz and Elżbieta Ptasińska-Denga",
   title        = "{Scattering of electrons from hydride molecules: PH3}",
   journal      = "Journal of Physics B: Atomic, Molecular and Optical Physics",
   volume       = "37",
   number       = "9",
   pages        = "1833--1840",
   month        = "apr",
   year         = "2004",
   doi          = "10.1088/0953-4075/37/9/005",
   url          = "https://doi.org/10.1088/0953-4075/37/9/005",
}

@ARTICLE{AriyasinghePRA2003,
   author       = "Ariyasinghe, W. M. and Wijerathna, T. and Powers, D.",
   title        = "{Total electron scattering cross sections of ${\mathrm{PH}}_{3}$ and ${\mathrm{SiH}}_{4}$ molecules in the energy range 90--3500 eV}",
   journal      = "Phys. Rev. A",
   volume       = "68",
   number        = "3",
   pages        = "032708",
   month        = "Sep",
   year         = "2003",
   publisher    = "American Physical Society",
   doi          = "10.1103/PhysRevA.68.032708",
   url          = "https://doi.org/10.1103/PhysRevA.68.032708",
   numpages     = "5",
}

@ARTICLE{WinsteadZPDAMC1992,
   author       = "Winstead, Carl and Sun, Qiyan and McKoy, Vincent and Luiz da Silva Lino, Jorge and Lima, Marco A. P.",
   title        = "{Low-energy electron scattering by methane, arsine and phosphine}",
   journal      = "Zeitschrift für Physik D Atoms, Molecules and Clusters",
   volume       = "24",
   number       = "2",
   pages        = "141--147",
   month        = "06",
   year         = "1992",
   doi          = "10.1007/BF01426698",
   url          = "https://doi.org/10.1007/BF01426698",
}

@ARTICLE{ShorifuddozaEPJD2025,
   author       = "Shorifuddoza, M. and Khandker, Mahmudul H. and Ragimkhanov, G. B. and Khalikova, Z. R. and Das, Pretam K. and Watabe, H. and Haque, A. K. Fazlul and Uddin, M. Alfaz",
   title        = "{Electron and positron scattering by water molecules: cross sections and transport characteristics}",
   journal      = "The European Physical Journal D",
   volume       = "79",
   number       = "6",
   pages        = "62",
   month        = "06",
   year         = "2025",
   doi          = "10.1140/epjd/s10053-025-01012-y",
   url          = "https://doi.org/10.1140/epjd/s10053-025-01012-y",
}

@ARTICLE{BuddeJPD2023,
   author       = "Budde, Maik and Dias, Tiago Cunha and Vialetto, Luca and Pinhão, Nuno and Guerra, Vasco and Silva, Tiago",
   title        = "{Electron-neutral collision cross sections for H2O: II. Anisotropic scattering and assessment of the validity of the two-term approximation}",
   journal      = "Journal of Physics D: Applied Physics",
   volume       = "56",
   number       = "25",
   pages        = "255201",
   month        = "apr",
   year         = "2023",
   publisher    = "IOP Publishing",
   doi          = "10.1088/1361-6463/accaf4",
   url          = "https://doi.org/10.1088/1361-6463/accaf4",
}

@ARTICLE{BuddeJPD2022,
   author       = "Budde, Maik and Cunha Dias, Tiago and Vialetto, Luca and Pinhão, Nuno and Guerra, Vasco and Silva, Tiago",
   title        = "{Electron-neutral collision cross sections for H2O: I. Complete and consistent set}",
   journal      = "Journal of Physics D: Applied Physics",
   volume       = "55",
   number       = "44",
   pages        = "445205",
   month        = "sep",
   year         = "2022",
   publisher    = "IOP Publishing",
   doi          = "10.1088/1361-6463/ac8da3",
   url          = "https://doi.org/10.1088/1361-6463/ac8da3",
}

@ARTICLE{MatsuiEPJD2016,
   author       = "Matsui, Midori and Hoshino, Masamitsu and Kato, Hidetoshi and Ferreira da Silva, Fillipe and Lim{\~a}o-Vieira, Paulo and Tanaka, Hiroshi",
   title        = "{Measuring electron-impact cross sections of water: elastic scattering and electronic excitation of the {$\tilde{a}^{3}B_{1}$} and {$\tilde{A}^{1}B_{1}$} states}",
   journal      = "The European Physical Journal D",
   volume       = "70",
   number       = "4",
   pages        = "77",
   month        = "04",
   year         = "2016",
   doi          = "10.1140/epjd/e2016-60473-6",
   url          = "https://doi.org/10.1140/epjd/e2016-60473-6",
}

@ARTICLE{FaureMNRAS2004,
   author       = "Faure, Alexandre and Gorfinkiel, Jimena D. and Tennyson, Jonathan",
   title        = "{Electron-impact rotational excitation of water}",
   journal      = "Monthly Notices of the Royal Astronomical Society",
   volume       = "347",
   number       = "1",
   pages        = "323-333",
   month        = "01",
   year         = "2004",
   issn         = "0035-8711",
   doi          = "10.1111/j.1365-2966.2004.07209.x",
   url          = "https://doi.org/10.1111/j.1365-2966.2004.07209.x",
}

@ARTICLE{ShynPRA1992,
   author       = "Shyn, T. W. and Grafe, Alan",
   title        = "{Angular distribution of electrons elastically scattered from water vapor}",
   journal      = "Phys. Rev. A",
   volume       = "46",
   issue        = "7",
   pages        = "4406--4409",
   month        = "Oct",
   year         = "1992",
   publisher    = "American Physical Society",
   doi          = "10.1103/PhysRevA.46.4406",
   url          = "https://link.aps.org/doi/10.1103/PhysRevA.46.4406",
   numpages     = "0",
}

@ARTICLE{DanjoJPSJ1985,
   author       = "Danjo ,Atsunori and Nishimura ,Hiroyuki",
   title        = "{Elastic Scattering of Electrons from H2O Molecule}",
   journal      = "Journal of the Physical Society of Japan",
   volume       = "54",
   number       = "4",
   pages        = "1224-1227",
   year         = "1985",
   doi          = "10.1143/JPSJ.54.1224",
   url          = "https://doi.org/10.1143/JPSJ.54.1224",
}

@INCOLLECTION{TakayanagiAAMP1970,
   author       = "Kazuo Takayanagi and Yukikazu Itikawa",
   title        = "{The Rotational Excitation of Molecules by Slow Electrons}",
   booktitle    = "{Advances in Atomic and Molecular Physics}",
   editor       = "D.R. Bates and Immanuel Esterrnan",
   pages        = "105-153",
   year         = "1970",
   publisher    = "Academic Press",
   doi          = "https://doi.org/10.1016/S0065-2199(08)60204-3",
   url          = "https://www.sciencedirect.com/science/article/pii/S0065219908602043",
   issn         = "0065-2199",
   series       = "Advances in Atomic and Molecular Physics",
   volume       = "6",
}

@INCOLLECTION{NorcrossAAMP1982,
   author       = "D.W. Norcross and L.A. Collins",
   title        = "{Recent Developments in the Theory of Electron Scattering by Highly Polar Molecules}",
   booktitle    = "{Advances in Atomic and Molecular Physics}",
   editor       = "David Bates and Benjamin Bederson",
   pages        = "341-397",
   year         = "1982",
   publisher    = "Academic Press",
   doi          = "https://doi.org/10.1016/S0065-2199(08)60245-6",
   url          = "https://www.sciencedirect.com/science/article/pii/S0065219908602456",
   issn         = "0065-2199",
   series       = "Advances in Atomic and Molecular Physics",
   volume       = "18",
}

@ARTICLE{ItikawaTCA2000,
   author       = "Itikawa, Yukikazu",
   title        = "{The Born closure approximation for the scattering amplitude of an electron-molecule collision}",
   journal      = "Theoretical Chemistry Accounts",
   volume       = "105",
   number       = "2",
   pages        = "123--131",
   month        = "12",
   year         = "2000",
   doi          = "10.1007/s002140000189",
   url          = "https://doi.org/10.1007/s002140000189",
}

@ARTICLE{JonesPRA2008,
   author       = "Jones, N. C. and Field, D. and Lunt, S. L. and Ziesel, J.-P.",
   title        = "{Scattering of cold electrons by ammonia, hydrogen sulfide, and carbonyl sulfide}",
   journal      = "Phys. Rev. A",
   volume       = "78",
   number        = "4",
   pages        = "042714",
   month        = "Oct",
   year         = "2008",
   publisher    = "American Physical Society",
   doi          = "10.1103/PhysRevA.78.042714",
   url          = "https://doi.org/10.1103/PhysRevA.78.042714",
   numpages     = "9",
}

@ARTICLE{MahatoA2020,
   author       = "Mahato, Dibyendu and Sharma, Lalita and Srivastava, Rajesh",
   title        = "{Study of Electron and Positron Elastic Scattering from Hydrogen Sulphide Using Analytically Obtained Static Potential}",
   journal      = "Atoms",
   volume       = "8",
   number       = "4",
   year         = "2020",
   issn         = "2218-2004",
   doi          = "10.3390/atoms8040083",
   url          = "https://doi.org/10.3390/atoms8040083",
   pages        = "83",
}

@ARTICLE{MeenaJPB2024,
   author       = "Meena, Sunil K and Purohit, Ghanshyam",
   title        = "{Study of electron and positron elastic scattering cross-sections of astro molecule H2S}",
   journal      = "Journal of Physics B: Atomic, Molecular and Optical Physics",
   volume       = "57",
   number       = "23",
   pages        = "235201",
   month        = "oct",
   year         = "2024",
   publisher    = "IOP Publishing",
   doi          = "10.1088/1361-6455/ad840f",
   url          = "https://doi.org/10.1088/1361-6455/ad840f",
}

@ARTICLE{ChenPSST2023,
   author       = "Chen, Yingqi and Jiang, Xianwu and Yao, Lufeng and Jiang, Wei and Liu, Hainan and Zhang, Ya",
   title        = "{Electron scattering cross sections from NH3: a comprehensive study based on R-matrix method}",
   journal      = "Plasma Sources Science and Technology",
   volume       = "32",
   number       = "4",
   pages        = "045017",
   month        = "may",
   year         = "2023",
   publisher    = "IOP Publishing",
   doi          = "10.1088/1361-6595/acca46",
   url          = "https://doi.org/10.1088/1361-6595/acca46",
}

@ARTICLE{AkterMP2022,
   author       = "Nira Akter and M. Nure Alam Abdullah and M. Shorifuddoza and M. Atiqur R. Patoary and M. Masum Billah and Mahmudul H. Khandker and M. Maaza and Hiroshi Watabe and A. K. F. Haque and M Alfaz Uddin",
   title        = "{Theoretical study of e±-NH3 scattering}",
   journal      = "Molecular Physics",
   volume       = "120",
   number       = "13",
   pages        = "e2097135",
   year         = "2022",
   publisher    = "Taylor \& Francis",
   doi          = "10.1080/00268976.2022.2097135",
   url          = "https://doi.org/10.1080/00268976.2022.2097135",
}

@ARTICLE{MunjalPRA2006,
   author       = "Munjal, Hema and Baluja, K. L.",
   title        = "{Low-energy electron scattering with polar molecule ${\mathrm{NH}}_{3}$ using the $R$-matrix method}",
   journal      = "Phys. Rev. A",
   volume       = "74",
   number        = "3",
   pages        = "032712",
   month        = "Sep",
   year         = "2006",
   publisher    = "American Physical Society",
   doi          = "10.1103/PhysRevA.74.032712",
   url          = "https://doi.org/10.1103/PhysRevA.74.032712",
   numpages     = "7",
}

@ARTICLE{JainPRA1989,
   author       = "Jain, Arvind Kumar and Tripathi, A. N. and Jain, Ashok",
   title        = "{Elastic electron collision cross sections for ammonia molecules in the energy range 0.1--1.0 keV}",
   journal      = "Phys. Rev. A",
   volume       = "39",
   number        = "3",
   pages        = "1537--1540",
   month        = "Feb",
   year         = "1989",
   publisher    = "American Physical Society",
   doi          = "10.1103/PhysRevA.39.1537",
   url          = "https://doi.org/10.1103/PhysRevA.39.1537",
   numpages     = "0",
}

@ARTICLE{BettegaJPB2004,
   author       = "M H F Bettega and M A P Lima",
   title        = "{Electron collisions with the hydrides PH3, AsH3 and SbH3}",
   journal      = "Journal of Physics B: Atomic, Molecular and Optical Physics",
   volume       = "37",
   number       = "19",
   pages        = "3859--3864",
   month        = "sep",
   year         = "2004",
   doi          = "10.1088/0953-4075/37/19/007",
   url          = "https://doi.org/10.1088/0953-4075/37/19/007",
}

@ARTICLE{MunjalJPB2007,
   author       = "Munjal, Hema and Baluja, K L",
   title        = "{Electron-impact study of PH3: an R-matrix approach}",
   journal      = "Journal of Physics B: Atomic, Molecular and Optical Physics",
   volume       = "40",
   number       = "10",
   pages        = "1713",
   month        = "apr",
   year         = "2007",
   doi          = "10.1088/0953-4075/40/10/006",
   url          = "https://doi.org/10.1088/0953-4075/40/10/006",
}

@ARTICLE{JainPRA1992,
   author       = "Jain, Ashok and Baluja, K. L.",
   title        = "{Total (elastic plus inelastic) cross sections for electron scattering from diatomic and polyatomic molecules at 10--5000 eV: ${\mathrm{H}}_{2}$, ${\mathrm{Li}}_{2}$, HF, ${\mathrm{CH}}_{4}$, ${\mathrm{N}}_{2}$, CO, ${\mathrm{C}}_{2}$${\mathrm{H}}_{2}$, HCN, ${\mathrm{O}}_{2}$, HCl, ${\mathrm{H}}_{2}$S, ${\mathrm{PH}}_{3}$, ${\mathrm{SiH}}_{4}$, and ${\mathrm{CO}}_{2}$}",
   journal      = "Phys. Rev. A",
   volume       = "45",
   number        = "1",
   pages        = "202--218",
   month        = "Jan",
   year         = "1992",
   publisher    = "American Physical Society",
   doi          = "10.1103/PhysRevA.45.202",
   url          = "https://doi.org/10.1103/PhysRevA.45.202",
   numpages     = "0",
}

@ARTICLE{Hitran2024,
   author       = "I.E. Gordon and L.S. Rothman and R.J. Hargreaves and F.M. Gomez and T. Bertin and C. Hill and R.V. Kochanov and Y. Tan and P. Wcisło and V. Yu. Makhnev and P.F. Bernath and M. Birk and V. Boudon and A. Campargue and A. Coustenis and B.J. Drouin and R.R. Gamache and J.T. Hodges and D. Jacquemart and E.J. Mlawer and A.V. Nikitin and V.I. Perevalov and M. Rotger and S. Robert and J. Tennyson and G.C. Toon and H. Tran and V.G. Tyuterev and E.M. Adkins and A. Barbe and D.M. Bailey and K. Bielska and L. Bizzocchi and T.A. Blake and C.A. Bowesman and P. Cacciani and P. Čermák and A.G. Császár and L. Denis and S.C. Egbert and O. Egorov and A. Yu. Ermilov and A.J. Fleisher and H. Fleurbaey and A. Foltynowicz and T. Furtenbacher and M. Germann and E.R. Guest and J.J. Harrison and J.-M. Hartmann and A. Hjältén and S.-M. Hu and X. Huang and T.J. Johnson and H. Jóźwiak and S. Kassi and M.V. Khan and F. Kwabia-Tchana and T.J. Lee and D. Lisak and A.-W. Liu and O.M. Lyulin and N.A. Malarich and L. Manceron and A.A. Marinina and S.T. Massie and J. Mascio and E.S. Medvedev and V.V. Meshkov and G. Ch. Mellau and M. Melosso and S.N. Mikhailenko and D. Mondelain and H.S.P. Müller and M. O’Donnell and A. Owens and A. Perrin and O.L. Polyansky and P.L. Raston and Z.D. Reed and M. Rey and C. Richard and G.B. Rieker and C. Röske and S.W. Sharpe and E. Starikova and N. Stolarczyk and A.V. Stolyarov and K. Sung and F. Tamassia and J. Terragni and V.G. Ushakov and S. Vasilchenko and B. Vispoel and K.L. Vodopyanov and G. Wagner and S. Wójtewicz and S.N. Yurchenko and N.F. Zobov",
   title        = "{The HITRAN2024 molecular spectroscopic database}",
   journal      = "Journal of Quantitative Spectroscopy and Radiative Transfer",
   volume       = "353",
   pages        = "109807",
   year         = "2026",
   issn         = "0022-4073",
   doi          = "https://doi.org/10.1016/j.jqsrt.2026.109807",
   url          = "https://www.sciencedirect.com/science/article/pii/S0022407326000014",
}

@ARTICLE{GreenerJPB1994,
   author       = "R Greer and D Thompson",
   title        = "{The scattering of low energy electrons by H2O and H2S}",
   journal      = "Journal of Physics B: Atomic, Molecular and Optical Physics",
   volume       = "27",
   number       = "15",
   pages        = "3533",
   month        = "aug",
   year         = "1994",
   issn         = "",
   publisher    = "",
   doi          = "10.1088/0953-4075/27/15/025",
   url          = "https://doi.org/10.1088/0953-4075/27/15/025",
}

@ARTICLE{JungJPB1982,
   author       = "K Jung and T Antoni and R Muller and K -H Kochem and H Ehrhardt",
   title        = "{Rotational excitation of N2, CO and H2O by low-energy electron collisions}",
   journal      = "Journal of Physics B: Atomic and Molecular Physics",
   volume       = "15",
   number       = "19",
   pages        = "3535",
   month        = "oct",
   year         = "1982",
   issn         = "",
   publisher    = "",
   doi          = "10.1088/0022-3700/15/19/020",
   url          = "https://doi.org/10.1088/0022-3700/15/19/020",
}

@ARTICLE{KataseJPB1986,
   author       = "A Katase and K Ishibashi and Y Matsumoto and T Sakae and S Maezono and E Murakami and K Watanabe and H Maki",
   title        = "{Elastic scattering of electrons by water molecules over the range 100-1000 eV}",
   journal      = "Journal of Physics B: Atomic and Molecular Physics",
   volume       = "19",
   number       = "17",
   pages        = "2715",
   month        = "sep",
   year         = "1986",
   issn         = "",
   publisher    = "",
   doi          = "10.1088/0022-3700/19/17/020",
   url          = "https://doi.org/10.1088/0022-3700/19/17/020",
}

@ARTICLE{ZeccaJPB1987,
   author       = "A Zecca and G Karwasz and S Oss and R Grisenti and R S Brusa",
   title        = "{Total absolute cross sections for electron scattering on H2O at intermediate energies}",
   journal      = "Journal of Physics B: Atomic and Molecular Physics",
   volume       = "20",
   number       = "4",
   pages        = "L133",
   month        = "feb",
   year         = "1987",
   issn         = "",
   publisher    = "",
   doi          = "10.1088/0022-3700/20/4/005",
   url          = "https://doi.org/10.1088/0022-3700/20/4/005",
}

@ARTICLE{GianturcoJCP1987,
   author       = "Gianturco, F. A. and Scialla, S.",
   title        = "{Low‐energy electron scattering from water molecules: A study of angular distributions}",
   journal      = "The Journal of Chemical Physics",
   volume       = "87",
   number       = "11",
   pages        = "6468-6473",
   month        = "12",
   year         = "1987",
   issn         = "0021-9606",
   publisher    = "",
   doi          = "10.1063/1.453428",
   url          = "https://doi.org/10.1063/1.453428",
}

@ARTICLE{ShynPRA1987,
   author       = "Shyn, T. W. and Cho, S. Y.",
   title        = "{Vibrationally elastic scattering cross section of water vapor by electron impact}",
   journal      = "Phys. Rev. A",
   volume       = "36",
   issue        = "11",
   pages        = "5138--5142",
   month        = "Dec",
   year         = "1987",
   issn         = "",
   publisher    = "American Physical Society",
   doi          = "10.1103/PhysRevA.36.5138",
   url          = "https://link.aps.org/doi/10.1103/PhysRevA.36.5138",
   numpages     = "0",
}

@ARTICLE{NishimuraJPSJ1988,
   author       = "Nishimura ,Hiroyuki and Yano ,Kyo",
   title        = "{Total Electron Scattering Cross Sections for Ar,N2, H2O and D2O}",
   journal      = "Journal of the Physical Society of Japan",
   volume       = "57",
   number       = "6",
   pages        = "1951-1956",
   month        = "",
   year         = "1988",
   issn         = "",
   publisher    = "",
   doi          = "10.1143/JPSJ.57.1951",
   url          = "https://doi.org/10.1143/JPSJ.57.1951",
}

@ARTICLE{SaglamJPB1990,
   author       = "Z Saglam and N Aktekin",
   title        = "{Absolute total cross section for electron scattering on water in the energy range 25-300 eV}",
   journal      = "Journal of Physics B: Atomic, Molecular and Optical Physics",
   volume       = "23",
   number       = "9",
   pages        = "1529",
   month        = "may",
   year         = "1990",
   issn         = "",
   publisher    = "",
   doi          = "10.1088/0953-4075/23/9/022",
   url          = "https://doi.org/10.1088/0953-4075/23/9/022",
}

@ARTICLE{SaglamJPB1991,
   author       = "Z Saglam and N Aktekin",
   title        = "{Absolute total cross sections for scattering of electrons by H2O in the energy range 4-20 eV}",
   journal      = "Journal of Physics B: Atomic, Molecular and Optical Physics",
   volume       = "24",
   number       = "15",
   pages        = "3491",
   month        = "aug",
   year         = "1991",
   issn         = "",
   publisher    = "",
   doi          = "10.1088/0953-4075/24/15/016",
   url          = "https://doi.org/10.1088/0953-4075/24/15/016",
}

@ARTICLE{RescignoZPD1992,
   author       = "Rescigno, T. N. and Lengsfield, B. H.",
   title        = "{A fixed-nuclei, ab initio treatment of low-energy electron-H2O scattering}",
   journal      = "Zeitschrift f{\"u}r Physik D Atoms, Molecules and Clusters",
   volume       = "24",
   number       = "2",
   pages        = "117-124",
   month        = "Jun",
   year         = "1992",
   issn         = "1431-5866",
   publisher    = "",
   doi          = "10.1007/BF01426695",
   url          = "https://doi.org/10.1007/BF01426695",
   day          = "01",
}

@ARTICLE{OkamotoJPB1993,
   author       = "Y Okamoto and K Onda and Y Itikawa",
   title        = "{Vibrationally elastic cross sections for electron scattering from water molecules}",
   journal      = "Journal of Physics B: Atomic, Molecular and Optical Physics",
   volume       = "26",
   number       = "4",
   pages        = "745",
   month        = "feb",
   year         = "1993",
   issn         = "",
   publisher    = "",
   doi          = "10.1088/0953-4075/26/4/013",
   url          = "https://doi.org/10.1088/0953-4075/26/4/013",
}

@ARTICLE{GianturcoJCP1998,
   author       = "Gianturco, F. A. and Meloni, S. and Paioletti, P. and Lucchese, R. R. and Sanna, N.",
   title        = "{Low-energy electron scattering from the water molecule: Angular distributions and rotational excitation}",
   journal      = "The Journal of Chemical Physics",
   volume       = "108",
   number       = "10",
   pages        = "4002-4012",
   month        = "03",
   year         = "1998",
   issn         = "0021-9606",
   publisher    = "",
   doi          = "10.1063/1.475349",
   url          = "https://doi.org/10.1063/1.475349",
}

@ARTICLE{Varella111JCP1999,
   author       = "Varella, Márcio T. do N. and Bettega, Márcio H. F. and Lima, Marco A. P. and Ferreira, Luiz G.",
   title        = "{Low-energy electron scattering by H2O, H2S, H2Se, and H2Te}",
   journal      = "The Journal of Chemical Physics",
   volume       = "111",
   number       = "14",
   pages        = "6396-6406",
   month        = "10",
   year         = "1999",
   issn         = "0021-9606",
   publisher    = "",
   doi          = "10.1063/1.480017",
   url          = "https://doi.org/10.1063/1.480017",
}

@ARTICLE{FaureJPB2004,
   author       = "A Faure and J D Gorfinkiel and Jonathan Tennyson",
   title        = "{Low-energy electron collisions with water: elastic and rotationally inelastic scattering}",
   journal      = "Journal of Physics B: Atomic, Molecular and Optical Physics",
   volume       = "37",
   number       = "4",
   pages        = "801",
   month        = "jan",
   year         = "2004",
   issn         = "",
   publisher    = "",
   doi          = "10.1088/0953-4075/37/4/007",
   url          = "https://doi.org/10.1088/0953-4075/37/4/007",
}

@ARTICLE{MachadoEPJD2005,
   author       = "Machado, L. E. and Brescansin, L. M. and Iga, I. and Lee, M.-T.",
   title        = "{Elastic and rotational excitation cross-sections for electron-water collisionsin the low- and intermediate-energy ranges}",
   journal      = "The European Physical Journal D - Atomic, Molecular, Optical and Plasma Physics",
   volume       = "33",
   number       = "2",
   pages        = "193-199",
   month        = "May",
   year         = "2005",
   issn         = "1434-6079",
   publisher    = "",
   doi          = "10.1140/epjd/e2005-00046-4",
   url          = "https://doi.org/10.1140/epjd/e2005-00046-4",
   day          = "01",
}

@ARTICLE{ItikawaJPCRD2005,
   author       = "Itikawa, Yukikazu and Mason, Nigel",
   title        = "{Cross Sections for Electron Collisions with Water Molecules}",
   journal      = "Journal of Physical and Chemical Reference Data",
   volume       = "34",
   number       = "1",
   pages        = "1-22",
   month        = "02",
   year         = "2005",
   issn         = "0047-2689",
   publisher    = "",
   doi          = "10.1063/1.1799251",
   url          = "https://doi.org/10.1063/1.1799251",
}

@ARTICLE{CurikPRL2006,
   author       = "\ifmmode \check{C}\else \v{C}\fi{}ur\'{\i}k, R. and Ziesel, J. P. and Jones, N. C. and Field, T. A. and Field, D.",
   title        = "{Rotational Excitation of ${\mathrm{H}}_{2}\mathrm{O}$ by Cold Electrons}",
   journal      = "Phys. Rev. Lett.",
   volume       = "97",
   issue        = "12",
   pages        = "123202",
   month        = "Sep",
   year         = "2006",
   issn         = "",
   publisher    = "American Physical Society",
   doi          = "10.1103/PhysRevLett.97.123202",
   url          = "https://link.aps.org/doi/10.1103/PhysRevLett.97.123202",
   numpages     = "4",
}

@ARTICLE{SilvaPRL2008,
   author       = "Silva, H. and Muse, J. and Lopes, M. C. A. and Khakoo, M. A.",
   title        = "{Low Energy Elastic Differential Electron Scattering from ${\mathrm{H}}_{2}\mathrm{O}$}",
   journal      = "Phys. Rev. Lett.",
   volume       = "101",
   issue        = "3",
   pages        = "033201",
   month        = "Jul",
   year         = "2008",
   issn         = "",
   publisher    = "American Physical Society",
   doi          = "10.1103/PhysRevLett.101.033201",
   url          = "https://link.aps.org/doi/10.1103/PhysRevLett.101.033201",
   numpages     = "4",
}

@ARTICLE{VinodkumarEPJD2011,
   author       = "Vinodkumar, M. and Limbachiya, C. G. and Joshipura, K. N. and Mason, N. J.",
   title        = "{Electron impact calculations of total elastic cross sections over a wide energy range -- 0.01 eV to 2 keV for CH4, SiH4 and H2O}",
   journal      = "The European Physical Journal D",
   volume       = "61",
   number       = "3",
   pages        = "579-585",
   month        = "Feb",
   year         = "2011",
   issn         = "1434-6079",
   publisher    = "",
   doi          = "10.1140/epjd/e2010-10368-7",
   url          = "https://doi.org/10.1140/epjd/e2010-10368-7",
   day          = "01",
}

@ARTICLE{KadokuraPRL2019,
   author       = "Kadokura, R. and Loreti, A. and K{\"o}v{\'e}r, {\'A}. and Faure, A. and Tennyson, J. and Laricchia, G.",
   title        = "{Angle-Resolved Electron Scattering from ${\mathrm{H}}_{2}\mathrm{O}$ near 0\ifmmode^\circ\else\textdegree\fi{}}",
   journal      = "Phys. Rev. Lett.",
   volume       = "123",
   issue        = "3",
   pages        = "033401",
   month        = "Jul",
   year         = "2019",
   issn         = "",
   publisher    = "American Physical Society",
   doi          = "10.1103/PhysRevLett.123.033401",
   url          = "https://link.aps.org/doi/10.1103/PhysRevLett.123.033401",
   numpages     = "5",
}

@ARTICLE{TriggianiFM2023,
   author       = "Triggiani, Francesca and Morresi, Tommaso and Taioli, Simone and Simonucci, Stefano",
   title        = "{Elastic scattering of electrons by water: An ab initio study}",
   journal      = "Frontiers in Materials",
   volume       = "10",
   issue        = "",
   pages        = "",
   month        = "",
   year         = "2023",
   issn         = "2296-8016",
   publisher    = "",
   doi          = "10.3389/fmats.2023.1145261",
   url          = "https://www.frontiersin.org/journals/materials/articles/10.3389/fmats.2023.1145261",
}

@ARTICLE{JainJPB1984,
   author       = "A Jain and D G Thompson",
   title        = "{Low-energy electron scattering by H2S molecules: elastic, rotational and vibrational excitation}",
   journal      = "Journal of Physics B: Atomic and Molecular Physics",
   volume       = "17",
   number       = "3",
   pages        = "443",
   month        = "feb",
   year         = "1984",
   issn         = "",
   publisher    = "",
   doi          = "10.1088/0022-3700/17/3/014",
   url          = "https://doi.org/10.1088/0022-3700/17/3/014",
}

@ARTICLE{GianturcoJPB1991,
   author       = "F A Gianturco",
   title        = "{Ab initio model calculations to treat electron scattering from polar polyatomic targets: H2S and NH3}",
   journal      = "Journal of Physics B: Atomic, Molecular and Optical Physics",
   volume       = "24",
   number       = "21",
   pages        = "4627",
   month        = "nov",
   year         = "1991",
   issn         = "",
   publisher    = "",
   doi          = "10.1088/0953-4075/24/21/014",
   url          = "https://doi.org/10.1088/0953-4075/24/21/014",
}

@ARTICLE{KarwaszRNC2001,
   author       = "Karwasz, Grzegorz P. and Brusa, Roberto S. and Zecca, Antonio",
   title        = "{One century of experiments on electron-atom and molecule scattering: a critical review of integral cross-sections}",
   journal      = "La Rivista del Nuovo Cimento",
   volume       = "24",
   number       = "1",
   pages        = "1-118",
   month        = "Jan",
   year         = "2001",
   issn         = "1826-9850",
   publisher    = "",
   doi          = "10.1007/BF03548893",
   url          = "https://doi.org/10.1007/BF03548893",
   day          = "01",
}

@ARTICLE{AouchicheNIMPRSB2014,
   author       = "H. Aouchiche and F. Medegga and C. Champion",
   title        = "{Doubly differential and integral cross sections for electron elastic scattering by hydrogen sulfide}",
   journal      = "Nuclear Instruments and Methods in Physics Research Section B: Beam Interactions with Materials and Atoms",
   volume       = "333",
   issue        = "",
   pages        = "113-119",
   month        = "",
   year         = "2014",
   issn         = "0168-583X",
   publisher    = "",
   doi          = "https://doi.org/10.1016/j.nimb.2014.04.018",
   url          = "https://www.sciencedirect.com/science/article/pii/S0168583X14005205",
}

@ARTICLE{MachadoJMS1995,
   author       = "Luiz E. Machado and Emerson P. Leal and Lee Mu-Tao and Luiz M. Brescansin",
   title        = "{Low energy elastic scattering of electrons by hydrogen sulphide molecules}",
   journal      = "Journal of Molecular Structure: THEOCHEM",
   volume       = "335",
   number       = "1",
   pages        = "37-43",
   month        = "",
   year         = "1995",
   issn         = "0166-1280",
   publisher    = "",
   doi          = "https://doi.org/10.1016/0166-1280(94)03980-Y",
   url          = "https://www.sciencedirect.com/science/article/pii/016612809403980Y",
   note         = "Proceedings of the VIIth Brazilian Symposium on Theoretical Chemistry",
}

@ARTICLE{JiangPRA1995,
   author       = "Jiang, Yuhai and Sun, Jinfeng and Wan, Linde",
   title        = "{Total cross sections for electron scattering by polyatomic molecules at 10--1000 eV: ${\mathrm{H}}_{2}$S, ${\mathrm{SiH}}_{4}$, ${\mathrm{CH}}_{4}$, ${\mathrm{CF}}_{4}$, ${\mathrm{CCl}}_{4}$, ${\mathrm{SF}}_{6}$, ${\mathrm{C}}_{2}$${\mathrm{H}}_{4}$, ${\mathrm{CCl}}_{3}$F, ${\mathrm{CClF}}_{3}$, and ${\mathrm{CCl}}_{2}$${\mathrm{F}}_{2}$}",
   journal      = "Phys. Rev. A",
   volume       = "52",
   issue        = "1",
   pages        = "398--403",
   month        = "Jul",
   year         = "1995",
   issn         = "",
   publisher    = "American Physical Society",
   doi          = "10.1103/PhysRevA.52.398",
   url          = "https://link.aps.org/doi/10.1103/PhysRevA.52.398",
   numpages     = "0",
}

@ARTICLE{NishimuraJPB1996,
   author       = "Tamio Nishimura and Yukikazu Itikawa",
   title        = "{Vibrationally elastic and inelastic scattering of electrons by hydrogen sulphide molecules}",
   journal      = "Journal of Physics B: Atomic, Molecular and Optical Physics",
   volume       = "29",
   number       = "18",
   pages        = "4213",
   month        = "sep",
   year         = "1996",
   issn         = "",
   publisher    = "",
   doi          = "10.1088/0953-4075/29/18/017",
   url          = "https://doi.org/10.1088/0953-4075/29/18/017",
}

@ARTICLE{JoshipuraZPD1997,
   author       = "Joshipura, K. N. and Vinodkumar, Minaxi",
   title        = "{Total cross sections of electron collisions with S atoms; H2S, OCS and SO2 molecules (Ei$\geq$50 eV)}",
   journal      = "Zeitschrift f{\"u}r Physik D Atoms, Molecules and Clusters",
   volume       = "41",
   number       = "2",
   pages        = "133-137",
   month        = "Jun",
   year         = "1997",
   issn         = "1431-5866",
   publisher    = "",
   doi          = "10.1007/s004600050301",
   url          = "https://doi.org/10.1007/s004600050301",
   day          = "01",
}

@ARTICLE{SzmytkowskiRPC2003,
   author       = "Czesław Szmytkowski and Paweł Możejko and Andrzej Krzysztofowicz",
   title        = "{Measurements of absolute total cross sections for electron scattering from triatomic polar molecules: SO2 and H2S}",
   journal      = "Radiation Physics and Chemistry",
   volume       = "68",
   number       = "1",
   pages        = "307-311",
   month        = "",
   year         = "2003",
   issn         = "0969-806X",
   publisher    = "",
   doi          = "https://doi.org/10.1016/S0969-806X(03)00306-2",
   url          = "https://www.sciencedirect.com/science/article/pii/S0969806X03003062",
   note         = "Proceedings of the 2nd Conference on Elementary Processess in Atomic Systems, CEPAS 2002. Gdansk, Poland, 2-6 September 2002",
}

@ARTICLE{ShiCPL2006,
   author       = "D.H. Shi and J.F. Sun and Y.F. Liu and Z.L. Zhu",
   title        = "{Total cross section for electron scattering by H2S and C2F6 at 30–5000eV: Considering the geometric shielding effect}",
   journal      = "Chemical Physics Letters",
   volume       = "429",
   number       = "1",
   pages        = "271-275",
   month        = "",
   year         = "2006",
   issn         = "0009-2614",
   publisher    = "",
   doi          = "https://doi.org/10.1016/j.cplett.2006.07.055",
   url          = "https://www.sciencedirect.com/science/article/pii/S0009261406010712",
}

@ARTICLE{GuptaEPJD2007,
   author       = "Gupta, M. and Baluja, K. L.",
   title        = "{Application of R-matrix method to electron-H2S collisions in the low energy range}",
   journal      = "The European Physical Journal D",
   volume       = "41",
   number       = "3",
   pages        = "475-483",
   month        = "Mar",
   year         = "2007",
   issn         = "1434-6079",
   publisher    = "",
   doi          = "10.1140/epjd/e2006-00245-5",
   url          = "https://doi.org/10.1140/epjd/e2006-00245-5",
   day          = "01",
}

@ARTICLE{ItikawaADNDT1974,
   author       = "Yukikazu Itikawa",
   title        = "{Momentum-transfer cross sections for electron collisions with atoms and molecules}",
   journal      = "Atomic Data and Nuclear Data Tables",
   volume       = "14",
   number       = "1",
   pages        = "1-10",
   month        = "",
   year         = "1974",
   issn         = "0092-640X",
   publisher    = "",
   doi          = "https://doi.org/10.1016/S0092-640X(74)80026-4",
   url          = "https://www.sciencedirect.com/science/article/pii/S0092640X74800264",
}

@ARTICLE{JainJPB1983,
   author       = "A Jain and D G Thompson",
   title        = "{Momentum transfer cross sections for the low-energy electron scattering by NH3 molecules}",
   journal      = "Journal of Physics B: Atomic and Molecular Physics",
   volume       = "16",
   number       = "14",
   pages        = "2593",
   month        = "jul",
   year         = "1983",
   issn         = "",
   publisher    = "",
   doi          = "10.1088/0022-3700/16/14/019",
   url          = "https://doi.org/10.1088/0022-3700/16/14/019",
}

@ARTICLE{JainJPB1988,
   author       = "A Jain",
   title        = "{Theoretical study of the total (elastic+inelastic) cross sections for electron -H2O (NH3) scattering at 10-3000 eV}",
   journal      = "Journal of Physics B: Atomic, Molecular and Optical Physics",
   volume       = "21",
   number       = "5",
   pages        = "905",
   month        = "mar",
   year         = "1988",
   issn         = "",
   publisher    = "",
   doi          = "10.1088/0953-4075/21/5/018",
   url          = "https://doi.org/10.1088/0953-4075/21/5/018",
}

@ARTICLE{SzmytkowskiJPB1989,
   author       = "C Szmytkowski and K Maciag and G Karwasz and D Filipovic",
   title        = "{Total absolute cross section measurements for electron scattering on NH3, OCS and N2O}",
   journal      = "Journal of Physics B: Atomic, Molecular and Optical Physics",
   volume       = "22",
   number       = "3",
   pages        = "525",
   month        = "feb",
   year         = "1989",
   issn         = "",
   publisher    = "",
   doi          = "10.1088/0953-4075/22/3/015",
   url          = "https://doi.org/10.1088/0953-4075/22/3/015",
}

@ARTICLE{PritchardPRA1989,
   author       = "Pritchard, H. P. and Lima, M. A. P. and McKoy, V.",
   title        = "{Studies of elastic e-${\mathrm{NH}}_{3}$ collisions}",
   journal      = "Phys. Rev. A",
   volume       = "39",
   issue        = "5",
   pages        = "2392--2396",
   month        = "Mar",
   year         = "1989",
   issn         = "",
   publisher    = "American Physical Society",
   doi          = "10.1103/PhysRevA.39.2392",
   url          = "https://link.aps.org/doi/10.1103/PhysRevA.39.2392",
   numpages     = "0",
}

@ARTICLE{RescignoPRA1992,
   author       = "Rescigno, T. N. and Lengsfield, B. H. and McCurdy, C. W. and Parker, S. D.",
   title        = "{Ab initio description of polarization in low-energy electron collisions with polar molecules: Application to electron-${\mathrm{NH}}_{3}$ scattering}",
   journal      = "Phys. Rev. A",
   volume       = "45",
   issue        = "11",
   pages        = "7800--7809",
   month        = "Jun",
   year         = "1992",
   issn         = "",
   publisher    = "American Physical Society",
   doi          = "10.1103/PhysRevA.45.7800",
   url          = "https://link.aps.org/doi/10.1103/PhysRevA.45.7800",
   numpages     = "0",
}

@ARTICLE{YuanPRA1992,
   author       = "Yuan, Jianmin and Zhang, Zhijie",
   title        = "{Low-energy electron scattering with ${\mathrm{H}}_{2}$O and ${\mathrm{NH}}_{3}$ molecules}",
   journal      = "Phys. Rev. A",
   volume       = "45",
   issue        = "7",
   pages        = "4565--4571",
   month        = "Apr",
   year         = "1992",
   issn         = "",
   publisher    = "American Physical Society",
   doi          = "10.1103/PhysRevA.45.4565",
   url          = "https://link.aps.org/doi/10.1103/PhysRevA.45.4565",
   numpages     = "0",
}

@ARTICLE{LiuZPD1997,
   author       = "Liu, Yufang and Sun, Jinfeng and Li, Zhenxin and Jiang, Yuhai and Wan, Lingde",
   title        = "{Total cross sections for electron scattering from molecules: NH3 and H2O}",
   journal      = "Zeitschrift f{\"u}r Physik D Atoms, Molecules and Clusters",
   volume       = "42",
   number       = "1",
   pages        = "45-48",
   month        = "Mar",
   year         = "1997",
   issn         = "1431-5866",
   publisher    = "",
   doi          = "10.1007/s004600050330",
   url          = "https://doi.org/10.1007/s004600050330",
   day          = "01",
}

@ARTICLE{Varella110JCP1999,
   author       = "Varella, Márcio T. do N. and Bettega, Márcio H. F. and da Silva, Antônio J. R. and Lima, Marco A. P.",
   title        = "{Cross sections for rotational excitations of NH3, PH3, AsH3, and SbH3 by electron impact}",
   journal      = "The Journal of Chemical Physics",
   volume       = "110",
   number       = "5",
   pages        = "2452-2464",
   month        = "02",
   year         = "1999",
   issn         = "0021-9606",
   publisher    = "",
   doi          = "10.1063/1.477951",
   url          = "https://doi.org/10.1063/1.477951",
}

@ARTICLE{RibeiroCPC2001,
   author       = "E.M.S. Ribeiro and L.E. Machado and M.-T. Lee and L.M. Brescansin",
   title        = "{Application of the method of continued fractions to electron scattering by polyatomic molecules}",
   journal      = "Computer Physics Communications",
   volume       = "136",
   number       = "1",
   pages        = "117-125",
   month        = "",
   year         = "2001",
   issn         = "0010-4655",
   publisher    = "",
   doi          = "https://doi.org/10.1016/S0010-4655(01)00151-5",
   url          = "https://www.sciencedirect.com/science/article/pii/S0010465501001515",
}

@ARTICLE{LinoRMF2005,
   author       = "Jorge L. S. Lino",
   title        = "{Elastic scattering of low-energy electrons from ammonia}",
   journal      = "Revista Mexicana de Física",
   volume       = "51",
   number       = "1",
   pages        = "100--103",
   month        = "",
   year         = "2005",
   issn         = "0035-001X",
   publisher    = "",
   doi          = "",
   url          = "https://classic.scielo.org.mx/scielo.php?script=sci_abstract&pid=S0035-001X2005000100015&lng=en&nrm=iso",
}

@ARTICLE{BrescansinIJQC2008,
   author       = "Brescansin, L. M. and Lee, M.-T. and Machado, L. E.",
   title        = "{A comparative study on low-energy elastic electron-NHx (x = 1,2,3) collisions}",
   journal      = "International Journal of Quantum Chemistry",
   volume       = "108",
   number       = "13",
   pages        = "2312-2317",
   month        = "",
   year         = "2008",
   issn         = "",
   publisher    = "",
   doi          = "https://doi.org/10.1002/qua.21607",
   url          = "https://onlinelibrary.wiley.com/doi/abs/10.1002/qua.21607",
}

@ARTICLE{SnoeckxPSST2023,
   author       = "Snoeckx, Ramses and Tennyson, Jonathan and Cha, Min Suk",
   title        = "{Theoretical cross sections for electron collisions relevant for ammonia discharges part 1: NH3, NH2, and NH}",
   journal      = "Plasma Sources Science and Technology",
   volume       = "32",
   number       = "11",
   pages        = "115020",
   month        = "nov",
   year         = "2023",
   issn         = "",
   publisher    = "IOP Publishing",
   doi          = "10.1088/1361-6595/ad0d07",
   url          = "https://doi.org/10.1088/1361-6595/ad0d07",
}

@ARTICLE{BettegaJCP1996,
   author       = "Bettega, M. H. F. and Lima, M. A. P. and Ferreira, L. G.",
   title        = "{Cross sections for collisions of low‐energy electrons with the hydrides PH3, AsH3, SbH3, SnH4, TeH2, and HI}",
   journal      = "The Journal of Chemical Physics",
   volume       = "105",
   number       = "3",
   pages        = "1029-1033",
   month        = "07",
   year         = "1996",
   issn         = "0021-9606",
   publisher    = "",
   doi          = "10.1063/1.471947",
   url          = "https://doi.org/10.1063/1.471947",
}

\end{document}